\documentclass[onecolumn]{emulateapj}
\usepackage{multirow}
\usepackage{lscape}
\newcommand{\be}{\begin{equation}}
\newcommand{\ee}{\end{equation}}
\newcommand{\bea}{\begin{eqnarray}}
\newcommand{\eea}{\end{eqnarray}}
\shortauthors{Yan \& Lazarian}
\begin{document}
\title{Atomic alignment and Diagnostics of Magnetic Fields in Diffuse Media}
\author{Huirong Yan\altaffilmark{1} \& A. Lazarian\altaffilmark{2}}
\altaffiltext{1}{Canadian Institute for Theoretical Astrophysics, 60 St. George, Toronto, ON M5S3H8, yanhr@cita.utoronto.ca}
\altaffiltext{2}{Department of Astronomy, University of Wisconsin, 475 N. Charter St., Madison, WI 53706, alazarian@wisc.edu}

\begin{abstract}
We continue our studies of alignment of atoms by radiation in diffuse media and their realignment by the ambient magnetic field.
We understand atomic alignment as the alignment of atoms or ions in their ground or metastable states that have more than two fine or
hyperfine sublevels. In particular we consider the alignment of in interstellar  and circumstellar media, with the goal  of developing new 
diagnostics of magnetic fields in these environments.  We provide predictions of the polarization that arises from astrophysically important aligned atoms (ions) with fine structure of the ground level, namely, OI and SII and Ti II. Unlike our earlier papers which dealt with weak fields only, a part of our current paper is devoted to the studies of atomic alignment when magnetic fields get strong enough to affect the 
emission from the excited level.  This is a regime of Hanle effect, but modified by the atomic alignment of the atomic ground state. We discuss also ground Hanle effect where the magnetic splitting is comparable to pumping rate. Using an example of emission and absorption lines of SII ion we  demonstrate how polarimetric studies can probe magnetic fields in circumstellar regions and accretion disks. In addition, we show that
atomic alignment induced by anisotropic radiation can induce substantial variations of magnetic dipole transitions within the ground state, thus affecting abundance studies based on this emission. Moreover, we show that the radio emission arising this way is polarized, which provides a new way to study magnetic fields, e.g. at the epoch of Universe reionization.

\end{abstract}


\section{Introduction}



Magnetic fields have important or dominant effects in many areas of astrophysics, but have been very difficult to quantify. We believe that
atomic alignment is a new promising way of studying magnetic fields in the radiation-dominated environments. In fact, this diagnostics is
far more sensitive then those based on the Zeeman effect.
The basic idea of atomic alignment is simple: consider {\em atoms} or {\em ions} irradiated by a nearby star and embedded in a magnetic field. Anisotropic radiation pump the atoms differentially from different magnetic sublevels, resulting in over- or under-populations of the atomic states of
various magnetic quantum numbers, M. The non-LTE populations produce observable polarization in the absorption or emission line involved in
the interaction. The sensitivity of aligned atoms to magnetic field arises from atoms precessing in the magnetic field. It is obvious that to be
sensitive to weak magnetic fields, the atomic level should be long-lived. The ground and metastable atomic levels with fine and hyperfine structure
correspond to this requirement. As a result, these species are the focus of our study.

This paper is the third in the series of papers where we study atomic alignment. Rigorously speaking, ``atomic alignment" means that the angular momentum of the atoms is not distributed isotropically: in ``fluorescent alignment" this is due to angular momentum pumping by anisotropic illuminating light; ``magnetic realignment" occurs when this alignment is modified by precession in a magnetic field. In the paper, however, we shall use, whenever this does not cause confusion, the terms "alignment" and "realignment" interchangeably. 
In Yan \& Lazarian (2006, henceforth YLa) we dealt with polarization of absorption lines of the atoms (ions) having fine splitting, while in Yan \& Lazarian (2007, henceforth YLb) we dealt with polarization of absorption and emission lines arising from
atoms (ions) with hyperfine splitting. Although both papers were focused on developing new techniques for studies of magnetic fields in diffuse media, in both papers we discussed other effects of atomic alignment, e.g. the variations of absorption and emission intensities induced by atomic alignment. We showed that atomic alignment opens wide avenues for studies of weak magnetic fields, i.e. fields $\lesssim10^{-5}$~G, which makes it a unique tool for magnetic field studies in many diffuse astrophysical environments. In YLab we discussed both cases of interplanetary,
interstellar, circumstellar and extragalactic studies employing atomic alignment. 

In this paper, we, first of all, fill in the gap and discuss the polarization of emission lines from atoms (ions) with fine splitting. 
As in our earlier papers, we show that this provides a technique to study the {\em geometry} of weak magnetic fields.
Similar to our earlier paper, henceforth, for the sake of simplicity,
we talk about {\em atomic alignment} and alignment of {\em atoms} even when we deal with {\em ions}.  More importantly,
we discuss particular regimes that are also sensitive to the magnetic field intensity.

It is well known, that magnetic fields play essential roles in the dynamics of circumstellar discs and have a strong influence on the evolution of pre-main sequence stars, on accretions, outflows, stellar winds etc. Magnetically controlled accretion has been confirmed for the low mass T Tauri stars (see Bertout 1989, Camenzind 1990; K\"onigl 1991; Cameron \& Campbell 1993; Shu et al. 1994; Johns-Krull et al. 1999). For intermediate mass stars, particularly Herbig Ae/Be stars, it is believed that a global magnetic field with a complex configuration is responsible for various peculiar phenomena observed for Herbig stars (e.g., Hubrig et al. 2007). For massive stars, magnetic field affects the stellar wind and causes substantial deviation from the spherical geometry (see Ignace, Nordsieck , \& Cassinelii 1999). Zeeman effect, including both Zeeman broadening and the circular polarization is the main diagnostics adopted by the community. However, the measurements based on Zeeman effect have serious limitations: first of all, Zeeman effect is only applicable to very strong magnetic field $\gtrsim$kG; secondly, the Zeeman splitting will be smeared out by various line-broadening    
processes, rotational, winds, accretion, turbulence, especially in the highly dynamical discs; and moreover Zeeman effect requires the geometry of the field to be particularly simple, as, otherwise, there will be significant cancellations of polarization between the Zeeman components. 

On the other hand, the spectropolarimetry based on atomic alignment and Hanle effect does not have these constrains. Hanle effect can measure magnetic field down to 1G. And both direction and strength can be measured\footnote{Note, that, unlike the weak magnetic field regime, which is uniquely relevant to diffuse medium, the stronger field regime, i.e. Hanle
regime, is relevant to Solar studies. 
The research into Hanle effect resulted in important change of the views on solar chromospheres (see Bommier \& Sahal-Brechot 1978, Landi Degl'Innocenti 1983, 1984, 1999, Stenflo \& Keller 1997, Trujillo Bueno 1999). Studies of magnetic fields with the Hanle effect that take into account atomic alignment have been successful
from the ground and resulted in calls for space-based mission to study the Sun polarization with 
more lines (Trujillo Bueno et al. 2005).}. 


However, if Hanle effect affects the ground state, again very {\em weak} magnetic fields can be studied. 
In the paper we also study ground state Hanle effect, which takes place when the precession rate gets comparable with the pumping rate of the source. As the result, polarization of both {\em absorption} and {\em emission} will be influenced according to the topology and the direction of magnetic field. As far as we know, no {\em emission} lines from species with fine structure of the ground level have been reported, to be observed to emanate from interstellar or circumstellar gas. {\em Absorption}
lines are readily available, however (see Morton 1975).  The polarization of absorption lines that we study here is thus currently look more
promising. Lower-level Hanle effect had been noticed in the analysis of HeI $D_3$ and Na $D_2$ polarization for solar prominences (Bommier 1980, Landolfi \& Landi Degl'Innocenti 1985). However, given the intense pumping in solar case, the lower level Hanle regime is not well separated from the Hanle regime. For the weaker pumping by more distant source we study here, the lower-level Hanle regime become distinctive. Moreover, we shall demonstrate a new diagnostic through the lower level Hanle effect with the polarization of {\em absorption} lines. 

Atomic alignment in diffuse media demonstrates a real profusion of interesting physical effects, all of which we cannot cover in the paper. Nevertheless, we briefly touch upon an issue of the influence of atomic alignment on the measurements of element abundances, in particular, using magnetic dipole lines. Although physically different, this effect can be calculated with the same
formalism as in the rest of the paper, which justifies its discussion within the same publication.  

In what follows we outline the different regimes of atomic alignment and Hanle effect in \S2. The basic formalism for studying the problem is described in \S3. Following this, we  discuss in \S4 the magnetic realignment regime where complementary studies to YLa,b, namely, on polarized absorption and emission are provided. The polarizations in both upper level and lower level Hanle regimes are presented in \S5. The atomic alignment is included there, and has a notable influence to the predictions of polarizations. Example studies for a circumstellar disk with poloidal or toroidal magnetic field are performed in \S6. In \S7, we discuss the influence of atomic alignment to radio lines. It is demonstrated that atomic alignment not only modulates the distortion of CMB arising from optically pumped lines, therefore affects, e.g. abundance estimates in early universe, but also causes the line to be polarized, thus suggesting a new promising technique to detect magnetic field in the epoch of reionization. Discussion and summary are provided in, respectively, \S\S8 and 9.     

\section{Basics of atomic alignment and Hanle effect}

\subsection{Relevant Rates}

Considering the various rates (see Table \ref{difftime})
involved. Those are 1) the rate of the Larmor precession, $\nu_L$, 2)
the rate of the optical pumping, $R_F$, 3) the rate of collisional
randomization, $\tau_c^{-1}$, 4) the rate of the transition within
ground state, $\tau^{-1}_T$.  In many cases $\nu_L>R_F>\tau_c^{-1},
\tau_T^{-1}$. Other relations are possible, however. If
$\tau_T^{-1}>R_F$, the transitions within the sublevels of ground
state need to be taken into account and relative distribution among
them will be modified. Since emission is spherically symmetric, the
angular momentum in the atomic system is preserved and thus alignment
persists in this case. In the case $\nu_L<R_F$, the magnetic field
does not affect the atomic occupations and atoms are aligned with
respect to the direction of radiation. From the expressions in
Table~\ref{difftime}, we see, for instance, that magnetic field can
realign CII at a distance $r> 0.8$Au from an O star if the magnetic
field strength $\sim 5\times 10^{-4}$G.

If the Larmor precession rate $\nu_L$ is comparable to any of the
other rates, the atomic line polarization becomes sensitive to the
strength of the magnetic field. In these situations, it is possible to
get information about the {\em magnitude} of magnetic field. If
$\nu_L$ gets comparable with the decay rate of the excited state, we
get into regime of the {\em upper state} Hanle effect. If the rate of
ion excitation $R_F$ gets comparable with $\nu_L$ we are in
the {\em ground state} Hanle effect regime. In the Hanle regimes the plane of polarization is being rotated and the degree of polarization is reduced.
Below we shall give brief descriptions for each regime (see Figure \ref{regimes}).

{\large
\begin{table}
\begin{tabular}{ccccc}
\hline
\hline
$\nu_L$(s$^{-1}$)&$\tau_R^{-1}$(s$^{-1}$)&A($s^{-1}$)&$A_m$(s$^{-1}$)&$\tau_c^{-1}$(s$^{-1}$)\\
\hline
$88(B/5\mu$ G)& $7.4\times 10^{5}\left(\frac{R_*}{r}\right)^2$&$2.8\times 10^8$&2.3$\times 10^{-6}$&$6.4\left(\frac{n_e}{0.1{\rm cm}^{-3}}\sqrt{\frac{8000{\rm K}}{T}}\right)\times 10^{-9}$\\
\hline
\hline
\end{tabular}
\caption{Relevant rates for atomic alignment. $\nu_L$ is the Larmor precession rate, $\tau_R^{-1}$ is the absorption rate, A is the emission rate of permitted line, $A_m$ is the magnetic dipole emission rate for transitions among J levels of the ground state of an atom. Example values for C II are given. $\tau_R^{-1}$ is calculated for an O type star, where $R_*$ is the radius of the star radius and r is the distance to the star.}
\label{difftime}
\end{table}
}
\subsection{Magnetic realignment regime}

We start by discussing discuss a toy model for
alignment in which the basic physics is displayed (Varshalovich 1971,
Trujillo Bueno et al. 2005, YLa): a two-level atom, with an upper $^1$S
state (quantum number J = 0) and a $^1$P ground state, so J = 1 (read
F for J, for hyperfine structure). Let M be the projection of the
atomic angular momentum onto an incident magnetic field, strong enough
to dominate the interactions of the atom. For the
ground state, M can be -1, 0, and 1, while for the upper state M = 0.
An incident unpolarized photon beam contains an equal number of left
and right circular polarized photons with projected angular momenta of
1 and -1 along the beam. Consider Figure \ref{???}, representing
atoms viewed along the beam. This radiation coming along the
quantization axis will induce transitions only from the M = -1 and M =
+1 states, leaving the M = 0 ground state overpopulated since no
radiative excitations out of it are possible, while de-excitations
into it are. The optical properties of the medium are changed for both
polarization and absorption.

For moderate pumping rate in circumstellar medium, the magnetic precession is much faster than the pumping (see Table \ref{difftime}), atoms are in this case realigned toward the magnetic field. This can be interpreted both in {\em classical} and {\em quantum} picture.

\begin{figure}
\includegraphics[%
  width=0.7\textwidth,
  height=0.3\textheight]{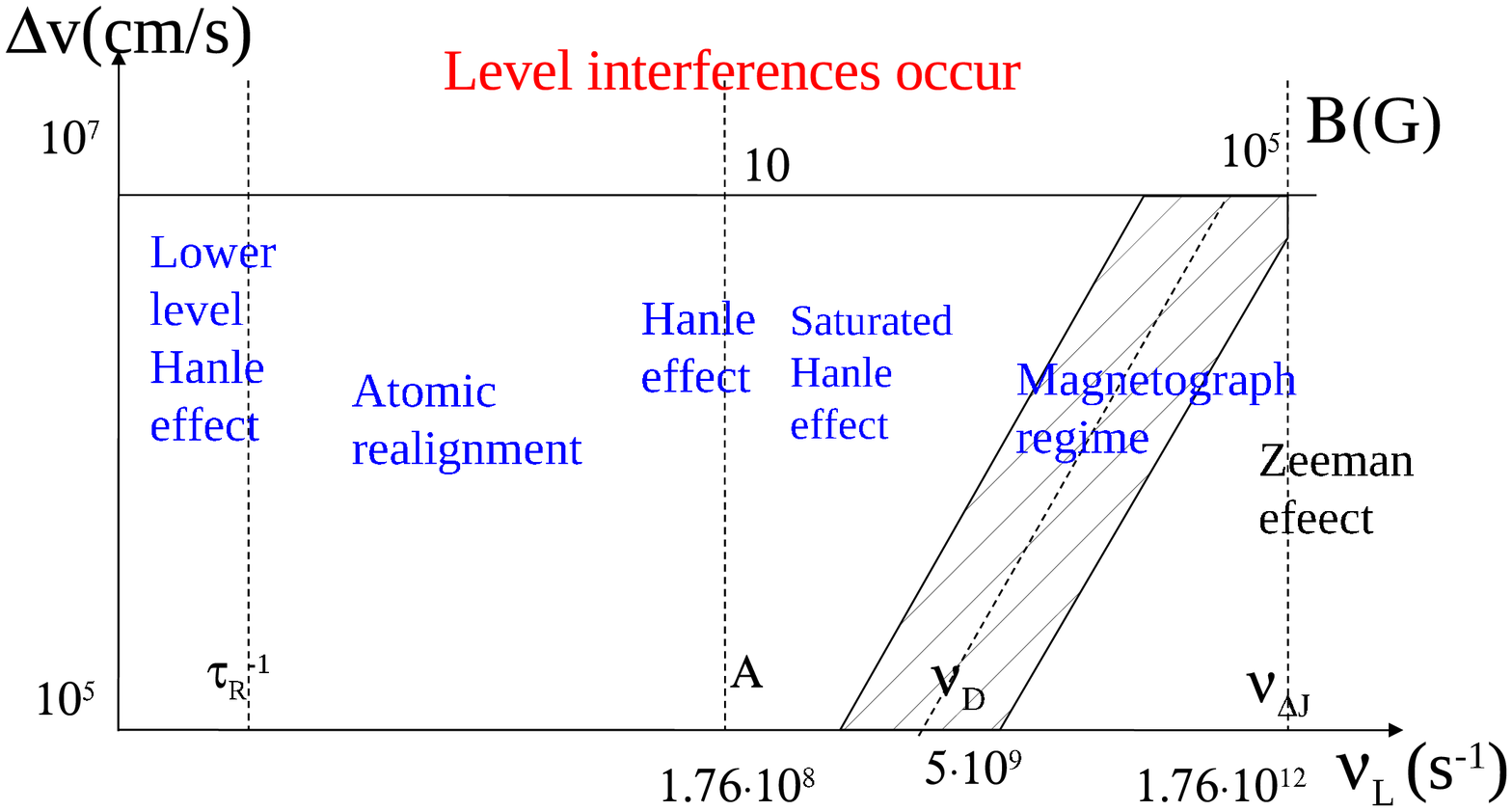}
\includegraphics[%
  width=0.25\textwidth,
  height=0.2\textheight]{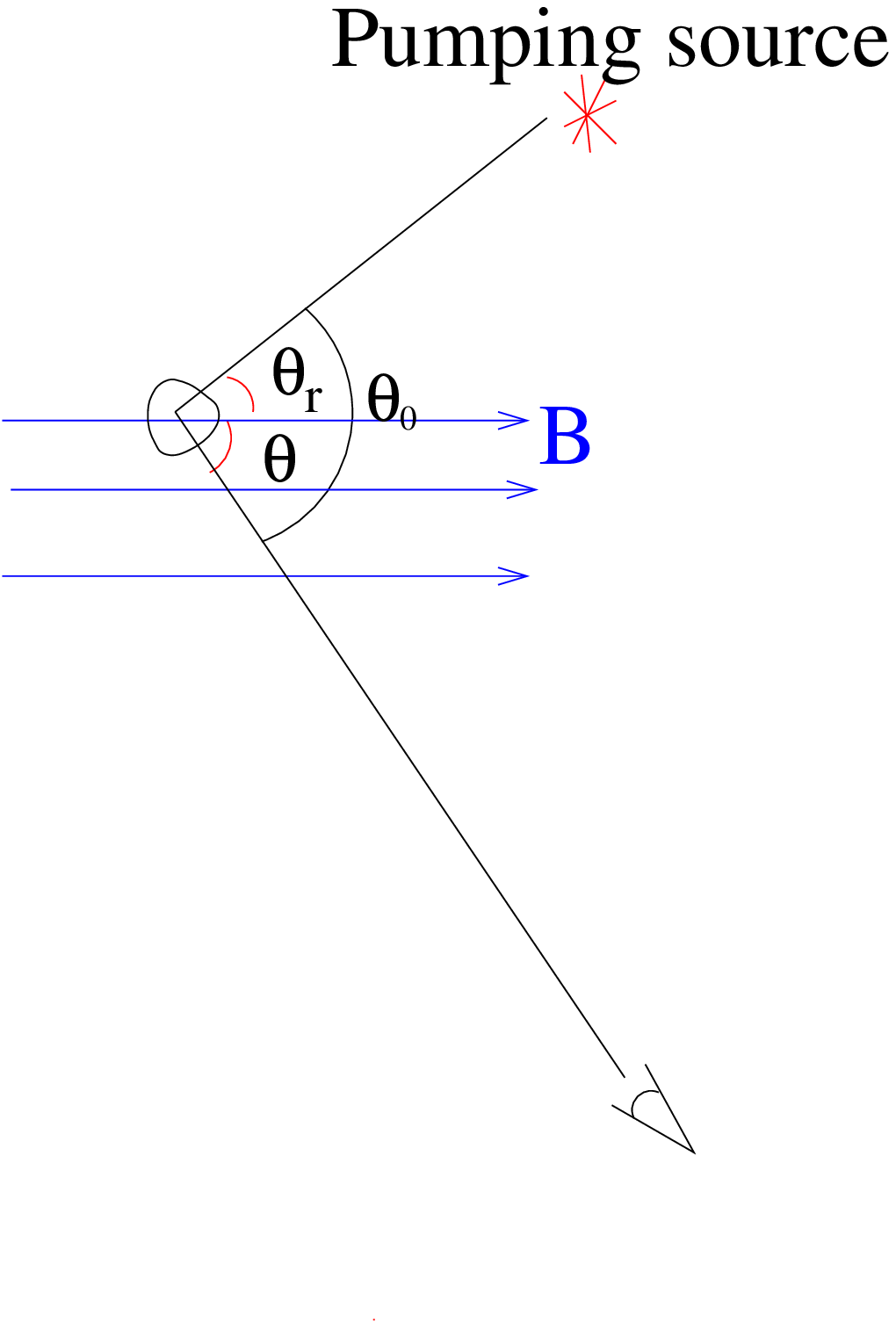}
\caption{Different regimes divided according to the strength of magnetic field and the Doppler line width. Atomic realignment is applicable to weak field ($<1G$) in diffuse medium. Level interferences are negligible unless the medium is substantially turbulent ($\delta v\gtrsim$ 100km/s) and the corresponding Doppler line width becomes comparable to the fine level splitting $\nu_{\Delta J}$. For strong magnetic field, Zeeman effect dominates. When magnetic splitting becomes comparable to the Doppler width, $\sigma$ and $\pi$ components (note: we remind the reader that $\sigma$ is the circular polarization and $\pi$ represents the linear polarization.) can still distinguish themselves through polarization, this is the magnetograph regime; Hanle effect is dominant if Larmor period is comparable to the lifetime of excited level $\nu_L^{-1}\sim A^{-1}$; similarly, for ground Hanle effect, it requires Larmor splitting to be of the order of photon pumping rate; for weak magnetic field ($<1G$) in diffuse medium, however, atomic alignment is the main effect provided that $\nu_L=17.6(B/\mu G){\rm s}^{-1}>\tau_R^{-1}$; {\em Right}: typical astrophysical environment where atomic alignment can happen. A pumping source deposits angular momentum to atoms in the direction of radiation and causes differential occupations on their ground states. In a magnetized medium where the Larmor precession rate $\nu_L$ is larger than the photon arrival rate $\tau_R^{-1}$, however, atoms are realigned with respect to magnetic field. Atomic alignment is then determined by $\theta_r$, the angle between the magnetic field and the pumping source. The polarization of scattered line also depends on the direction of line of sight, $\theta$ and $\theta_0$ (or $\phi$, defined afterward).} 
\label{regimes}
\end{figure}

Owing to the precession, the atoms with different projections of angular momentum will be mixed up. As the result, angular momentum is redistributed among the atoms, and the alignment is altered according to the angle between the magnetic field and radiation field $\theta_r$ (see Figure\ref{regimes}{\em right}). This is the {\em classical} picture. 

In {\em quantum} picture, if magnetic precession is dominant, then the natural quantization axis will be the magnetic field, which in general is different from the symmetry axis of the radiation. The radiative pumping is to be seen coming from different directions according to the angle between the magnetic field and radiation field $\theta_r$, which results in different alignment.

In the magnetic realignment regime, there is no interference among the magnetic sublevels. And atoms are aligned either $\parallel$ or $\bot$ to the magnetic field and so does the linear polarization of absorptions from the ground level. The switch of the polarization between the two cases $\parallel$ {\bf B} ($P>0$) and $\bot$ {\bf B} ($P<0$) happens at the Van-Vleck angle $\theta_r=54.7^o$.

\subsection{Hanle effect}

If the magnetic precession (or splitting) is comparable to the emission rate (or line-width A), the polarization is modified according to both the {\em strength} and the direction of the magnetic field. This is the so-called {\em Hanle effect} (see, e.g. Stenflo 1994).

{\em Classically}, the dipole oscillator will be rotating due to the precession. If the Zeeman splitting $\nu_L$ is significantly larger than the linewidth $A$ (or the inverse of the life time of the upper level), the dipole oscillator rotate several times before it gets damped, there is no polarization. However, if the magnetic precession happens at a timescale comparable to the lifetime of the upper level, the dipole oscillator rotates only a limited angle within the life (or damping) time. As the result, the degree of polarization is both reduced and the polarization plane is rotated compared to the case without magnetic field. The degree and the direction of polarization are determined by both the magnetic field and the direction of observation if the incident light is assumed to be fixed.

In {\em quantum} picture, Zeeman splitting lifts the degeneracy among the different magnetic sublevels and modifies the interference among the sublevels therefore. This results in the characteristic dependence on magnetic field of the linear polarizations of the emission lines. 

Different from the classical picture, the lines have nonzero polarizations in the case of $\nu_L\gg A$, this is a regime that atoms are realigned toward the magnetic field on the excited state before decaying to their ground state. Similar to the ``Magnetic realignment regime", the polarization of emission is either $\|$ or $\bot$ to the magnetic field and insensitive to the direction of magnetic field. This regime is ``saturated Hanle regime".  Note that in the Hanle regime, atoms are also aligned in the ground state as the magnetic precession period is longer than the
life-time of the ground state, and the ground state alignment affects the polarized signal of the scattered light. 

\subsection{Lower level Hanle regime}

An analogy can be made between the lower level Hanle effect and the Hanle effect. In places close to a pumping source, the radiative pumping can get comparable to the magnetic precession provided that magnetic field is weak enough. In this case, similar scenario occurs on the ground state in analogy to the upper level in the Hanle regime. In this situation, coherence appears in the ground state and as the result, the polarization of absorption does not have a preferential direction unlike in the alignment regime. Polarization of scattered lines is modulated by the magnetic field as well. 

\section{Formalism}

The basic equations describing the statistics of atomic density tensors in the aforementioned different regimes are the same (see YLa),

\bea
\dot{\rho^k_q}(J_u)&+&i 2\pi\nu_Lg_uq\rho^k_q(J_u) = -\sum_{J_l}A(J_u\rightarrow J_l)\rho^k_q(J_u)+\sum_{J_lk'}[J_l]\left[\delta_{kk'}p_{k'}(J_u,J_l)B_{lu}\bar{J}^0_0+\sum_{Qq'}r_{kk'}(J_u,J_l,Q,q')B_{lu}\bar{J}^2_Q\right]\rho^{k'}_{-q'}(J_l),
\label{evolution}
\eea
\bea
\dot{\rho^k_q}(J_l)&+&i 2\pi\nu_Lg_lq\rho^k_q(J_l) = \sum_{J_u}p_k(J_u,J_l)[J_u]A(J_u\rightarrow J_l)\rho^k_q(J_u)-\sum_{J_u,k'}\left[\delta_{kk'}B_{lu}\bar{J}^0_0+\sum_{Q,q'}s_{kk'}(J_u,J_l,Q,q')B_{lu}\bar{J}^2_Q \right]\rho^{k'}_{-q'}(J_l),
\label{evolutiong}
\eea

where
\be
p_k(J_u,J_l)=(-1)^{J_u+J_l+1}\left\{\begin{array}{ccc}
J_l & J_l & k\\J_u & J_u &1\end{array}\right\},\,
p_0(J_u,J_l)=\frac{1}{\sqrt{[J_u, J_l]}},
\label{pk}
\ee
\be
r_{kk'}(J_u,J_l,Q,q)=(3[k,k',2])^{1/2}\left\{\begin{array}{ccc} 
1 & J_u & J_l\\1& J_u & J_l\\ 2 &k& k' \end{array}\right\}\left(\begin{array}{ccc}
k & k' & 2\\ q & q' & Q \end{array}\right),
\label{rkk}
\ee
\be
s_{kk'}(J_u,J_l)=(-1)^{J_l-J_u+1}[J_l](3[k,k',2])^{1/2}\left(\begin{array}{ccc}
k & k' & 2\\ q & q'& Q\end{array}\right)\left\{\begin{array}{ccc} 
1 & 1 & 2\\J_l& J_l & J_u\end{array}\right\}\left\{\begin{array}{ccc} 
k & k' & 2\\J_l& J_l & J_l\end{array}\right\}.
\label{skk}
\ee
In Eqs.(\ref{evolution},\ref{evolutiong}), $\rho^k_q$ and ${\bar J}^K_Q$ are the irreducible form of density matrix of atoms and radiation, respectively. The evolution of upper state ($\rho^k_q(J_u)$) is represented by Eq.(\ref{evolution}) and the ground state ($\rho^k_q(J_l)$) is described by Eq.(\ref{evolutiong}). The second terms on the left side of Eqs.(\ref{evolution},\ref{evolutiong}) represent mixing by magnetic field, where $g_u$ and $g_l$ are the Land\'e factors for the upper and ground level.  The two terms on the right side of Eq.(\ref{evolution}, \ref{evolutiong}) are due to spontaneous emissions and the excitations from ground level. Transitions to all upper states are taken into account by summing over $J_u$ in Eq.(\ref{evolutiong}). Vice versa, for an upper level, transitions to all ground sublevels ($J_l$) are summed up in Eq.(\ref{evolution}). Let us remind our reader, this multi-level treatment is well justified as level crossing interferences do not exist in the weak field regime unless the medium gets strongly turbulent so that Doppler width becomes comparable to the energy separation between levels ($\delta v\gtrsim 100$km/s, see Fig.\ref{regimes}). The matrix with big ``\{ \}" represents the 6-j or 9-j symbol, depending on the size of the matrix. Throughout this paper, we define $[j]\equiv 2j+1$, which means $[J_l]=2J_l+1$ and $[J_u]=2J_u+1$, etc.  

The excitation depends on  
\be
\bar{J}^K_Q=\int d\nu \frac{\nu_0^2}{\nu^2}\Psi(\nu-\nu_0)\oint \frac{d\Omega}{4\pi}\sum_{i=0}^3{\cal J}^K_Q(i,\Omega)S_i(\nu, \Omega),
\label{incidentrad}
\ee
which is the radiation tensor of the incoming light averaged over the whole solid-angle and the 
line profile $\Psi(\nu-\nu_0)$. $S_i=[I,\, Q,\, U,\, V]$ represent the Stokes parameters. In many cases, the radiation source is so far that it can be treated as a point source. The irreducible unit tensors for Stokes parameters $I,Q,U$ are:
\bea
{\cal J}^0_{0}(i,\Omega)&=&\left(\begin{array}{l}1\\0\\0\end{array}\right),\;~\;\; {\cal J}^2_{0}(i,\Omega)=\frac{1}{\sqrt {2}}\left[\begin{array}{l} ( 1-1.5\sin^2\theta )\\-3/2 \sin^2\theta\cos2\gamma\\3/2 \sin^2\theta\sin2\gamma\end{array}\right],\;~
{\cal J}^2_{\pm1}(i,\Omega)=\sqrt{3}e^{\pm i\phi}\left[\begin{array}{l}\mp\sin2 \theta/4 \\\mp(\sin 2 \theta\cos2\gamma-2i\sin\theta\sin2\gamma)/4 \\\pm(\sin 2 \theta\sin2\gamma-2i\sin\theta\cos2\gamma)/4 
\end{array}\right],\nonumber\\
{\cal J}^2_{\pm2}(i,\Omega)&=&\sqrt{3}e^{\pm 2i\phi}\left\{\begin{array}{l}\sin^2\theta/4\\-\left[( 1+\cos^2\theta )\cos2\gamma\mp 2i\cos\theta\sin2\gamma\right]/4\\\left[(1+\cos^2\theta)\sin2\gamma\mp 2 i\cos\theta\cos2\gamma\right]/4  \end{array}\right\}.
\label{irredrad}
\eea
They are determined by its direction $\Omega$ and the reference chosen to measure the polarization (see Fig.\ref{radiageometry}{\em right}). We consider here unpolarized incident light, thus $Q=U=V=0$. For the incoming radiation from ($\theta_r,\phi_r$), the nonzero elements of the radiation tensor are:

\be
\bar{J}^0_0=I_*, \bar{J}^2_0=\frac{W_a}{2\sqrt{2}W}(2-3\sin^2\theta_r)I_*, \bar{J}^2_{\pm2}=\frac{\sqrt{3}W_a}{4W}\sin^2\theta_rI_*{e^{\pm 2i\phi_r}}, \bar{J}^2_{\pm1}=\mp\frac{\sqrt{3}W_a}{4W}\sin2\theta_r I_*{e^{\pm i\phi_r}} 
\label{irredradia}
\ee
where W is the dilution factor of the radiation field, which can be divided into anisotropic part $W_a$ and isotropic part $W_i$ (Bommier \& Sahal-Brechot 1978). The solid-angle averaged intensity for a black-body radiation source is 
\be
I_*=W\frac{2h\nu^3}{c^2}\frac{1}{e^{h\nu/k_BT}-1}.
\label{Idilu}
\ee

By setting the first terms on the left side of Eq.(\ref{evolution},\ref{evolutiong}) to zeros, we can obtain the following linear equations for the steady state density tensors in the ground states of atoms: 
\bea
i\Gamma\rho^k_q(J_l)q &-&\sum_{J_u,k'}\left\{p_k(J_u,J_l)\frac{[J_u]}{\sum_{J''_l}A''/A+i\Gamma'q}\sum_{J'_l}[J'_l]\left[\delta_{kk'}p_{k'}(J_u,J'_l)+\sum_{Qq'}r_{kk'}(J_u,J_l,Q,q')\bar{J}^2_Q/\bar{J}^0_0\right]\right.\nonumber\\
&-&\left.\left[\delta_{kk'}+\sum_{Q,q'}s_{kk'}(J_u,J_l,Q,q')\bar{J}^2_Q/\bar{J}^0_0\right]\right\}\rho^{k'}_{-q'}(J_l)=0
\label{lowlevel}
\eea
where $\Gamma=2\pi\nu_Lg_l/B_{lu}\bar{J}^0_0,~ \Gamma'=2\pi\nu_Lg_u/A$.

The multiplet effect is counted for by summing over $J_u$. The alignment of atoms induces polarization for absorption lines. Depending on the comparison between the first term (due to magnetic mixing) and the other terms (due to optical pumping), there could be different regimes. In our previous studies (YLa,b), we dealt with the regime where magnetic mixing is the fastest process, this is the regime for atomic realignment (see Fig.\ref{regimes}). When magnetic splitting $2\pi\nu_L g_l$ is comparable to the photon pumping rate $B{\bar J}^0_0$, the coherence terms $\rho^k_{q\neq 0}$ appear. Physically, this is because magnetic field cannot remain a good quantization axis when the two processes magnetic precession and optical pumping happen on comparable timescales (unless the magnetic field is aligned along the symmetry axis of the radiation field). In such a situation, the density matrix of the ground level becomes sensitive to the strength of the magnetic field. This is the {\em lower-level Hanle regime}. Absorption from the ground level will be polarized and the polarization gets the imprint of the magnetic field. The corresponding absorption coefficients $\eta_i$ are (Landi Degl'Innocenti 1984, see also YLa)
\be
\eta_i(\nu, \Omega)=\frac{h\nu_0}{4\pi}Bn(J_l)\Psi(\nu-\nu_0)\sum_{KQ}(-1)^Kw^K_{J_lJ_u}\sigma^K_Q(J_l ){\cal J}^K_Q(i, \Omega),
\label{Mueller0}
\ee
where $n(J_l)=n\sqrt{[J_l]}\rho_0^0(J_l)$ is the total atomic population on level $J_l$, $\Psi(\nu-\nu_0)$ is the line profile, $\sigma^K_Q\equiv\rho^K_Q/\rho^0_0$, and
\be
w^K_{J_lJ_u}\equiv\left\{\begin{array}{ccc}1 & 1 & K\\J_l&J_l& J_u\end{array}\right\}/\left\{\begin{array}{ccc}1 & 1 & 0\\J_l&J_l& J_u\end{array}\right\}.
\label{w2}
\ee
The polarization produced by absorption through optical depth $\tau=\eta_0d$ is
\be
\frac{Q}{I\tau}=\frac{-\eta_1d I_0}{(1-\eta_0d)I_0\eta_0d}\simeq-\frac{\eta_1}{\eta_0},\, \frac{U}{I\tau}\simeq-\frac{\eta_2}{\eta_0}.
\label{genericabs}
\ee
In our choice of reference, only coherence density components contribute to U (see Eq.\ref{Mueller0},\ref{irredradia}). Therefore unlike the case for atomic realignment, U is not zero in the {\em ground Hanle regime}. Since incoming light is unpolarized, there is no orientation for the atoms. The circular polarization component is thus zero $V=\epsilon_3=0$. 

The differential occupation on the ground state (Eq.\ref{lowlevel}) can be transferred to the upper level of an atom by excitation. 
\bea
\rho^k_q(J_u)&=& \frac{1}{\sum_{J''_l}A''+iA\Gamma'q}\sum_{J'_lk'}[J'_l]\left[\delta_{kk'}p_{k'}(J_u,J'_l)B_{lu}\bar{J}^0_0+\sum_{Qq'}r_{kk'}(J_u,J'_l,Q,q')B_{lu}\bar{J}^2_Q\right]\rho^{k'}_{-q'}(J_l)
\label{uplevel}
\eea 
Emission from such a differentially populated state is polarized, the corresponding emission coefficients of the Stokes parameters are (Landi Degl'Innocenti 1984):
\be
\epsilon_i(\nu, \Omega)=\frac{h\nu_0}{4\pi}An(J_u, \theta_r)\Psi(\nu-\nu_0)\sum_{KQ}w^K_{J_uJ_l}\sigma^K_Q(J_u, \theta_r){\cal J}^K_Q(i, \Omega),
\label{emissivity}
\ee
where $n(J_u,\theta_r)=n\sqrt{[J_u]}\rho_0^0(J_u,\theta_r)$ is the total population on level $J_u$. For optically thin case,  the linear polarization degree $p=\sqrt{Q^2+U^2}/I=\sqrt{\epsilon_2^2+\epsilon_1^2}/\epsilon_0$, the positional angle $\chi=\frac{1}{2}\tan^{-1}(U/Q)=\frac{1}{2}\tan^{-1}(\epsilon_2/\epsilon_1)$. Obviously, unlike the case or pure magnetic realignment, in the presence of stronger magnetic fields, the positional angle  $\chi$
experiences variations with the magnetic field strength.  

\section{Polarization in magnetic realignment regime}
\begin{figure}
\includegraphics[%
  width=0.95\textwidth,
  height=0.3\textheight]{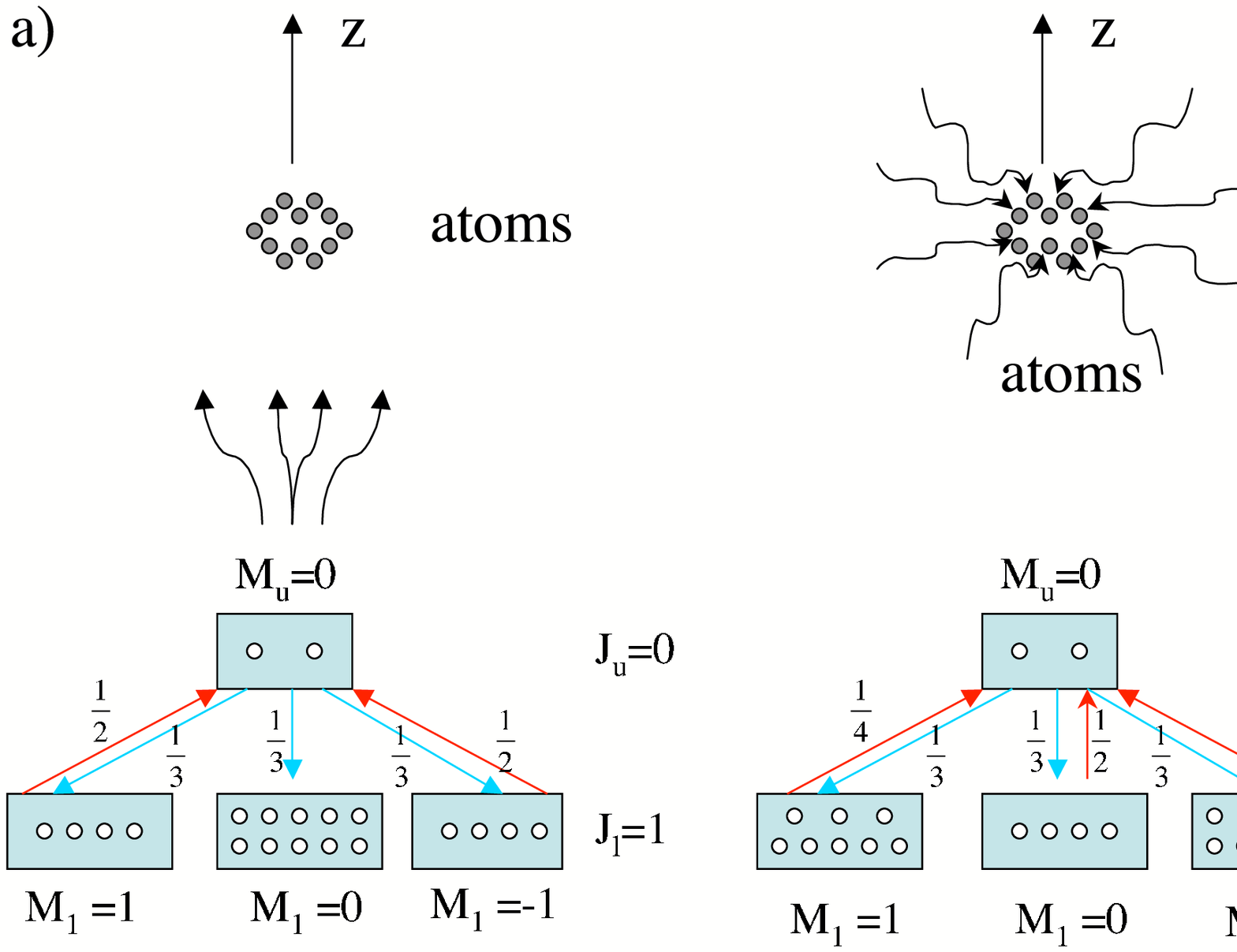}
\caption{\small A toy model to illustrate how a simple two-level atom is aligned by
an incident beam of light. M is the projection of the angular momentum
along an incident magnetic field. (a) When beam is coming along the direction of {\bf B} field,
atoms accumulate in the ground sublevel $M=0$ because radiation
removes them from the ground states $M=1$ and $M=-1$. Hence, M = 0 is
overpopulated; b) the incident beam is $\perp$ to the {\bf B} field,
$M=\pm 1$ is more populated because the absorptions from sublevels
$M=0$ have higher probabilities because both right- and left-handed
incident polarizations can have projected M = 0.}
\end{figure}
\subsection{Polarization of absorption lines}

In the regime of the weak field, polarization of absorption lines arising  from aligned atoms was dealt with for atoms with fine structure in YLa, while with
hyperfine structure in YLb. Since that time, we have provided calculations for more atomic lines. In particular, we show below the results for Ti II, which has transitions in the optical band and those can be observed from the {\em ground} (see Table~\ref{TiII}).
The calculations follow the same approach that we discussed in detail in YLa. 

\begin{figure}
\includegraphics[width=0.33\textwidth,
  height=0.25\textheight]{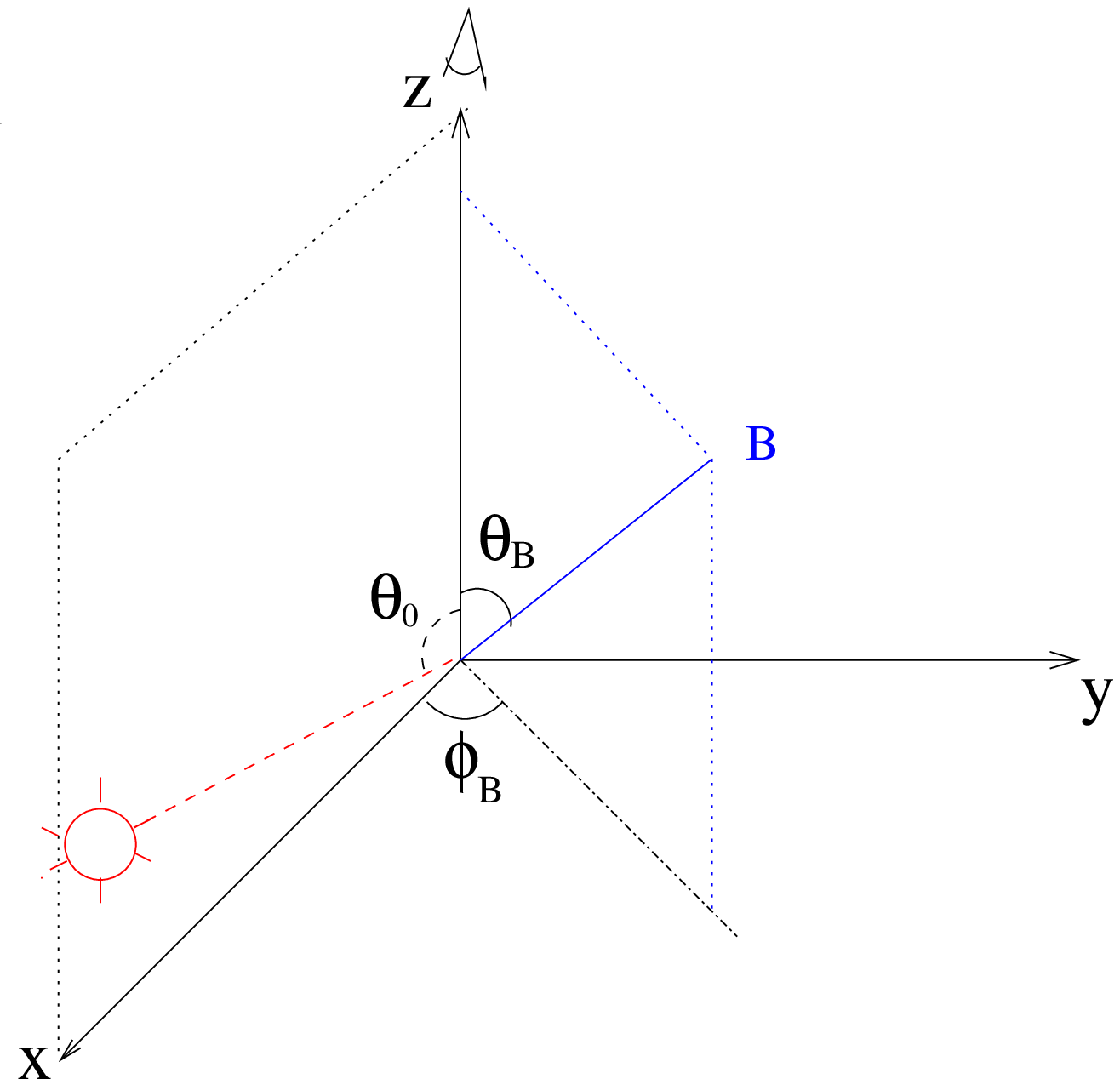}\includegraphics[width=0.33\textwidth,
  height=0.25\textheight]{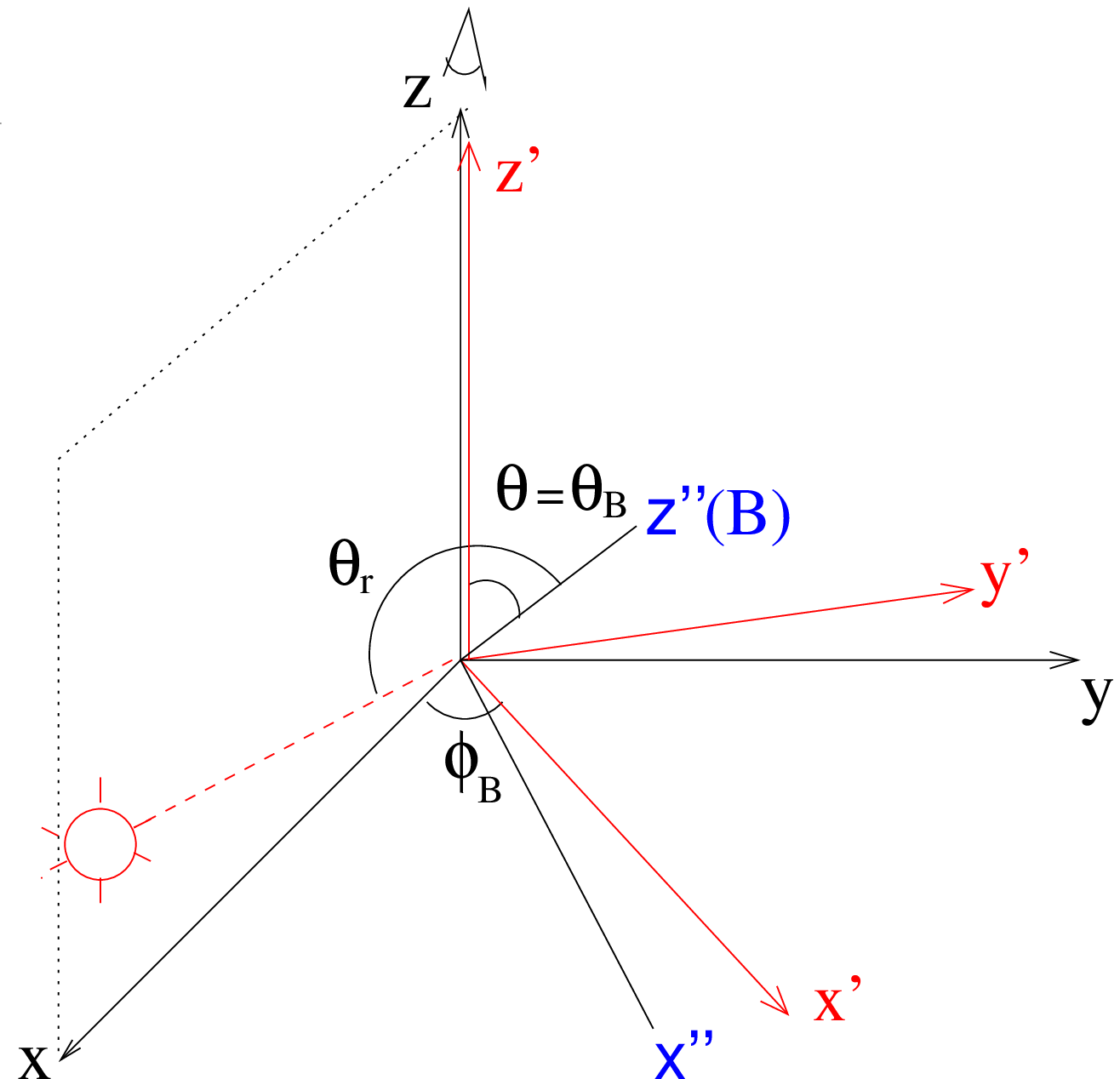}
\includegraphics[%
  width=0.34\textwidth,
  height=0.28\textheight]{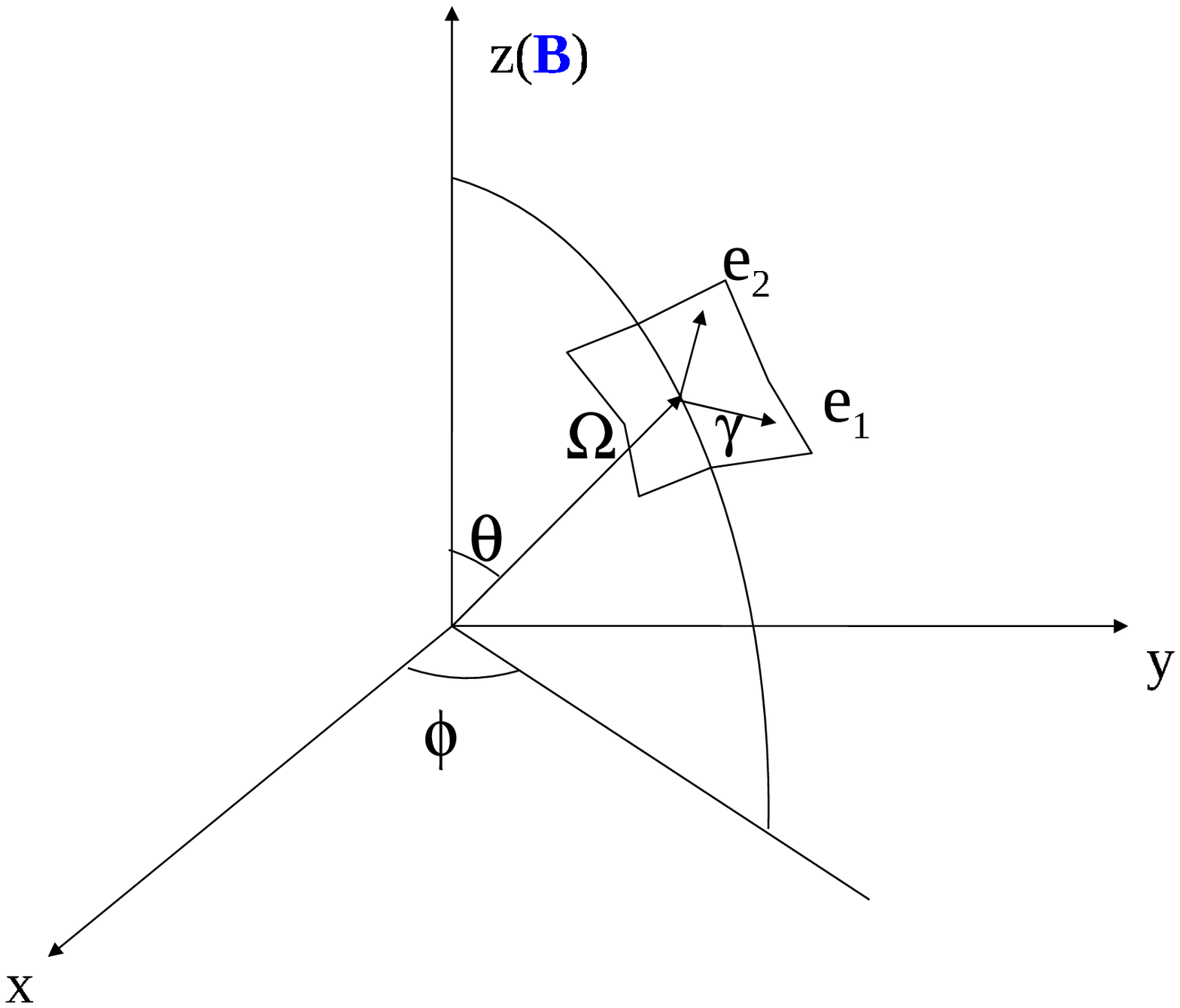}
\caption{{\em Left}: geometry of the observational frame. In this frame, the line of sight is z axis, together with the incident light, they specify the x-z plane. Magnetic field is in ($\theta_B, \phi_B$) direction; {\em Middle}: transformation to the ``theoretical frame" where magnetic field defines z" axis. This can be done by two successive rotations specified by Euler angles ($\phi_B, \theta_r$) (see YLa for details). The first rotation is from xyz coordinate system to x'y'z' coordinate system by an angle $\phi_B$ about the z-axis, the second is from x'y'z' coordinate system to x"y"z" coordinate system by an angle $\theta$  about the y'-axis. Atomic transitions are treated in the ``theoretical" frame where the line of sight is in ($\theta, \pi$) direction and the incident radiation is in ($\theta_r,\phi_r$) direction; {\em Right}: Radiation geometry and the polarization vectors in a given coordinate system. ${\bf \Omega}$ is the direction of radiation.}
\label{radiageometry}
\end{figure}

The geometry of the radiation system is illustrated by Fig.\ref{radiageometry}{\em left}. The origin of this frame is defined as the location of the atomic cloud. The line of sight defines z axis, and together with direction of radiation, they specify x-z plane. The x-y plane is thus the plane of sky. In this frame, the incident radiation is coming from  ($\theta_0, 0$), and the magnetic field is in the  direction ($\theta_B, \phi_B$). 

The magnetic field is chosen as the quantization axis ($z"$) for the atoms. In this ``theoretical'' reference frame, the line of sight is in ($\theta, \pi$) direction (i.e., the x"-z" plane is defined by the magnetic field and the line of sight, see Fig.\ref{radiageometry}{\em right}), and the radiation source is directed along ($\theta_r, \phi_r$). 
In YLa we discussed the resonance absorption lines in the appropriate for studying
magnetic fields in diffuse, low column density (AV $\sim$ few tenths)
neutral clouds in the interstellar medium are NI, OI, SII, MnII, and FeII. These are all in the ultraviolet.  
At higher column densities, the above lines become optically thick,
and lines of lower abundance, as well as excited states of the above
lines become available.  Significantly, some of these alignable species, e.g., TiII, FeI  are in the visible.

\begin{figure}
\includegraphics[%
  width=0.33\textwidth,
  height=0.25\textheight]{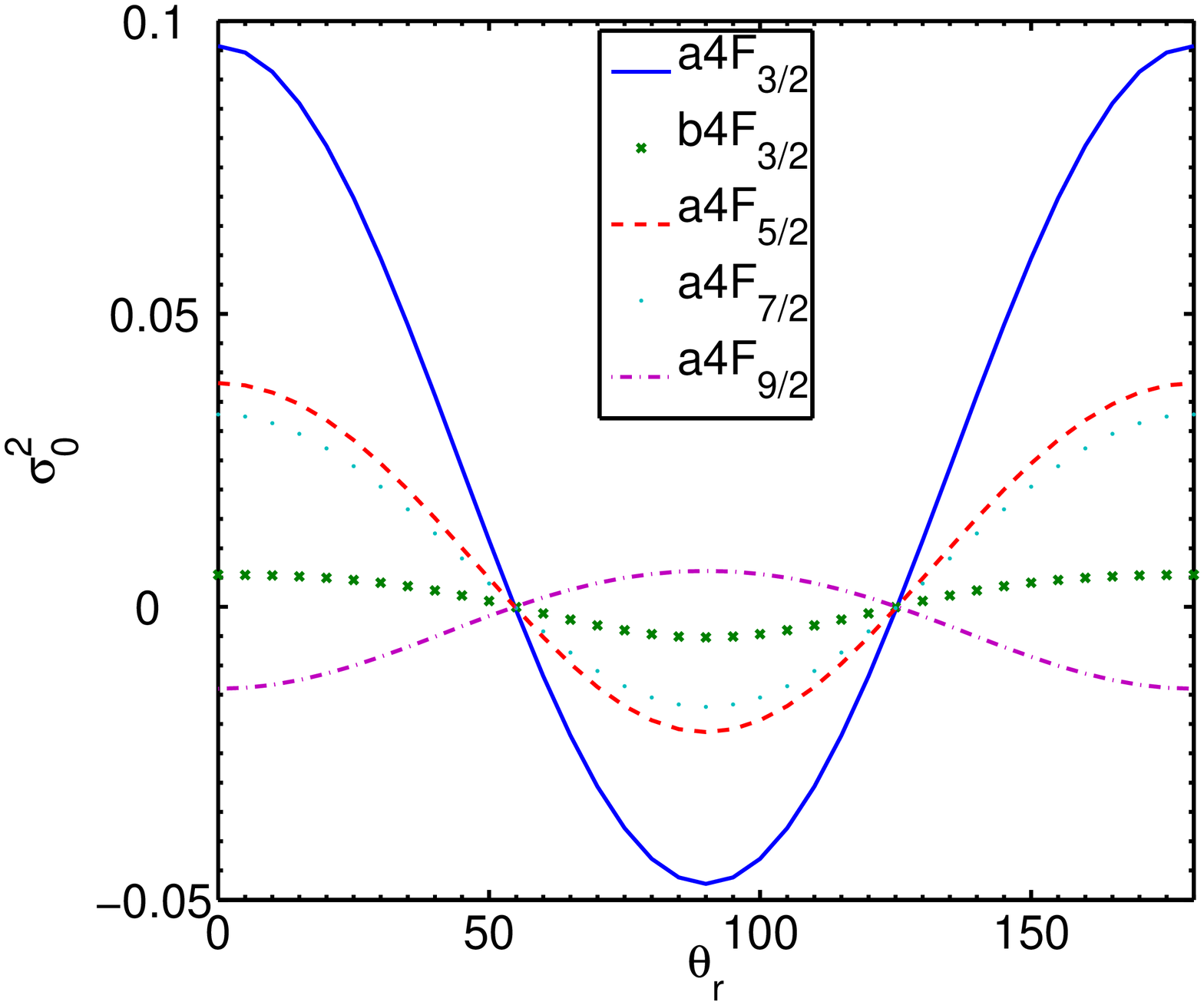}
\includegraphics[%
  width=0.33\textwidth,
  height=0.25\textheight]{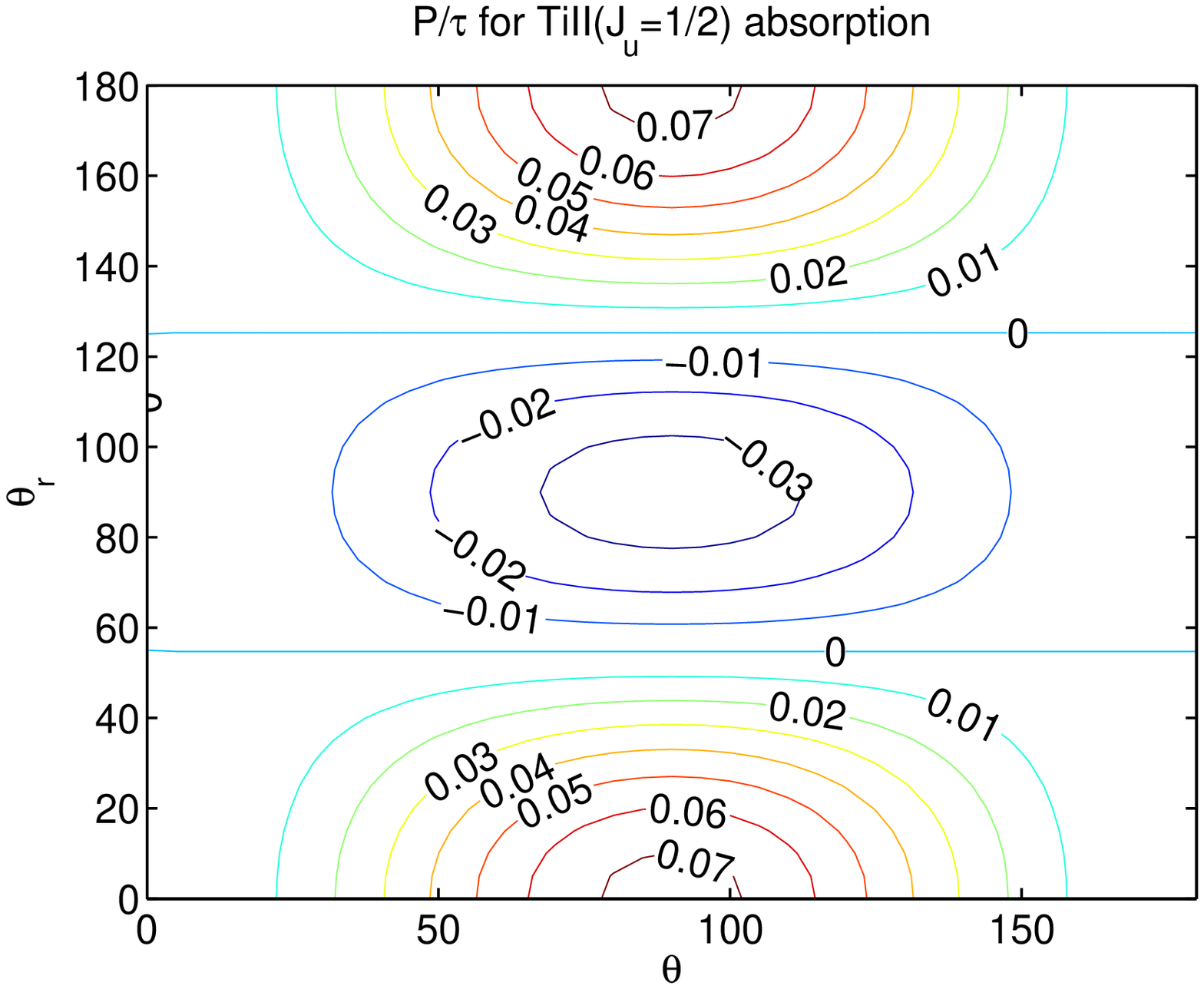}
\includegraphics[%
  width=0.33\textwidth,
  height=0.25\textheight]{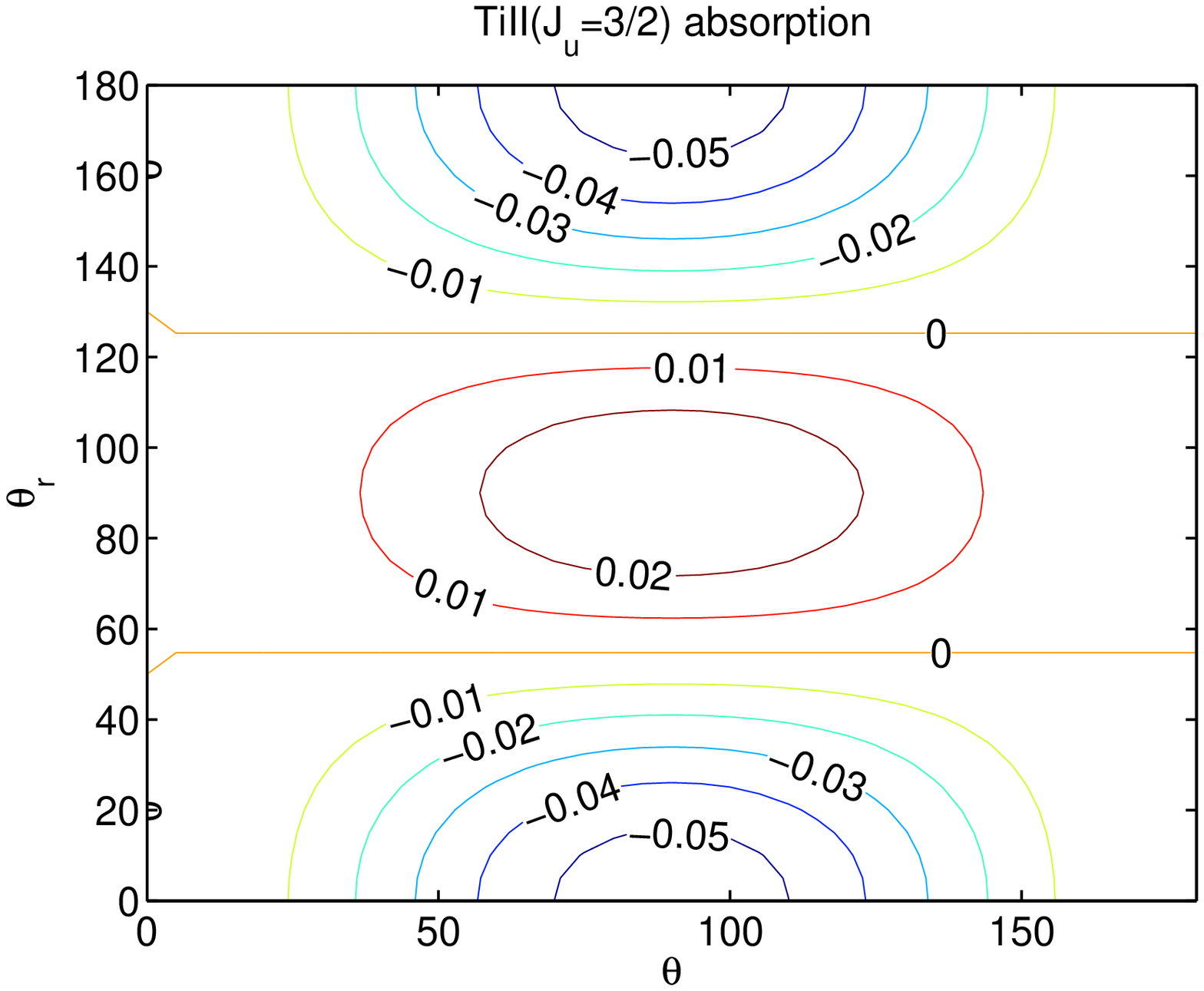}
\caption{{\em left}: the alignment of the ground state $a4F$ and metastable level $b4F_{3/2}$ of Ti II; {\em middle} and {\em right}: the contour of equal degree of polarization of Ti II absorption lines ($J_l=1/2 \rightarrow J_u=1/2,3/2$). $\theta_r$ and $\theta$ are the angles of incident radiation and line of sight from the magnetic field (see Fig.\ref{regimes}{\em right}). In the case of pumping source coincident with the background source, we have the degeneracy and polarization will be determined by one parameter $\theta_r=\theta$. }
\label{Tipol}
\end{figure}

The TiII lines provide a fairly accessible test of the magnetic
alignment diagnostic.
Ti II has a metastable state $b4F_{3/2,5/2,7/2,9/2}$ in addition to the ground state $a4F_{3/2,5/2,7/2,9/2}$ and the excited state $z4D^o_{1/2,3/2,5/2}, z4F^o_{3/2,5/2,7/2,9/2}, z4G^o_{5/2,7/2,9/2}$. By including all the transition between the excited and the metastable state, we obtain the alignment of Ti II (Fig.\ref{Tipol}). The polarization of absorption line can then be obtained according to Eq.(\ref{genericabs}):
\be
\frac{P}{\tau}\simeq -\frac{\eta_1}{\eta_0}=\frac{1.5\sigma^2_0(J_l)\sin^2\theta w^2_{J_lJ_u}}{\sqrt{2}+\sigma^2_0(J_l)(1-1.5\sin^2\theta) w^2_{J_lJ_u}}
\ee

\begin{figure}
\plottwo{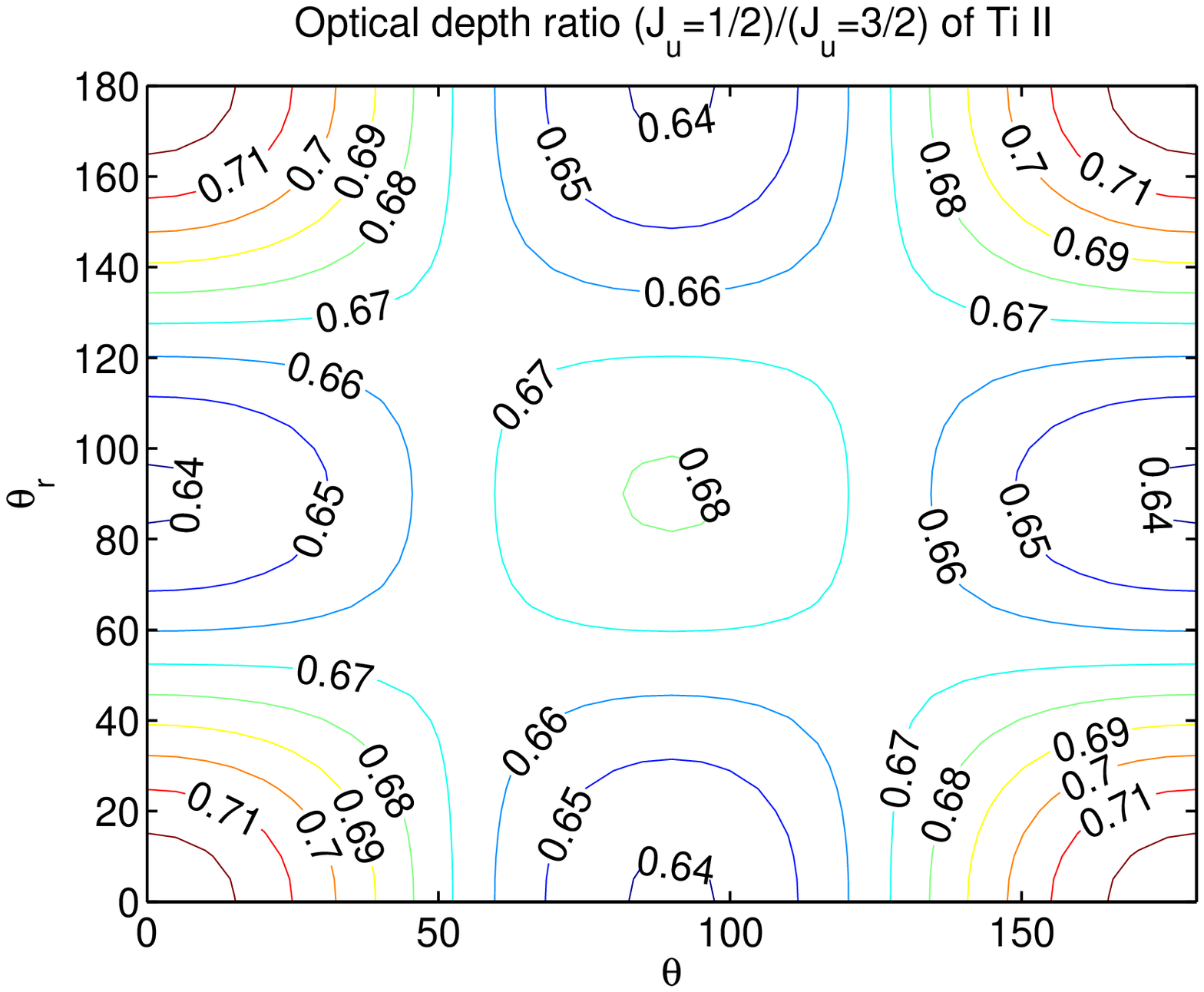}{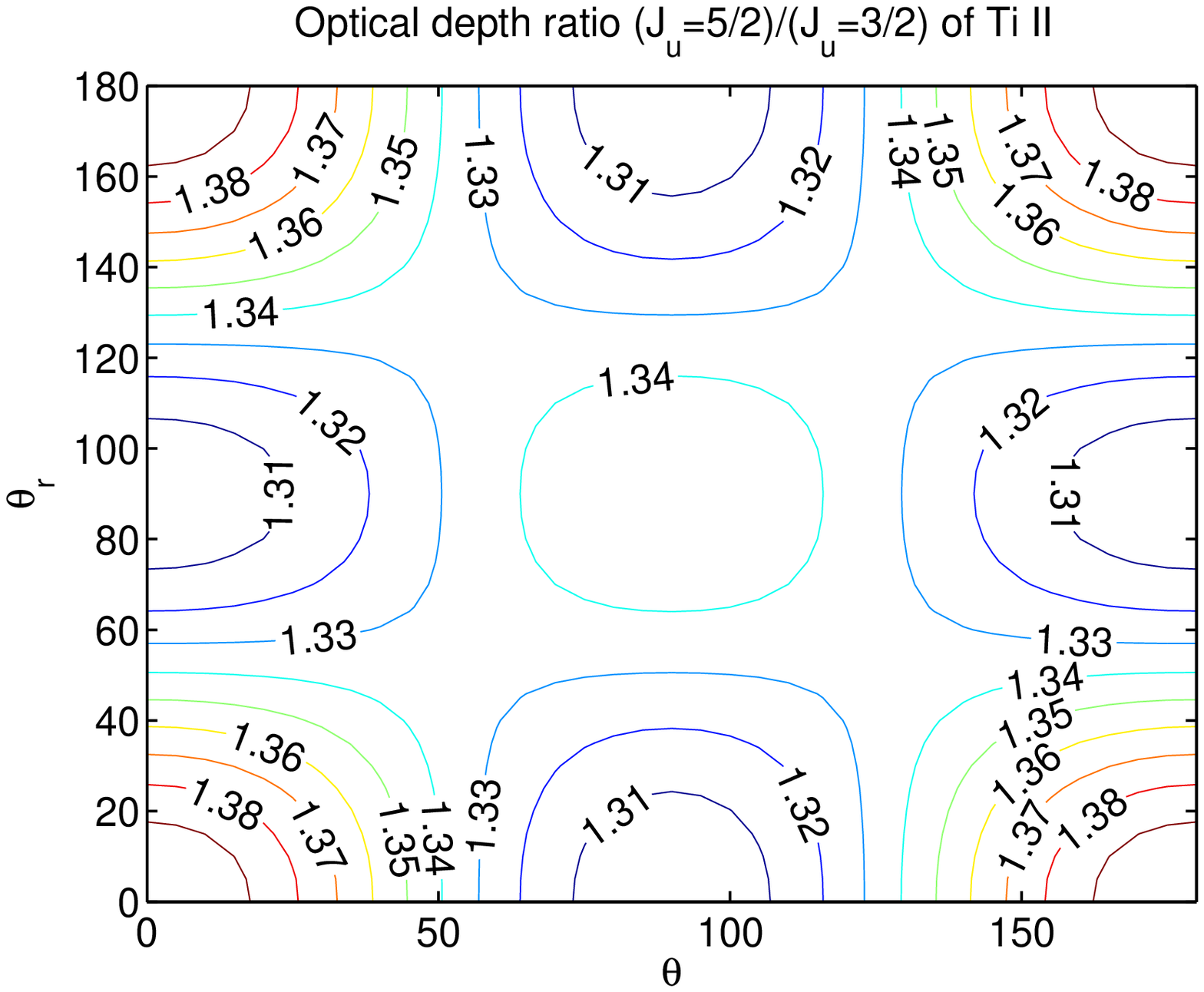}
\caption{The variations of the Ti II line ratios caused by the atomic alignment. }
\label{Tiratio}
\end{figure}

The predictions for the polarization are shown in Fig \ref{Tipol}. In addition, the line ratios are also modulated by the alignment (Fig.\ref{Tiratio}). 
Table \ref{TiII} summarizes the list of lines with predictions of the polarization arising from realignment of TiII in its ground state.
\begin{table*}
\begin{tabular}{cccccp{0.5in}ccccc}
\multicolumn{5}{c}{Absorption lines}&&\multicolumn{5}{c}{Emission lines}\\[0.5ex]
\cline{1-5} \cline{7-11}
\cline{1-5} \cline{7-11}
Species&Ground state&excited state&Wavelength(\AA)&$P_{max}$&&Species&Lower state&Upper state&Wavelength(\AA)&$|P_{max}|$\\[0.5ex]
\cline{1-5} \cline{7-11}
\multirow{5}{*}{TiII}&\multirow{5}{*}{$a4F_{3/2}$}&$z4G^o_{5/2}$&3384.74&-0.7\%&&\multirow{2}{*}{S II}&\multirow{2}{*}{$4S^o_{3/2}$}&$4P_{3/2}$&1253.81&30.6\%\\
&&$z4F^o_{5/2}$&3230.13&-0.7\%&&&&$4P_{5/2}$&1259.52&31.4\%\\
\cline{7-11}
&&$z4F^o_{3/2}$&3242.93&2.9\%&&&$3P_{0}$&$3S^{o}$&1306&16\%\\
&&$z4D^o_{3/2}$&3067.25&2.9\%&&&$3P_{1}$&$3S^{o}$&1304&8.5\%\\
&&$z4D^o_{1/2}$&3073.88&7.3\%&&O I&$3P_{2}$&$3S^{o}$&1302&1.7\%\\
\cline{1-5} \cline{8-11}
\cline{1-5}
&&&&&&&$3P$&$3S^{o}$&5555,6046,7254&2.3\%\\
\cline{8-11}
&&&&&&&$3P_{0}$&$3D^o$&1028&4.29\%\\
&&&&&&&$3P_{1}$&$3D^o$&1027&7.7\%\\
&&&&&&&$3P_{2}$&$3D^o$&1025&10.6\%\\
\cline{8-11}
&&&&&&&$3P$&$3D^o$&5513,5958,7002&1.3\%\\
\cline{7-11}
\cline{7-11}
\end{tabular}
\caption{}
\label{TiII}
\end{table*}

\subsection{Polarization of emission lines of atoms with fine structure}
The magnetic realignment diagnostic can also be used in emission, for
resonant or fluorescent scattering lines. 
Weak field regime for the atoms with fine structure was not covered completely in YLa.
While the calculations of the polarization arising from absorption of aligned atoms with fine structure were provided in YLa,
there we did not provide the calculations for the emission line polarization arising from such atoms. We, however, proved in YLb, that
emission polarization studies may be important for aligned atoms with hyperfine structure. Below we correct the deficiency
of our earlier studies and discuss the polarization arising from the emission of atoms with fine structure.

\begin{figure}
\includegraphics[%
  width=0.33\textwidth,
  height=0.25\textheight]{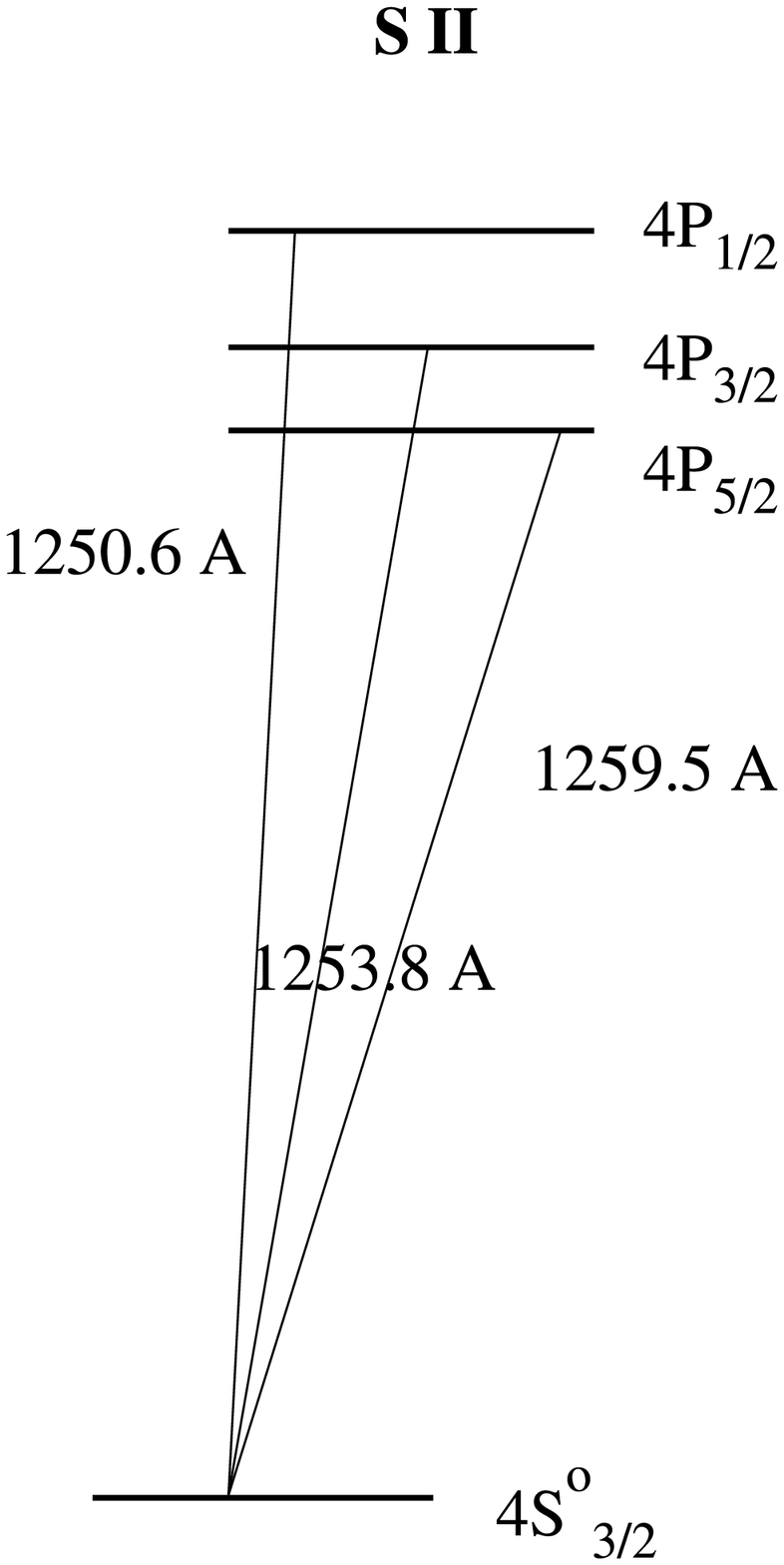}
\includegraphics[%
  width=0.33\textwidth,
  height=0.25\textheight]{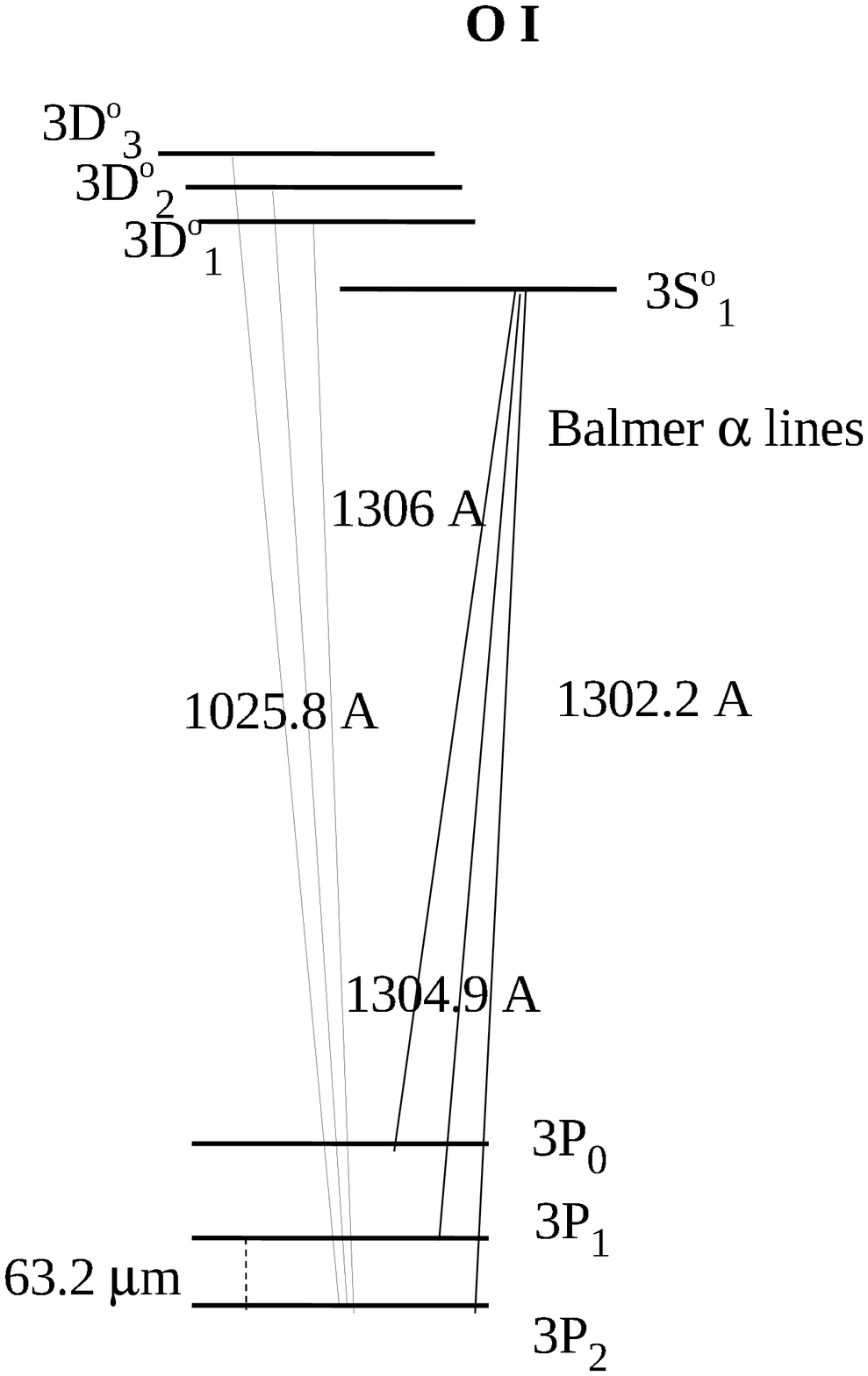}
\includegraphics[%
  width=0.33\textwidth,
  height=0.25\textheight]{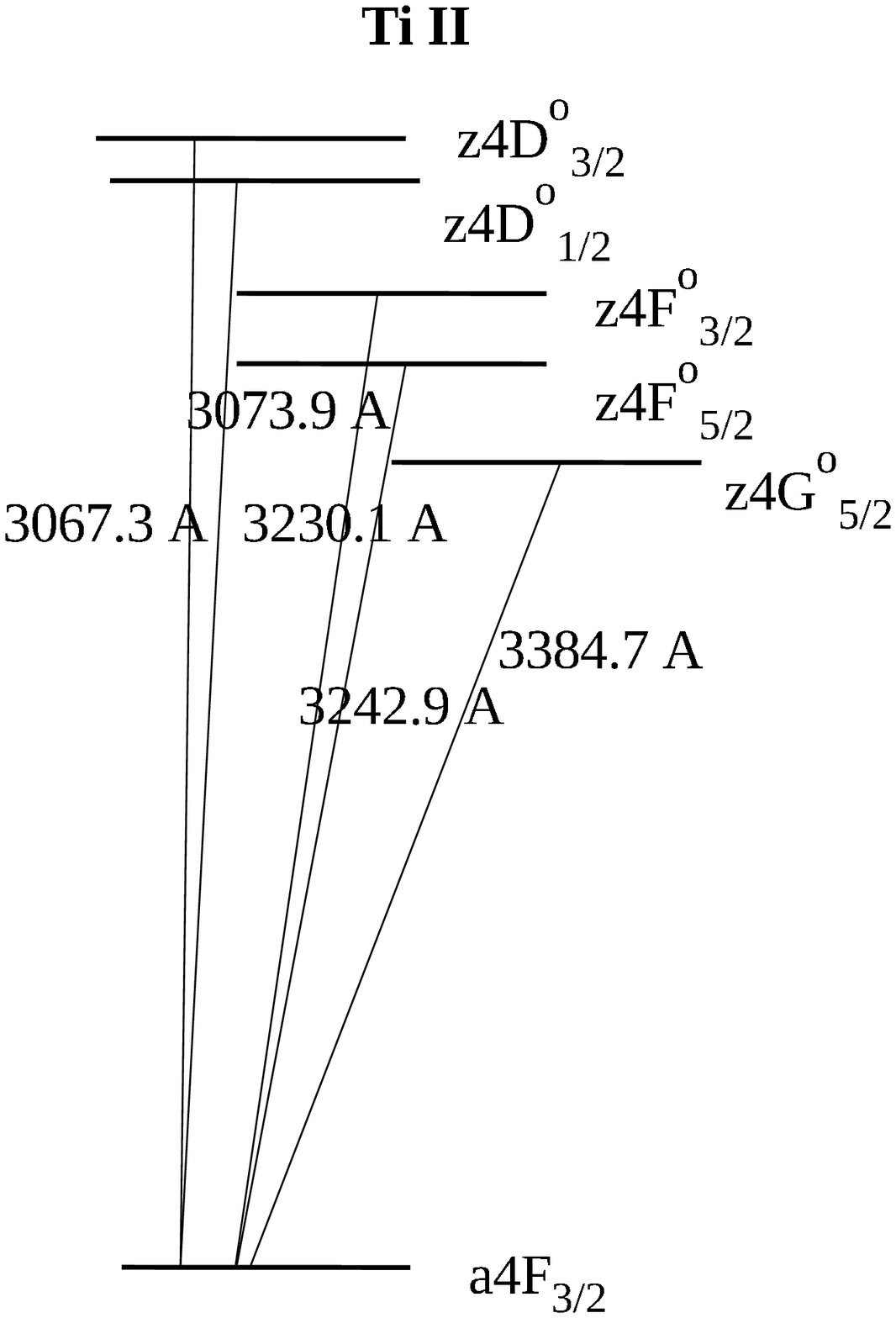}
\caption{{\em left}: schematics of the transitions within the fine structure of SII; {\em right}: schematics of the transitions within the fine structure of OI.}
\label{S2OI}
\end{figure}

When the magnetic precession rate becomes less than the emission rate of the upper level, the effect of magnetic field on the upper level is negligible. The only influence of magnetic field is on the ground state through the alignment of atoms. The atoms are aligned either parallel or perpendicular to the magnetic field. The photons absorbed or scattered by such aligned atoms are polarized in accordance with geometrical relation of the magnetic field with respect to the incoming light and line of sight (l.o.s.). Through the detection of the polarizations of the emission and/or absorption thus provide 3D information of the magnetic field. In YLa, we studied the polarizations of absorption lines arising from the atomic alignment. Here we shall use a couple of examples (S II and O I, see table~\ref{TiII} and Fig.\ref{S2OI}) to illustrate the emission line polarimetry induced by the atomic alignment.  

\begin{table}
\begin{tabular}{||c|c|c|c|c|c|c|c|c|c|c|c|c|c|c||}
\hline
\hline
$J_i\rightarrow J_f$&$p_2$&$r_{02}$&$r_{20}$& $r_{20}$&$r_{22}$& $r_{22}$& $r_{22}$&$r_{24}$& $r_{24}$& $r_{24}$&$s_{02}(s_{20})$&$s_{22}$&$s_{24}(s_{42})$\\
&&&$(q=0,\pm2)$&$(q=\pm 1)$&$(q=0)$&$(q=\pm2)$&$(q=\pm1)$&$(q=0)$&$(q=\pm2)$&$(q=\pm1)$&&&\\
\hline
$2\rightarrow 1$&0.15275&0.15275&0.02582&-0.02582&0.0309&-0.0309&-0.0154&0.24842&0.0414&0.1656&0.5916&0.1515&0.2711\\
$2\rightarrow 2$&0.1&-0.11832&-0.11832&0.11832&-0.10102&0.1010&0.0505&-5.4210e-2&-0.0090&-0.0361&-0.5916&-0.1515&-0.2711
\\$2\rightarrow 3$&0.14&2.8571e-02&0.0828&-0.0828&2.8278e-02
&-0.0283&-0.0141&6.3232e-03&0.0011&0.0042&0.169&4.329e-2&7.7444e-2\\
\hline
$\frac{3}{2}\rightarrow \frac{1}{2}$&0&0.25&\multicolumn{5}{c|}{0}&\multicolumn{3}{c|}{NA}&0.28284&0&0\\
$\frac{3}{2}\rightarrow \frac{3}{2}$&0.05&-0.14142&-0.14142&0.14142&-0.14142&0.14142&0.0707&\multicolumn{3}{c|}{NA}& -0.22627&0&0\\
$\frac{3}{2}\rightarrow \frac{5}{2}$&1.5275e-1&2.8868e-2&0.10801&-0.10801&3.0861e-2&-3.0861e-02&-1.543e-02&\multicolumn{3}{c|}{NA}&5.657e-2&0&0\\
\hline
\hline
\end{tabular}
\caption{The numerical coefficients for computing the density matrices of OI (1st--2nd rows) and Ti II, S II (3rd--5th rows) (see Eqs.\ref{evolution}-\ref{skk}).}
\label{OIparm}
\end{table}

\begin{figure}
\includegraphics[%
  width=0.33\textwidth,
  height=0.25\textheight]{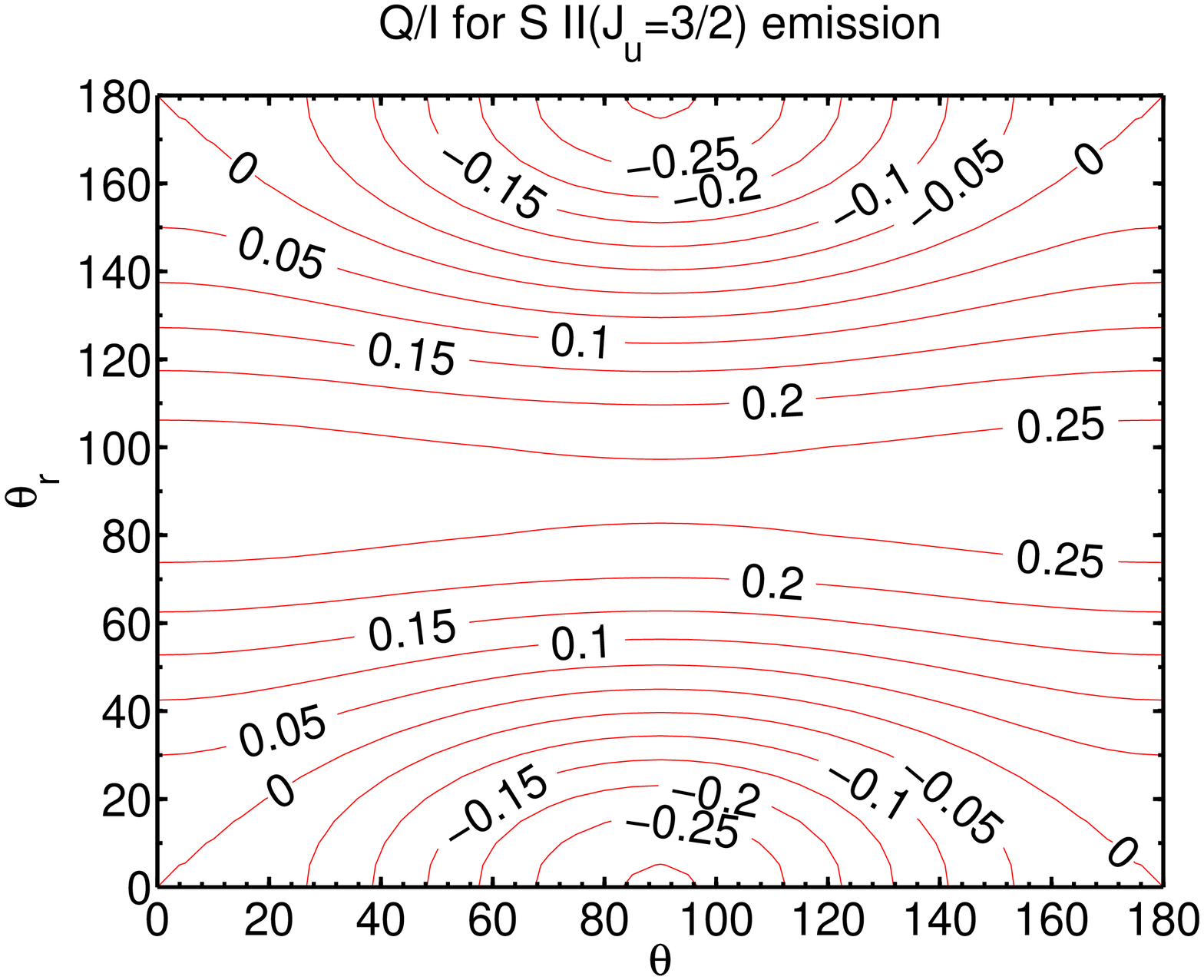} 
\includegraphics[%
  width=0.33\textwidth,
  height=0.25\textheight]{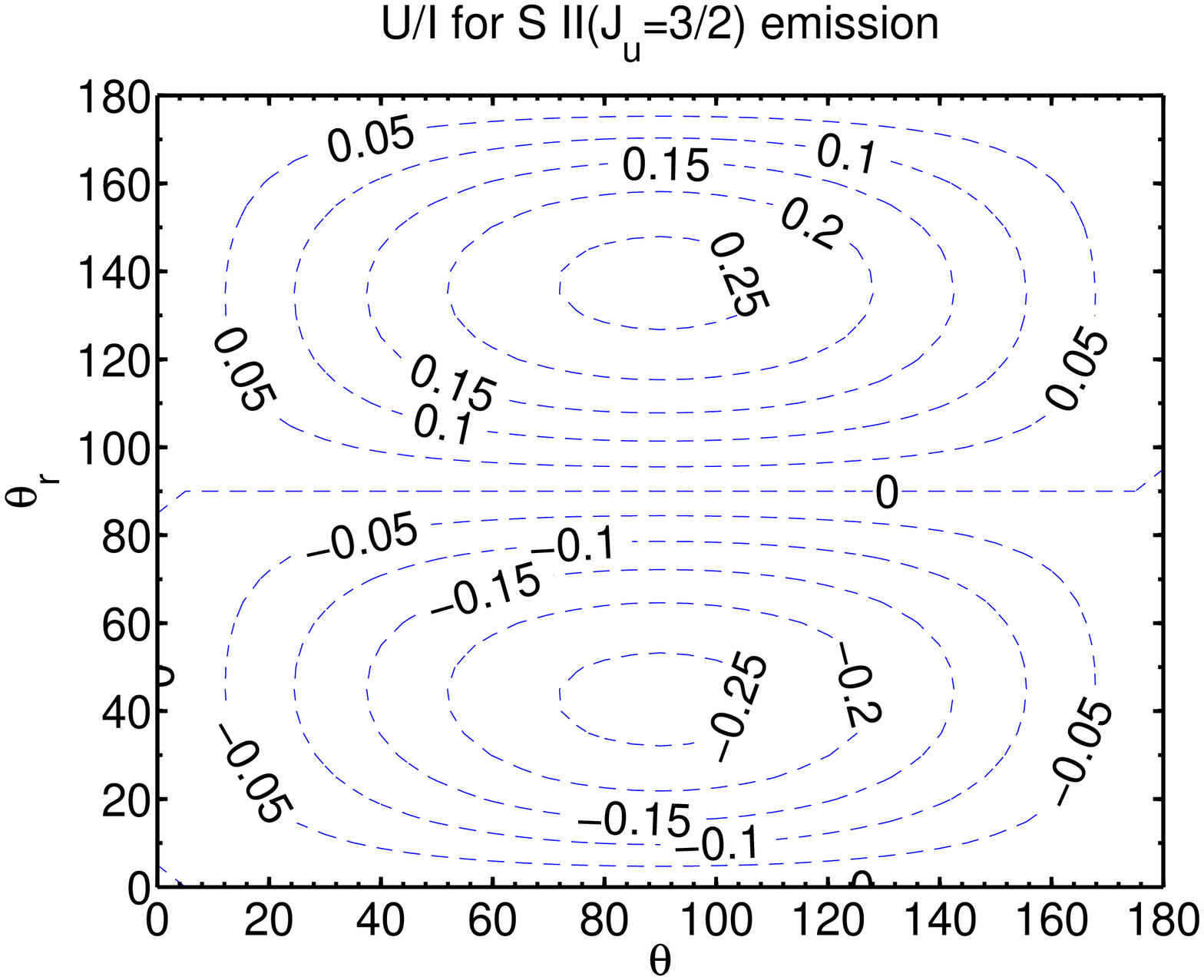} 
\includegraphics[%
  width=0.33\textwidth,
  height=0.25\textheight]{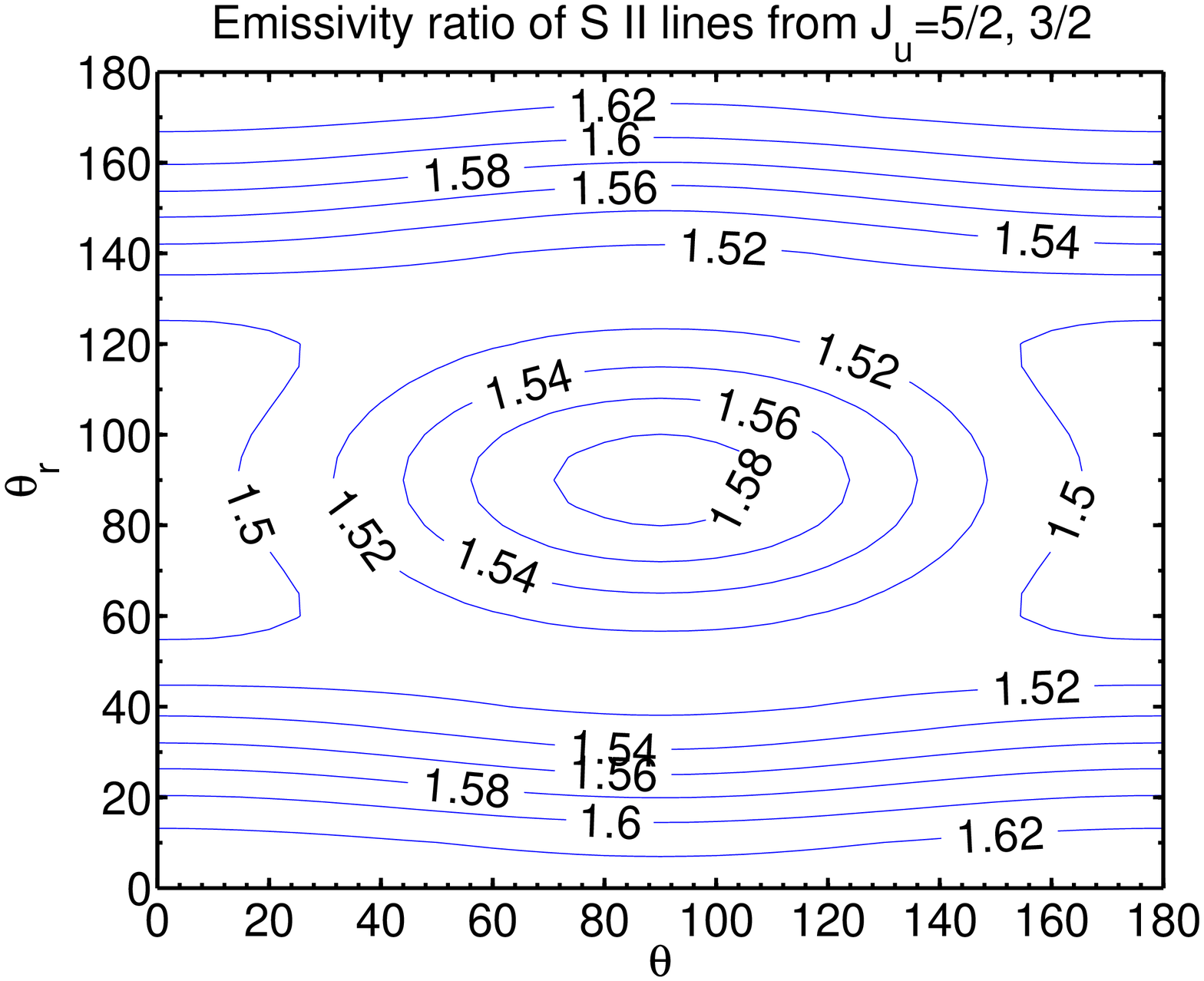} 
\caption{\emph{Left \& middle}: the contour graphs of S II emission line polarization from $J_u=5/2$; \emph{Right}: the contour graph of emissivity ratio of lines from $4P_{5/2,3/2}$ to $4S^o$. Parameters adopted: $R_{SD}=2.36$ (Eq.\ref{rSD}), $\phi_r=\pi/2$ (see Fig.\ref{radiageometry}\emph{right}).}  
\label{SIIaln}
\end{figure}

To remind our readers, SII has a ground level $4S^o_{3/2}$ and upper levels $4P_{1/2,3/2,5/2}$, which means $J_l=3/2$, $J_u=1/2,3/2,5/2$. The irreducible density tensor of the ground state $\rho^k_q(J_l)$ has components with $k=3/2-3/2, 3/2-1/2,...,3/2+3/2=0,1,2,3$. The ground state alignment, we obtained in YLa for SII is
\be
\sigma^2_0(J_l)=\frac{1.0481-3.1442 \cos^2\theta_r}{0.264 \cos^2\theta_r-13.161}
\label{S2ground}
\ee
The basic formalism is described in the previous section (Eq.\ref{evolution}-\ref{lowlevel},\ref{uplevel}-\ref{emissivity}) except that we are now in a limiting case $\Gamma'\rightarrow 0$ and $\Gamma \gg 1$. By inserting Eq.(\ref{S2ground}) into Eq.(\ref{uplevel}), we obtain
\bea
\rho^{0}_0\left(\begin{array}{c}J_u=3/2\\J_u=5/2\end{array}\right)&=&\frac{BI_*}{A}\rho^0_0(J_l)\left(
\begin{array}{l}
 1.887\cos ^4\theta_r-0.886 \cos
   ^2\theta_r-12.987\\
 1.6634 \cos ^4\theta_r+6.1058\cos
   ^2\theta_r-2.220 
\end{array}
\right)/\nonumber\\
&/&(1.487\cos^2\theta_r-52.787)\nonumber\\
\rho^{2}_0\left(\begin{array}{c}J_u=3/2\\J_u=5/2\end{array}\right)&=&\frac{BI_*}{A}\rho^0_0(J_l)\left(
\begin{array}{l} -0.385 \cos
   ^4\theta_r+0.560\cos ^2\theta_r-10.818 \\
  -0.2413\cos^4\theta_r-7.7510 \cos ^2\theta_r+2.6105
\end{array}
\right)/\nonumber\\
&/&(1.487\cos^2\theta_r-52.787)\nonumber\\
\rho^2_2\left(\begin{array}{c}J_u=3/2\\J_u=5/2\end{array}
\right)&=&\frac{BI_*}{A}\rho^0_0(J_l)e^{-2 i \phi_r} \sin^2\theta_r \left(\begin{array}{c}3.4892-0.86125\cos^2\theta_r \\
0.23763\cos ^2\theta_r-2.5249\end{array}
\right)/\left[(1.487+2.975i\Gamma')\cos^2\theta_r-52.787-105.57i\Gamma'\right]\nonumber\\
\rho^2_1\left(\begin{array}{c}J_u=3/2\\J_u=5/2\end{array}
\right)&=&\frac{BI_*}{A}\rho^0_0(J_l)e^{-i \phi_r}\sin2\theta_r \left(\begin{array}{c}-0.2940\cos
   ^2\theta_r-3.1041\\   0.0145\cos^2\theta_r+2.4409\end{array} \right)/\left[(1.487\cos^2\theta_r-52.787)(1+i\Gamma')\right].
\label{updensity}
\eea
In the magnetic realignment regime, we can directly get the density matrices by setting $\Gamma'=0$ in the equation. Then for optically thin case the Stokes parameters can be readily obtained by inserting the density matrices into the Eq.(\ref{emissivity}). The results in the optically thin case $Q/I\simeq \epsilon_1/\epsilon_0\,U/I\simeq \epsilon_2/\epsilon_0$ are illustrated in Fig.\ref{SIIaln} and table \ref{TiII}.

Another atomic species, namely, OI can be used as a magnetic field diagnostics. It can be used, for instance, in
reflection nebulae, since the lack of ionizing flux limits the number of levels being
pumped, and especially since common the atom is not ionized, which eliminates the interference of
the recombination radiation. The fluorescent line OI 8446 can be seen in spectra by Gordon, et al
(2000) in NGC 7023.

\begin{figure}
\includegraphics[%
  width=0.33\textwidth,
  height=0.25\textheight]{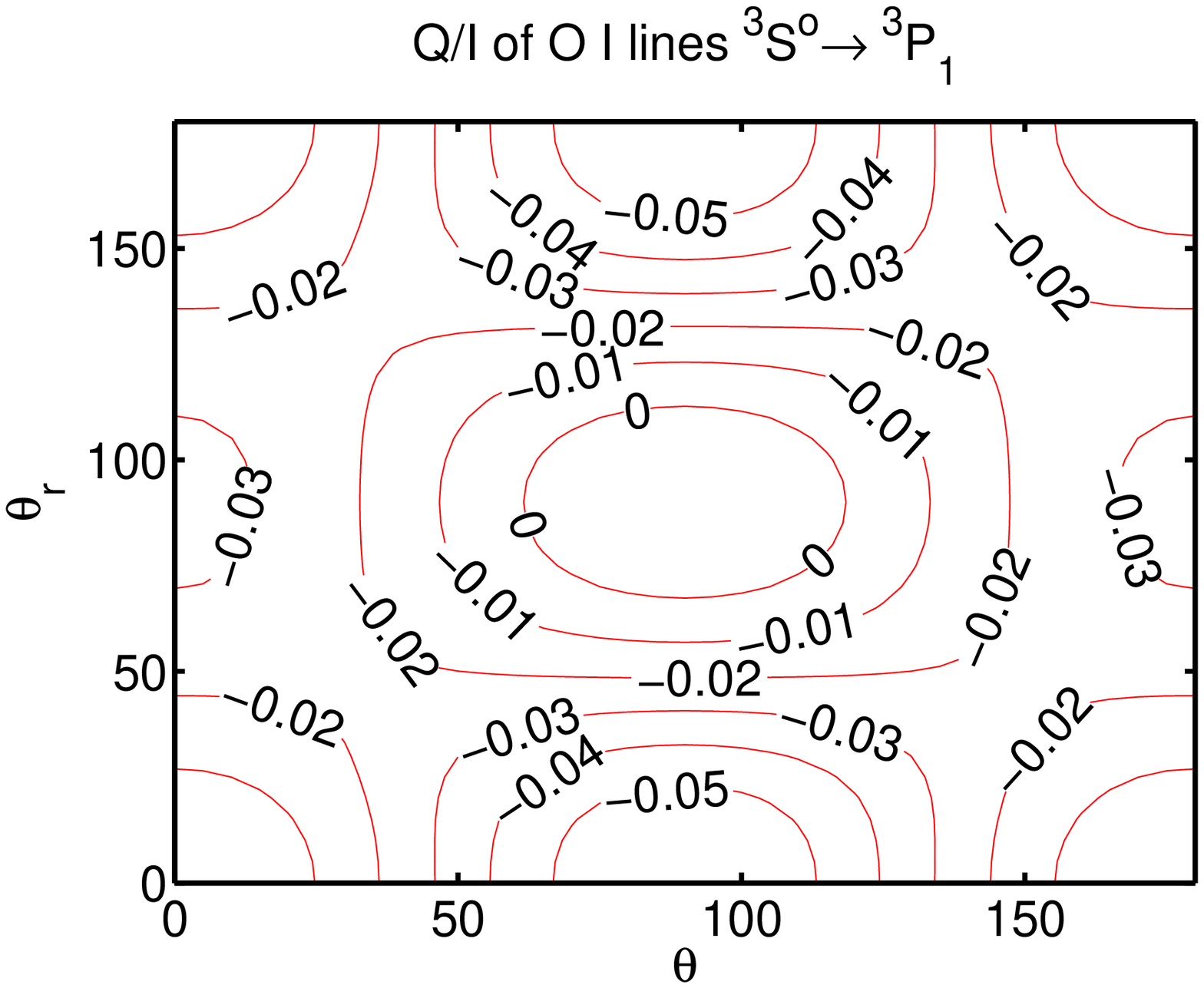}\includegraphics[%
  width=0.33\textwidth,
  height=0.25\textheight]{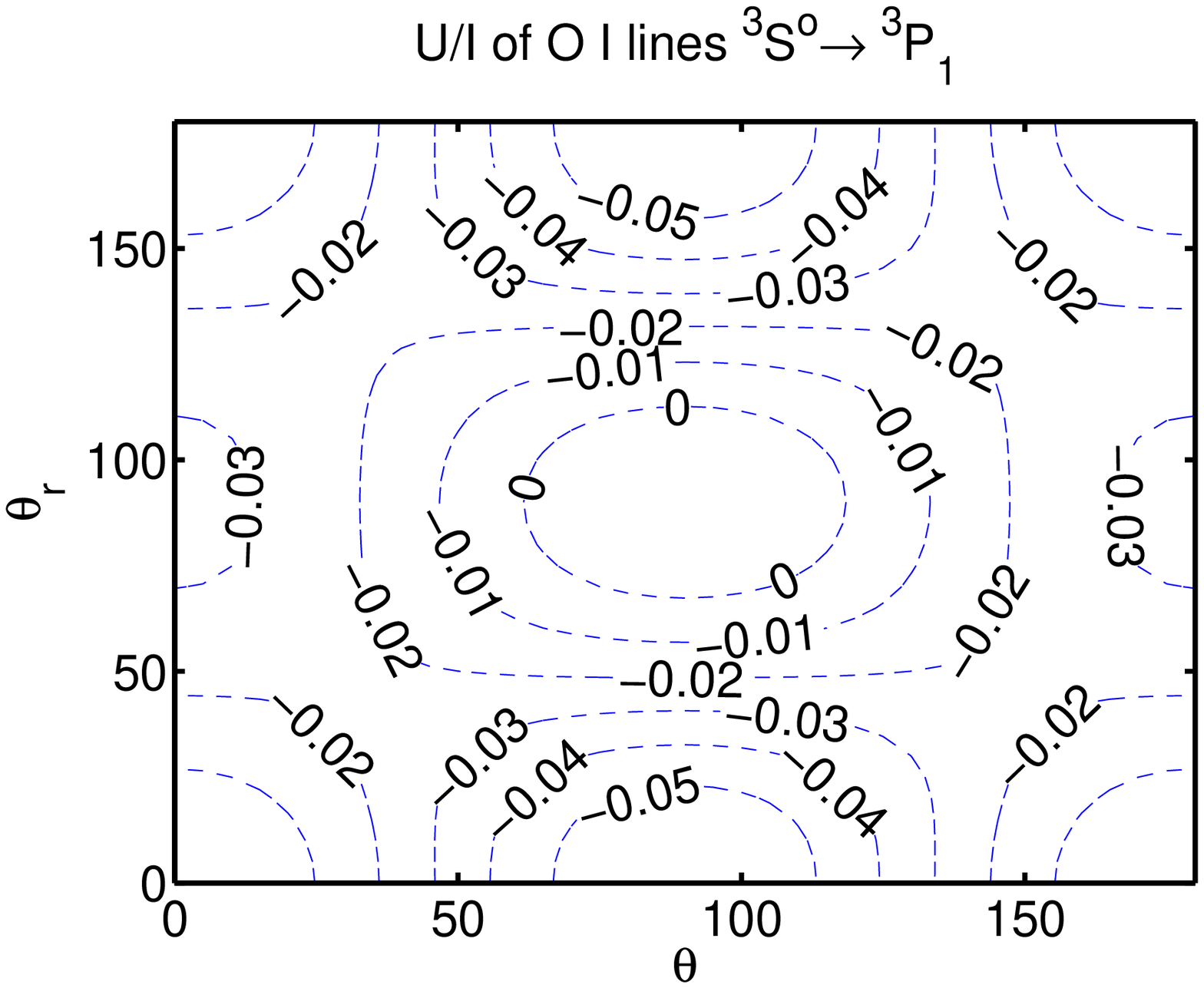}
\includegraphics[%
  width=0.33\textwidth,
  height=0.25\textheight]{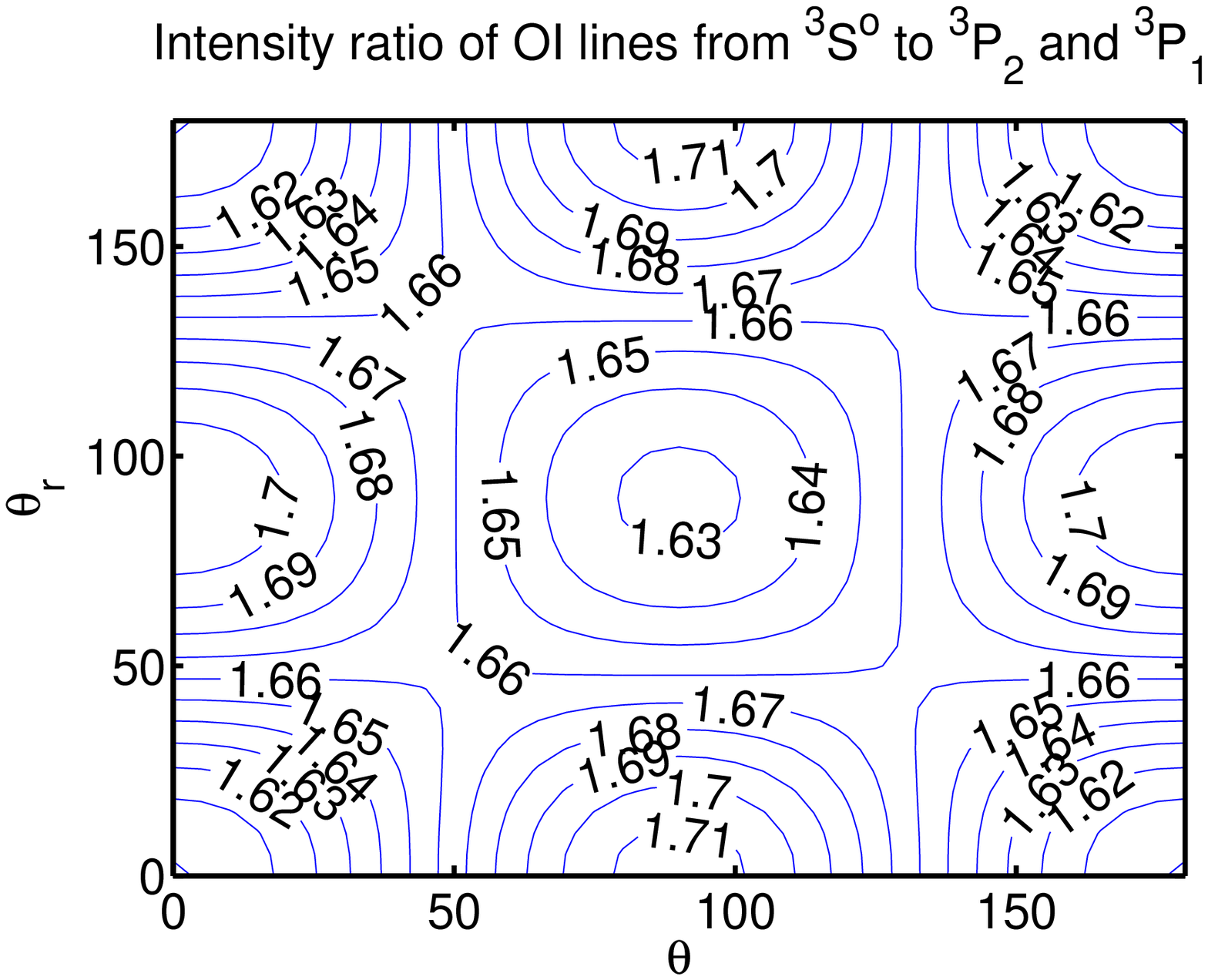}
\caption{\emph{Left \& Middle}: the contour graphs of polarization of OI lines $^3S^o\rightarrow ^3P_{1}$, \emph{Right}: the contour graph of emissivity ratio of OI lines from $^3S^o$ to $^3P_{2,1}$. $\phi_r=\pi/2$. We consider here the weak pumping regime, where only pumping from the ground level is accounted for (YLa).}
\label{OIaln}
\end{figure}

We attained the density matrices of the ground state in YLa. The density matrices on the upper level occupation are then obtained by inserting them into Eq.\ref{uplevel}. 
\bea
\rho^0_0(J_u)&=&\frac{\rho^0_0(J_l)[J_l]B}{\sum A''}\left[p_0(J_u, J_l){\bar J}^0_0+r_{02}(J_u, J_l){\bar J}^2_0\sigma^2_0(J_l)\right]\nonumber\\
\rho^2_0(J_u)&=&\frac{\rho^0_0(J_l)[J_l]B}{\sum A''}\left\{\left[p_2(J_u, J_l){\bar J}^0_0+r_{22}(J_u, J_l,0,0){\bar J}^2_0\right]\sigma^2_0(J_l)+\left[r_{20}(J_u, J_l,0,0)+r_{24}(J_u, J_l,0,0)\sigma^4_0(J_l)\right]{\bar J}^2_0\right\},\nonumber\\
\rho^2_{\pm2}(J_u)&=&\frac{\rho^0_0(J_l)[J_l]B}{\sum A''\pm i2A\Gamma'}\left[r_{20}(J_u, J_l,-2,2)+\sigma^2_0(J_l)r_{22}(J_u, J_l,-2,2)+r_{24}(J_u, J_l,-2,2)\sigma^4_0(J_l)\right]{\bar J}^2_{\mp2}\nonumber\\
\rho^2_{\pm1}(J_u)&=& \frac{\rho^0_0(J_l)[J_l]B}{\sum A''\pm iA\Gamma'}\left[r_{20}(J_u, J_l,-1,1)+\sigma^2_0(J_l)r_{22}(J_u, J_l,-1,1)+r_{24}(J_u, J_l,-1,1)\sigma^4_0(J_l)\right]{\bar J}^2_{\mp1}
\label{rhouOI}
\eea
where
\bea
\left[\begin{array}{l}\sigma^2_0(J_l)\\\sigma^4_0(J_l)\end{array}
\right]=\left[
\begin{array}{l}
  \frac{\left(\text{5.135} +\text{13.183} R_{SD}-\text{44.666} R_{SD}^2\right) \cos
   ^4\theta_r+\left(-\text{46.389} +\text{31.12} R_{SD}+\text{113.15} R_{SD}^2\right) \cos
   ^2\theta_r+\text{14.892} -\text{32.755}
   R_{SD}^2-\text{11.838} R_{SD}}{\left(-\text{2.35} -\text{5.654} R_{SD}+\text{11.837} R_{SD}^2\right) \cos^4\theta_r+\left(\text{21.789} +\text{47.27} R_{SD}+\text{37.903} R_{SD}^2\right) \cos^2\theta_r-\text{71.847} -\text{133.38}R_{SD}^2-\text{199.188} R_{SD}} \\
 \frac{\left(1 -\text{2.976} R_{SD}+\text{2.017} R_{SD}^2\right) \left(-\text{11.885} \cos
   ^4\theta_r+\text{7.923} \cos
   ^2\theta_r-\text{1.321}\right)}{\left(-\text{2.35} -\text{5.654} R_{SD}+\text{11.837} R_{SD}^2\right) \cos^4\theta_r+\left(\text{21.789} +\text{47.27} R_{SD}+\text{37.903} R_{SD}^2\right) \cos^2\theta_r-\text{71.847} -\text{133.38}R_{SD}^2-\text{199.188} R_{SD}}
\end{array}
\right],
\eea
\be
R_{SD}=\sum_{S21} I_\nu S_\nu/\sum_{D23} I_\nu S_\nu\label{rSD}\ee is a measure of the ratio of transitional probabilities from $^3P_2$ to $^3S^o_1$ states and $^3D^o_3$ states (see also YLa), where the line strength $S_\nu$ can be found, e.g., in the NIST Atomic Spectra Database\footnote{Available at http://physics.nist.gov/PhysRefData/ASD/index.html}. We emphasize that there are actually many states of different energies for the $^3S^o$ term and $^3D^o$ term. All of the transitions should be taken into account when calculating $R_{SD}$. Inserting Eqs.(\ref{irredradia},\ref{rhouOI}) and $w^2_{J_uJ_l}$ (Eq.\ref{w2}, see also table 4 in YLa) into Eq.(\ref{emissivity}), using the coefficients given in table~\ref{OIparm}, we get the polarization of the emission lines (see table~\ref{TiII}). Fig.\ref{OIaln} shows the polarization of lines $^3S^o\rightarrow ^3P_{1}$ for weak pumping\footnote{Weak pumping means only pumping from the ground level is considered (see YLa).} in the case of $R_{SD}\simeq 2.36$. Note that lines with the same initial and final terms have the same polarization regardless of their energy state. This is because  is is the angular momentum that is the key quantity for the atomic alignment process. Therefore, many optical lines (see Grandi 1975) arising from transitions at the higher excited states can be easily inferred from the transitions to the ground state (see table \ref{TiII}). From Eq.(\ref{rhouOI}), we see that the density tensors $\rho^2_q(J_u)$ have finite values even if the atomic alignment ($\sigma^2,4_0$) is completely neglected. Indeed the polarization of scattering lines are frequently discussed in the literatures without accounting for what happens on the ground state (see, e.g., Stenflo 1994). This oversimplified approach, however, brings erroneous results as we show here.

Table~\ref{OIparm} lists the numerical coefficients needed for the calculations of alignment of the three species, Ti II, O I, and S II. 

\section{Polarization in Hanle regime}

In what follows, we shall use SII as an example to discuss the aforementioned different polarization regimes. SII absorption lines are observed in interstellar medium (Morton 1975). Since we have discussed the polarization of SII in atomic alignment regime, having the same species studied in different regimes enable us to understand better how the atomic polarization changes with the magnetic field strength and the environment.

\subsection{Upper level Hanle regime}

Hanle effect has been known since 20's of the last century and directly associated with the coherence of pure atomic states (see Bohr 1924; Hanle 1924, 1925; Heisenberg 1925). Observationally it is rotation and depolarization of resonant scattered light in a weak magnetic field compared to non-magnetized case. Although semi-classical theory can give us a qualitative picture, generic theory based on quantum electrodynamics (QED) and density matrix has to be invoked to give an accurate descriptions (see Landi Degl'Innocenti 1999, Trujillo Bueno 1999). 


One of the major changes introduced by the QED approach is the ability to include atomic polarization on the ground level. This effect was ignored in most cases for a long time (A few exceptions can be found in Bommier 1977, Landi Degl'Innocenti 1982).  Although Hanle effect has been noticed and applied to stellar winds, the effect of ground level polarization, however, was still excluded (see e.g. Ignace, Nordsieck, \& Cassinelli 1999).     

In Hanle regime, the influence of magnetic field to the ground state is exactly the same as in the atomic alignment regime and so the absorptions from the ground state. 
The emission, however, is different. This is because the upper state is also influenced by the magnetic field directly unlike in the atomic alignment regime. Since the magnetic splitting is comparable to the line-width, the density tensors of the upper level are affected by the strength of the magnetic field as well and they are determined by Eq.(\ref{updensity}).

\begin{figure}
\includegraphics[%
  width=0.33\textwidth,
  height=0.25\textheight]{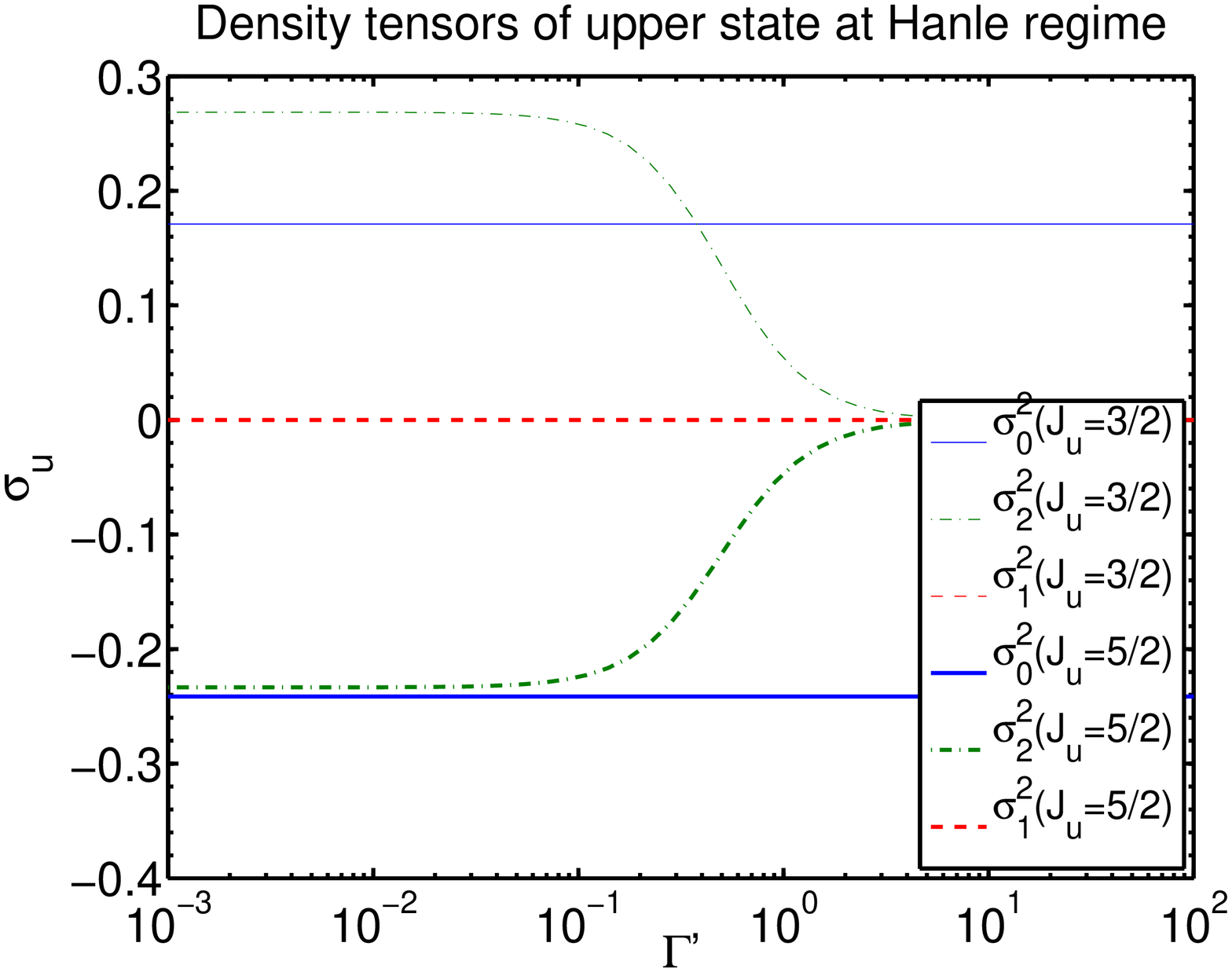}
\includegraphics[%
  width=0.33\textwidth,
  height=0.25\textheight]{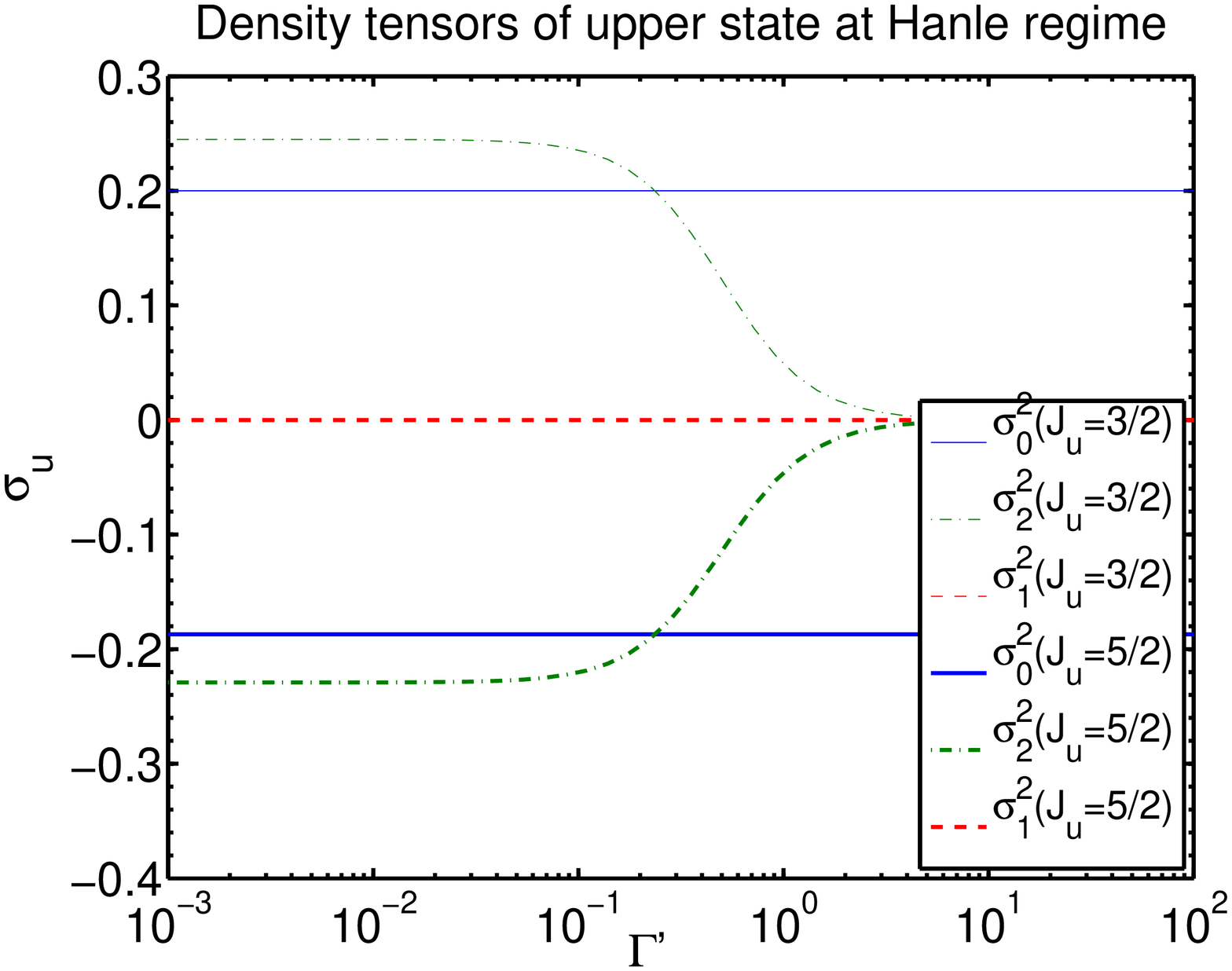}
\includegraphics[%
  width=0.33\textwidth,
  height=0.25\textheight]{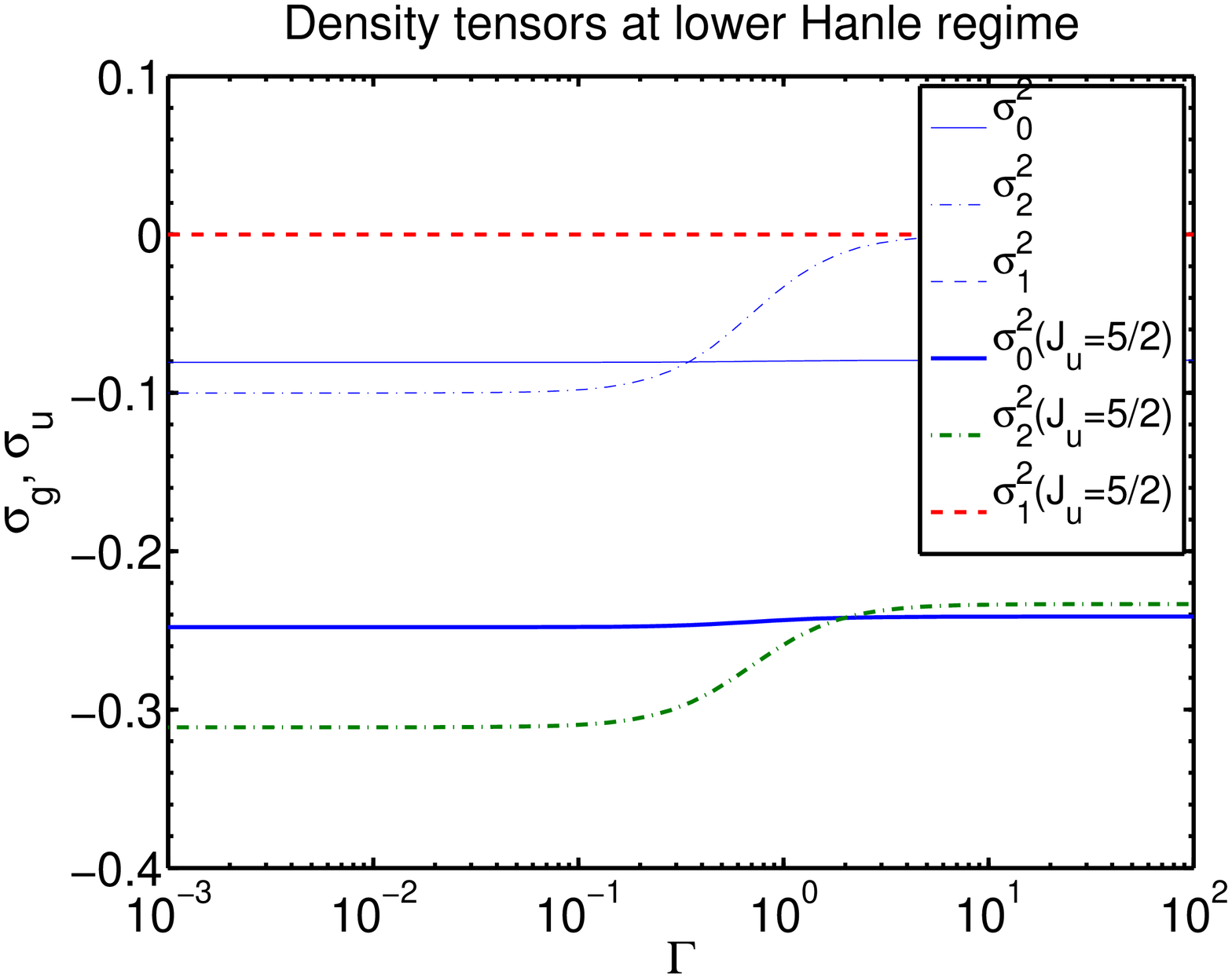}
\caption{The density matrix components vs. the ratio of magnetic splitting to the inverse of the life time of a level for $90^\circ$ scattering. Magnetic field is perpendicular to the radiation field, $\phi_B=90^\circ$. {\em Left}: The density components of upper states vs. $\Gamma'=2\pi\nu_Lg_u/A$ in the Hanle regime. As we see, coherence components $\sigma^2_2$ are reduced with the increase of the magnetic field strength. In the limit of $\Gamma'\gg 1$, the coherence disappears, corresponding to the saturated Hanle regime; {\em Middle}: the density components of upper states without account for ground level alignment in the Hanle regime; {\em Right}: The density components of upper and ground states vs. $\Gamma=2\pi\nu_Lg_l/B(3/2\rightarrow 5/2){\bar J}^0_0$ in the lower level Hanle regime. For $\Gamma\gg 1$, the coherence components become zeros on the ground state, approaching to the atomic alignment regime (see YLa,b).}
\label{density}
\end{figure}

\begin{figure}
\includegraphics[%
  width=0.45\textwidth,
  height=0.3\textheight]{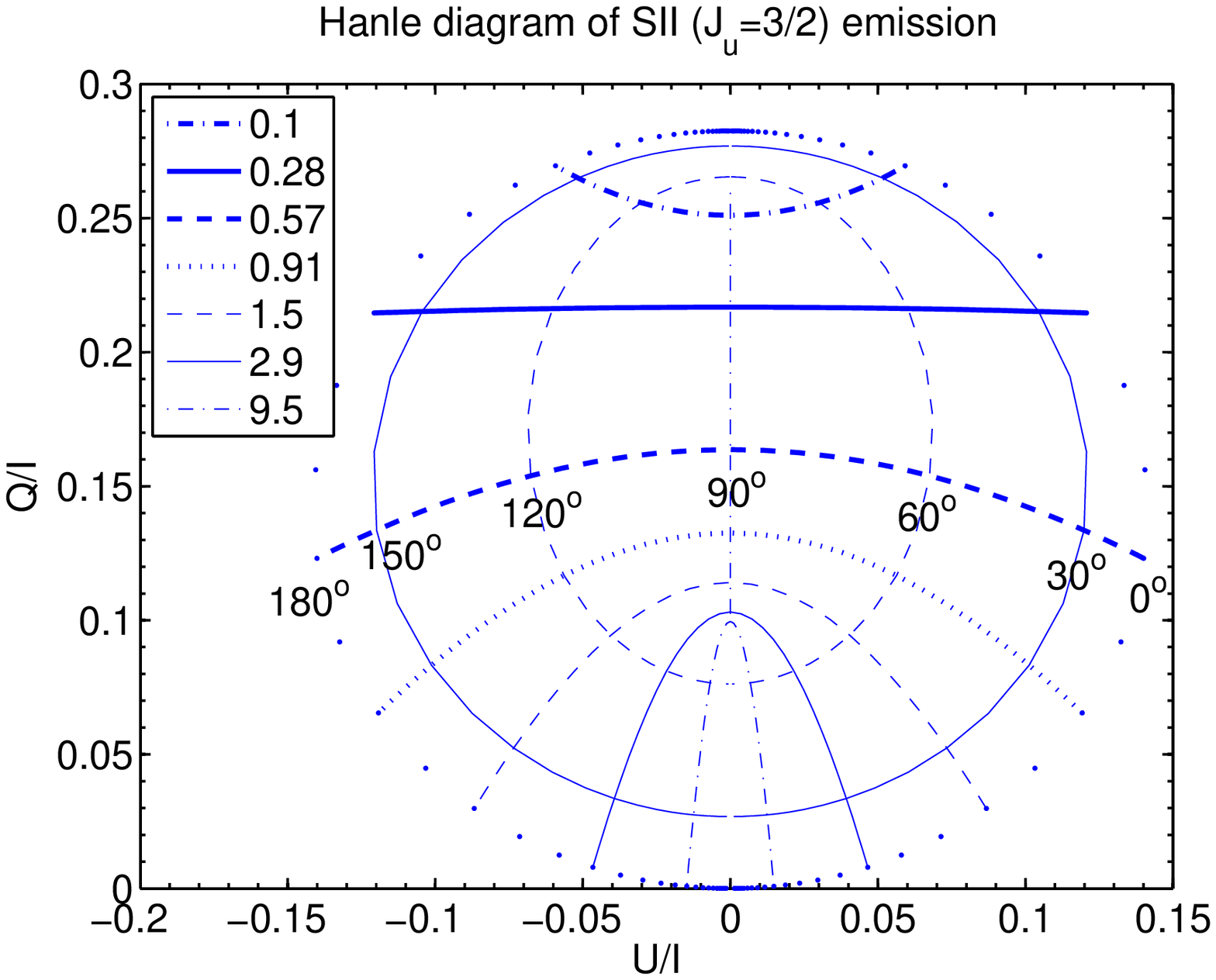} 
\includegraphics[%
  width=0.45\textwidth,
  height=0.3\textheight]{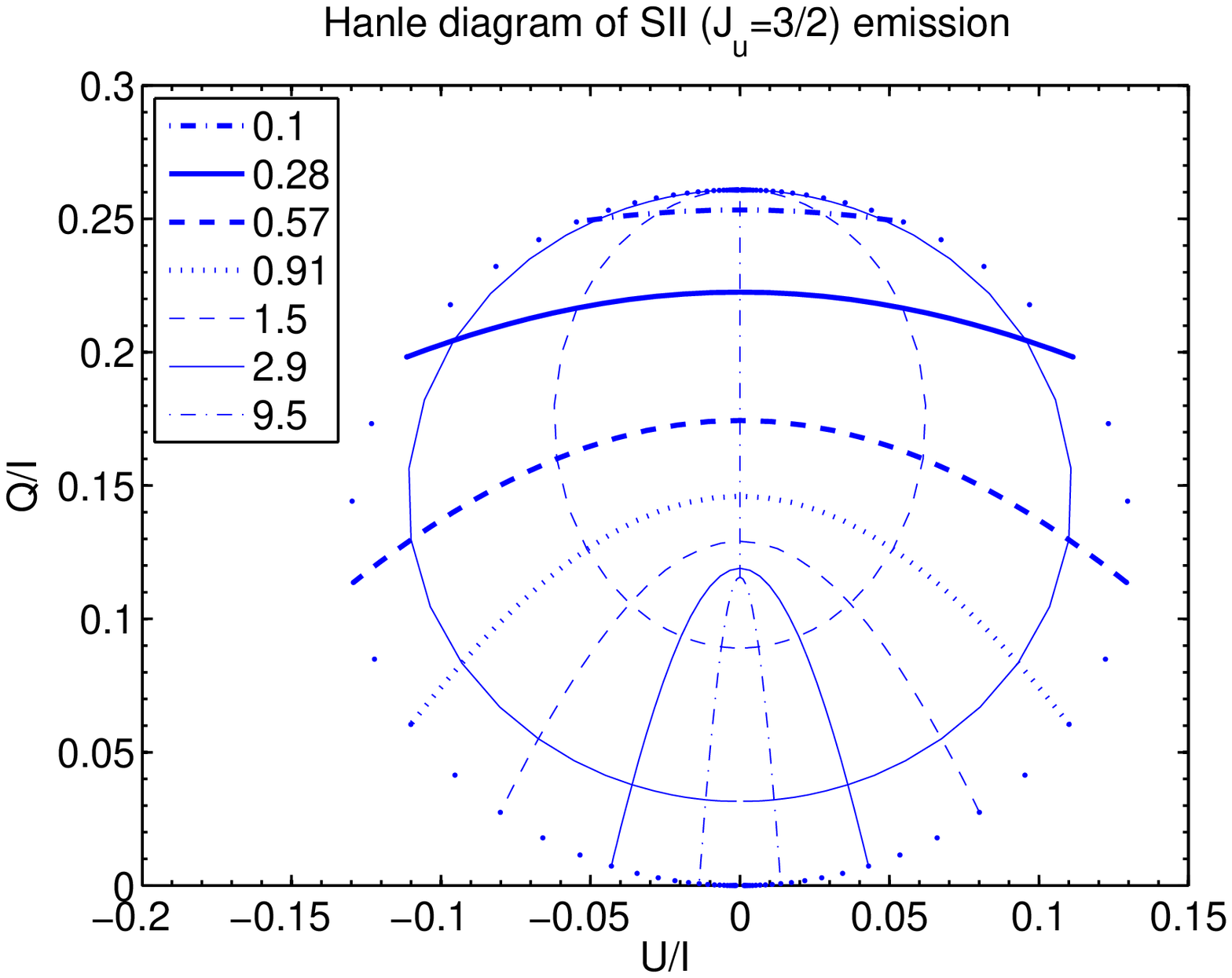} 
\includegraphics[%
  width=0.45\textwidth,
  height=0.3\textheight]{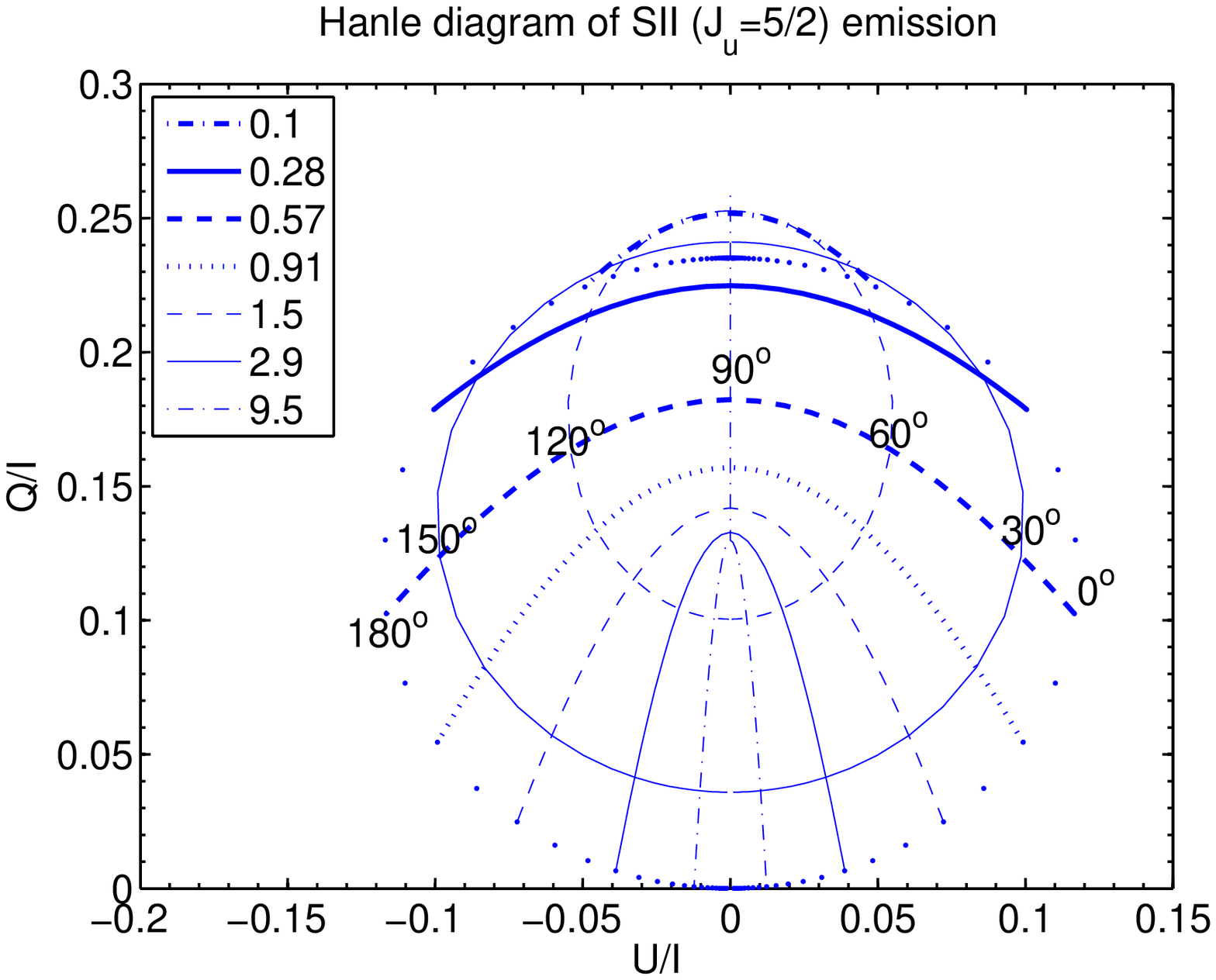} 
\includegraphics[%
  width=0.45\textwidth,
  height=0.3\textheight]{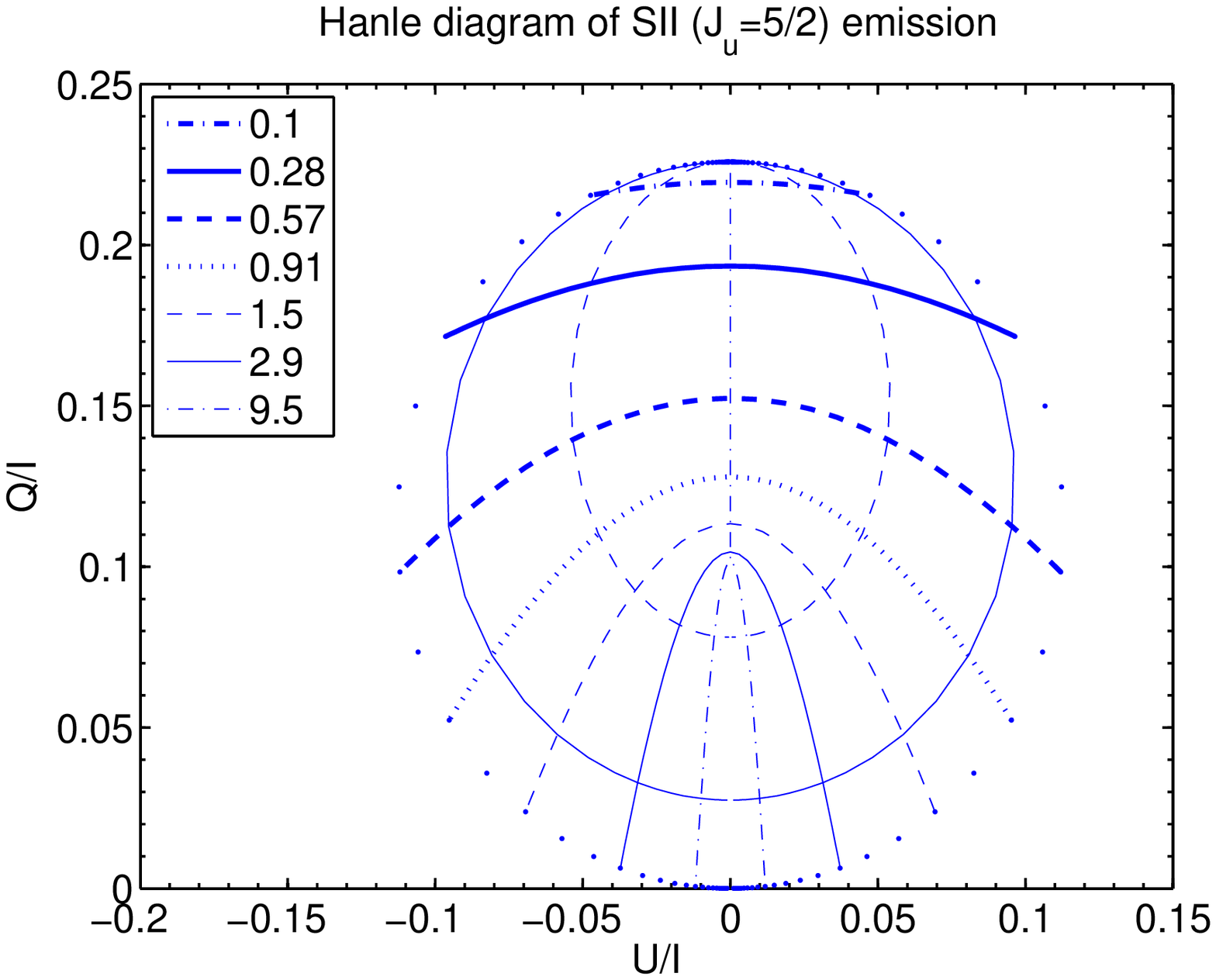}  
\caption{The polarization diagram of SII emission lines {\em left}: in the Hanle regime; {\em right}: in the Hanle regime, but without account for atomic alignment on the ground state. Same as in Fig.\ref{density}, $\theta_0=90^\circ, \phi_B=90^\circ$. The corresponding values of $\theta$ are marked in the plots, and the relative strength of magnetic field $\Gamma=2\pi\nu_Lg_u/A$ are illustrated in the legend. As $A\simeq 4.4\times 10^7{\rm s}^{-1}$, unity 
in the legend corresponds to the magnetic field of $0.25-25$G.}
\label{emission}
\end{figure}

The dependence of the density tensor components of the upper level $J_u=5/2$ are shown in the left panel of Fig.\ref{density}. Insert these density matrix components into Eq.(\ref{emissivity}), one can easily obtain the Stokes parameters in optically thin case. For the $90^\circ$ scattering ($\theta_0=90^\circ,\,\phi_B=90^\circ$ or $\theta_r=90^\circ,\,\phi_r=270^\circ$ as shown in Fig.\ref{radiageometry}), the polarization diagram are demonstrated in Fig.\ref{emission}. For comparison, we also show the density tensors and the Hanle diagrams without ground state alignment ($\sigma^2_0(J_l)=0$) in Fig.\ref{density}, and \ref{emission}. There is a notable difference between them, indicating ground state alignment plays an important role and should be accounted for. In addition to increasing the polarization, the main effect of ground state alignment is to reduce the degeneracy of $\theta$ (the angle between l.o.s. and magnetic field) when $\Gamma<1$. In other words, the atomic alignment carries on additional information of the direction of magnetic field, which would be veiled otherwise. In the opposite limit, when $\Gamma\gg 1$, the upper state is realigned by the fast magnetic mixing. Similar to the ``magnetic realignment regime", $U\rightarrow 0$ in this case and the polarization is either parallel or perpendicular to the magnetic field. This can be important for studies of magnetic fields in the intermediate range (10G$<B\lesssim 100$G).

\subsection{Lower level Hanle regime}   
  
\begin{figure}
\plottwo{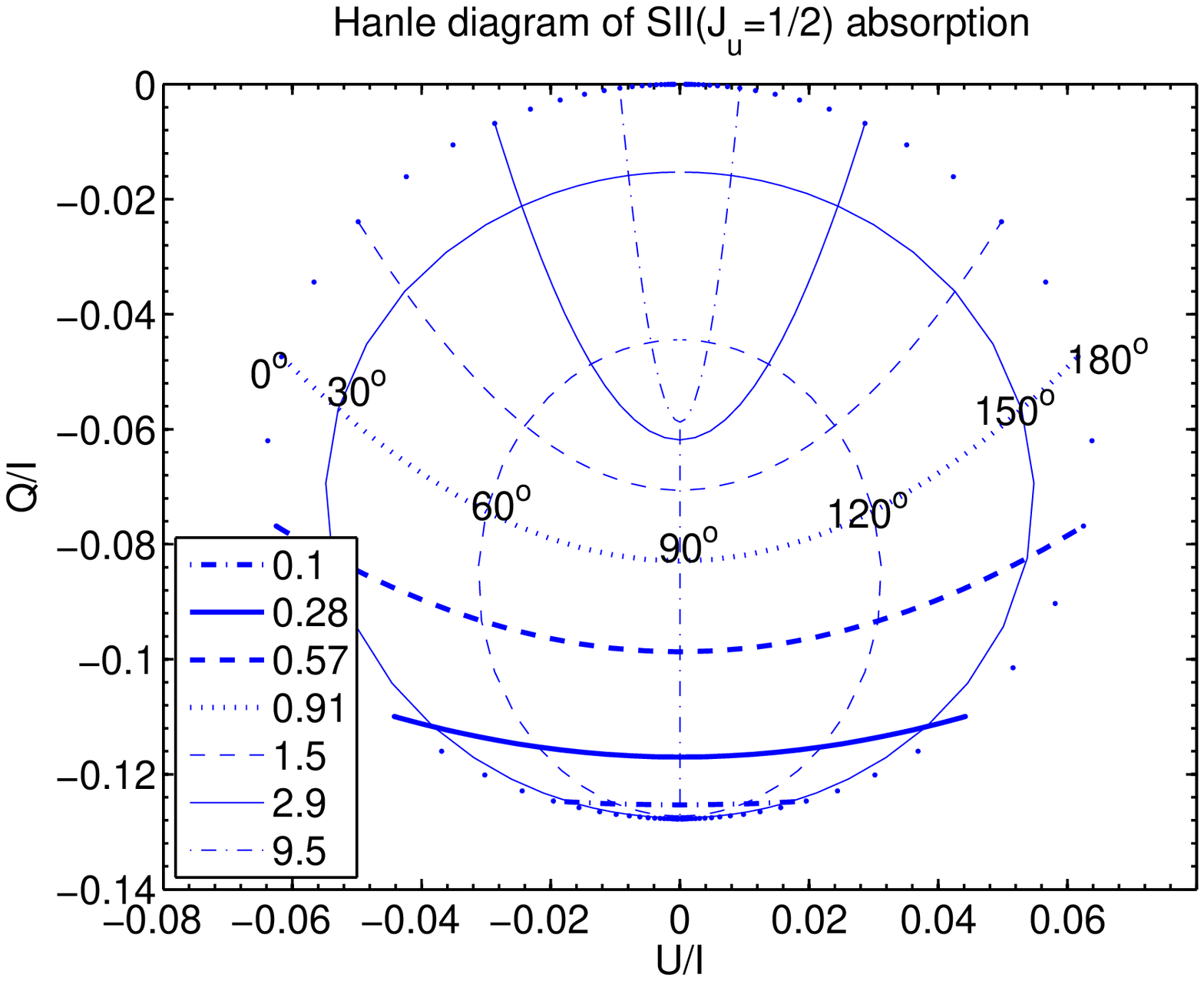}{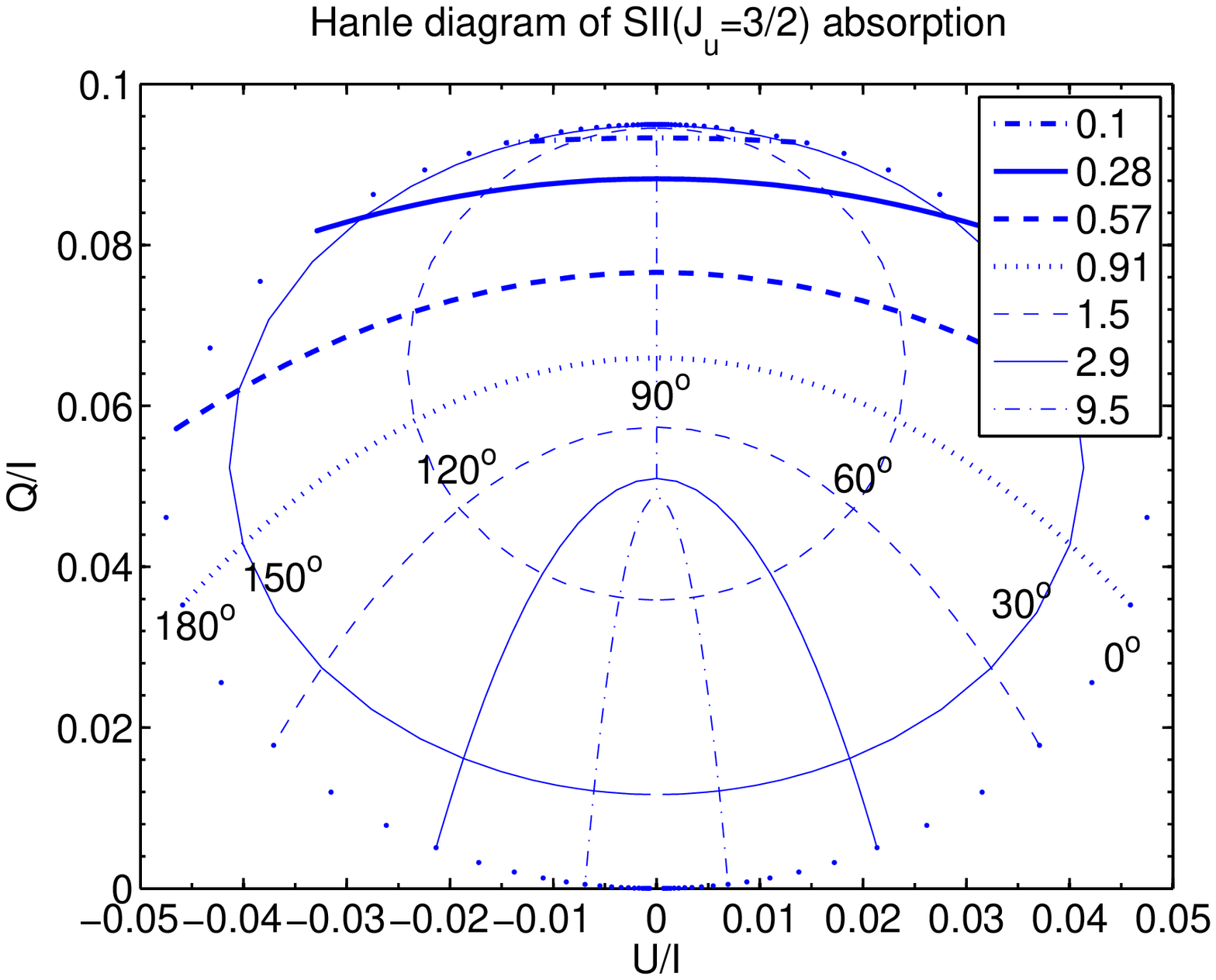}
\caption{The polarization diagram of SII absorption lines in the lower Hanle regime. Same as in Fig.\ref{density}, $\theta_0=90^\circ,\, \phi_B=90^\circ$. The corresponding values of $\theta$ are marked in the plots, and the relative strength of magnetic field $\Gamma=2\pi\nu_Lg_u/B(3/2\rightarrow 5/2){\bar J}^0_0$ are given in the legend.}
\label{absorb}
\end{figure}

If magnetic splitting is comparable to the optical pumping rate, 
which is the inverse of life-
time on the ground state, an analogy can be made between the ground state in this situation and the upper level in Hanle regime. In this case, coherence appears and ground state density matrix is modulated according to the strength and direction of the magnetic field. By solving Eq.(\ref{lowlevel}), we obtain the following density matrix components for the ground state, 
{\scriptsize
\bea
\frac{\rho^2_0}{\varrho^0_0}&=&-10825-(31990 \cos2 \theta_r+10663)
   \Gamma^4-\cos6 \phi_r \Gamma^2+2
   \cos6 \phi_r \cos2 \theta_r \Gamma^2-79494 \cos2
   \theta_r \Gamma^2-\cos6 \phi_r \cos4 \theta_r
   \Gamma^2+397 \cos4 \theta_r \Gamma^2\nonumber\\
&-&26393 \Gamma^2- 428\left(-0.5 \Gamma^2+\left(\Gamma^2+0.5\right) \cos2\theta_r+\left(-0.5 \Gamma^2-0.25\right) \cos4 \theta_r\right) \sin2\phi_r \Gamma-107\left(-0.75 \Gamma^2+\left(\Gamma^2+2\right) \cos2 \theta_r\right.\nonumber\\
&-&\left.\left(0.25\Gamma^2+0.54\right) \cos4 \theta_r+0.02\cos6 \theta_r-1.47\right) \sin4 \phi_r\Gamma-4.3 (\cos2 \theta_r-0.4\cos4 \theta_r-0.66)\sin6 \phi_r \Gamma+\cos6\phi_r\nonumber\\
&-&2 \cos6 \phi_r \cos2\theta_r-32222 \cos2 \theta_r+\cos6
   \phi_r \cos4 \theta_r-113 \cos4
   \theta_r-\cos6 \theta_r+\cos4 \phi_r\left[-\cos6 \theta_r \Gamma^2+55\Gamma^2-\left(74\Gamma^2+151\right) \cos2\theta_r\right.\nonumber\\
&+&\left.\left(20\Gamma^2+38\right) \cos4\theta_r+113\right]+\cos2 \phi_r
   \left[3\Gamma^2+\left(-607\Gamma^2-306\right) \cos2 \theta_r+\left(607\Gamma^2+306\right) \cos4 \theta_r-3(\Gamma^2-0.95^2) \cos6\theta_r\right],\nonumber
\eea
\bea
\frac{\Re(\rho^2_2)}{\varrho^0_0}&=&369 \Gamma^2 \cos4
   \phi_r \sin ^4\theta_r-374\cos4 \phi_r \sin
   ^4\theta_r-8 \Gamma^2 \cos4 \phi_r \cos2 \theta_r
   \sin ^4\theta_r-760 \Gamma \sin4 \phi_r \sin
   ^4\theta_r+39 \Gamma \cos2 \theta_r \sin4 \phi_r
   \sin ^4\theta_r\nonumber\\
&-&12215 \Gamma^2 \cos2 \phi_r \sin
   ^2\theta_r-26433 \cos2 \phi_r \sin
   ^2\theta_r+922 \Gamma^2 \cos2 \phi_r \cos2 \theta_r
   \sin ^2\theta_r+3\cos2 \phi_r \sin
   ^2\theta_r(\cos2 \theta_r-\cos4\theta_r)\nonumber\\
&+&\left[(524\cos2 \theta_r-18264) \Gamma^2-34687+3112 \cos2 \theta_r -15 \cos4 \theta_r\right] \Gamma\sin2 \phi_r \sin ^2\theta_r-140
   \Gamma^2+188 \Gamma^2 \cos2\theta_r\nonumber\\
&-&48\cos4\theta_r(\Gamma^2 +1)+186 \cos2 \theta_r-139,
\label{maindensity}
\eea
\bea
\frac{\Im(\rho^2_2)}{\varrho^0_0}&=&(9263 \cos2 \phi_r+131 \cos2\phi_r\cos4\theta_r -9394 \cos2\phi_r\cos2\theta_r) \Gamma^3-6707 \sin2\phi_r\Gamma^2+140 \sin4 \phi_r \Gamma^2+139\sin2 \phi_r\cos4 \theta_r
   \Gamma^2\nonumber\\
&-&188 \sin4 \phi_r\cos2 \theta_r
   \Gamma^2+6570 \sin2 \phi_r\cos2\theta_r
   \Gamma^2+48\sin4 \phi_r\cos4\theta_r
   \Gamma^2-\sin2\phi_r\cos6 \theta_r\Gamma^2+18903 \cos2 \phi_r \Gamma\nonumber\\
&+&298\cos4 \phi_r \Gamma- \cos4 \phi_r\cos6 \theta_r
   \Gamma-\cos2 \phi_r\cos6 \theta_r
   \Gamma-399 \cos4 \phi_r\cos2 \theta_r
   \Gamma+101\cos4 \phi_r\cos4\theta_r
   \Gamma-134 \cos2 \theta_r \Gamma\nonumber\\
&+&36\cos4 \theta_r \Gamma-\cos6 \theta_r
   \Gamma-18905\cos2 \phi_r\cos2\theta_r
   \Gamma+3\cos2 \phi_r\cos4\theta_r
   \Gamma+99 \Gamma-12848\sin2\phi_r-138 \sin4 \phi_r\nonumber\\
&+&\sin4\phi_r\cos6 \theta_r+\sin2 \phi_r\cos6\theta_r-371 \sin2 \phi_r\cos4\theta_r+186\sin4 \phi_r\cos2\theta_r-49 \sin4\phi_r\cos4\theta_r+13219 \sin2\phi_r\cos2\theta_r,\nonumber
\eea

\bea
\frac{\Re(\rho^2_1)}{\varrho^0_0}&=&
\left\{\Gamma \left[-45 \sin2\phi_r\left(\cos
   ^2\theta_r-18\right) \sin^2\theta_r-7.5\sin4\phi_r\left(\cos ^2\theta_r-1\right)^2\right]  \cos\phi_r+\left[(\sin ^2\phi_r \cos
   ^4\theta_r- \sin ^2\phi_r \cos
   ^2\theta_r+5 \cos2
   \theta_r\right.\right.\nonumber\\
&-&\left.\left.\cos4 \theta_r-4) \cos
   ^2 2 \phi_r+\left(21 \Gamma^2 \sin ^2\phi_r \sin^4\theta_r+185 \Gamma^2-183 \Gamma^2\cos2 \theta_r-\Gamma^2 \cos4
   \theta_r+738\sin^2\theta_r\right) \cos2 \phi_r+51821\Gamma^2\right.\right.\nonumber\\
&-&\left.\left.369 \Gamma^2 \sin ^2\phi_r+3
   \Gamma^2 \cos4 \theta_r \sin ^2\phi_r-738 \sin
   ^2\phi_r+\Gamma^2 \cos4
   \theta_r+\cos4 \theta_r+\cos2 \theta_r
   \left(-737\Gamma^2+\left(367
   \Gamma^2+738\right) \sin^2\phi_r+365\right)\right.\right.\nonumber\\
&+&\left.\left.26064\right] \cos\phi_r+\Gamma \left[-6980
   \Gamma^2+\left(262 \Gamma^2-396\right) \cos2
   \theta_r+\cos ^2 2 \phi_r (-4 \cos2
   \theta_r+ \cos4\theta_r+23)+5 \cos4 \theta_r\right.\right.\nonumber\\
&+&\left.\left.\cos2\phi_r (-400 \cos2 \theta_r+6 \cos4\theta_r+394)-18507\right] \sin\phi_r\right\}
   \sin2 \theta_r,\nonumber
\eea

\bea
\frac{\Im(\rho^2_1)}{\varrho^0_0}&=&
\left[(473 \cos2 \theta_r \sin\phi_r
   -\cos4 \theta_r \sin\phi_r
   -52085 \sin\phi_r +295
   \cos2 \theta_r \sin3 \phi_r -4\cos4
   \theta_r \sin3 \phi_r -820\sin3 \phi_r-\cos2 \theta_r \sin5 \phi_r
   \right.\nonumber\\
&+&\left.\sin5 \phi_r) \Gamma^2-\cos5 \phi_r \Gamma+\cos5 \phi_r \cos2
   \theta_r \Gamma+\cos3 \phi_r (397 \cos2
   \theta_r-5 \cos4 \theta_r-392)
   \Gamma+\cos\phi_r \left(-36790 \Gamma^2\right.\right.\nonumber\\
&+&\left.\left.\left(262 \Gamma^2-390\right) \cos2 \theta_r+3\cos4 \theta_r-18511\right) \Gamma-373 \cos2\theta_r \sin\phi_r+\cos4 \theta_r \sin\phi_r-26058 \sin\phi_r-4\cos2
   \theta_r \sin3 \phi_r\right.\nonumber\\
&+&\left.4\cos4 \theta_r \sin3
   \phi_r-\cos2\theta_r \sin5 \phi_r+\sin5 \phi_r\right] \sin2 \theta_r,\nonumber
\eea
and
\be
\rho^2_{-1}=-(\rho^2_1)^*,\,  \rho^2_{-2}=(\rho^2_2)^*,
\label{conj}
\ee
In Eq.(\ref{maindensity}),
\bea
\varrho^0_0&=&\rho^0_0/\left[(3783\cos2 \theta_r -264748)
   \Gamma^4-(5360\sin2\phi_r
   -5392\cos2 \theta_r \sin2\phi_r
   +31.5 \cos4 \theta_r \sin2\phi_r+13\cos2 \theta_r \sin4 \phi_r
   -3\cos4 \theta_r \sin4 \phi_r\right.\nonumber\\
&-&\left.9\sin4 \phi_r)\Gamma^3+ \left(13140\cos2 \theta_r-20 \cos4 \theta_r\right)
   \Gamma^2-658546-5317/2\sin2\phi_r \Gamma+5271/2\cos2
   \theta_r \sin2\phi_r \Gamma+23\cos4\theta_r \sin2\phi_r \Gamma\right.\nonumber\\
&-&\left.25 \cos2 \theta_r \sin4
   \phi_r \Gamma+6 \cos4 \theta_r \sin4 \phi_r
   \Gamma+19 \sin4 \phi_r \Gamma+52 \cos2 \theta_r-13 \cos4
   \theta_r+\cos4 \phi_r
   \left(7\Gamma^2+\left(-9 \Gamma^2-18\right) \cos2
   \theta_r\right.\right.\nonumber\\
&+&\left.\left.\left(2 \Gamma^2+5\right) \cos4 \theta_r+13\right)+\cos2 \phi_r
   \left(-7558 \Gamma^2+\left(7647 \Gamma^2+3738\right) \cos2\theta_r+\left(-89 \Gamma^2+9\right) \cos4 \theta_r-3747\right)-264112\right]
\label{rho00ground}
\eea}
$\rho^{0,2}_0(\Gamma^4)$ is of higher order of $\Gamma$ compared to other coherent components. Therefore in the limit $\Gamma\rightarrow \infty$, 
\be
\sigma^2_{q\neq 0}=\frac{\rho^2_{q\neq 0}}{\rho^2_0}=0, \, \sigma^2_0=\rho^2_0/\rho^2_0=-\frac{31990+10663\cos2\theta_r}{264748-3783 \cos2\theta_r}
\ee
agreeing with our earlier result in the atomic alignment regime (see Eq.17 in YLa).
Since the coherent components of the density matrix are not zeros, absorptions in the ground Hanle regime are different from the realignment case\footnote{Let us remind our reader, in atomic alignment regime, $U\equiv 0$, thus polarization can only be parallel or perpendicular to the magnetic field in the plane of sky, and the switch between the two direction always happens at Van-Vleck angle $\theta_r=54.7^o$.}. Although the amplitude of the polarization of the absorption lines are comparable (see Fig.\ref{absorb}, and Fig.8 in YLa), the direction of polarization becomes a complex function of both the strength and direction of the magnetic field as there are both Q and U components. It is similar to emission in this sense. Fig.\ref{absorb} is the Hanle diagram of SII absorption for the $90^\circ$ scattering ($\theta_0=90^\circ,\phi_B=90^\circ$, see Fig.\ref{radiageometry}{\em left}). For other geometries, the result can be readily obtained by inserting the density components (Eqs.\ref{maindensity}-\ref{rho00ground}) into Eq.(\ref{Mueller0}).

Apparently, the absorption from any lower level including metastable level would be affected in the same way as long as the absorption rate is comparable to the magnetic precession rate. For instance, the absorptions from the metastable level $b4F$ of TiII we discussed in \S4.1 and the metastable level $a6D$ of CrII we studied in YLa are polarized in a same fashion as SII when $\nu_L\sim \tau_R^{-1}$.

\begin{figure}
\includegraphics[%
  width=0.45\textwidth,
  height=0.3\textheight]{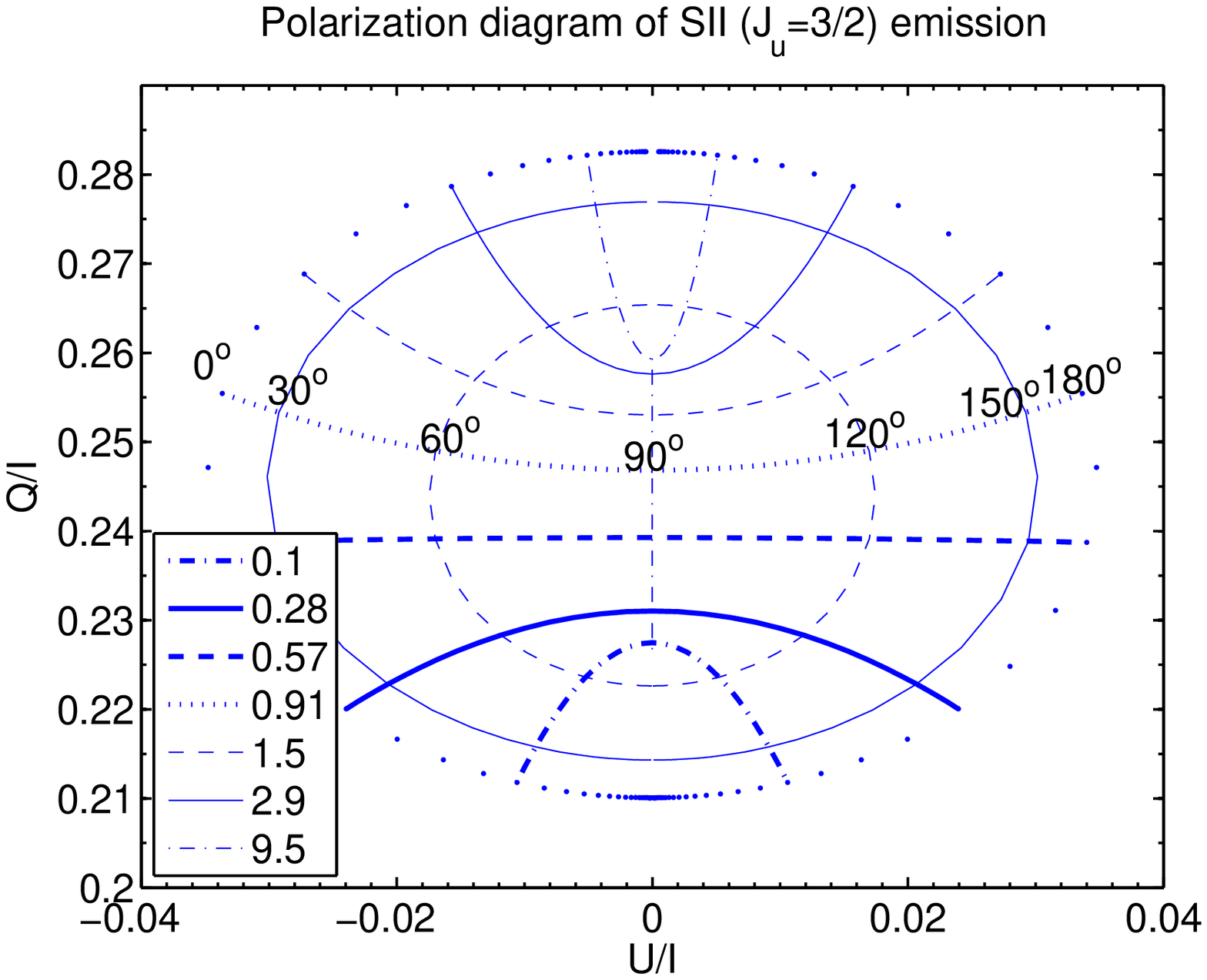} 
\includegraphics[%
  width=0.45\textwidth,
  height=0.3\textheight]{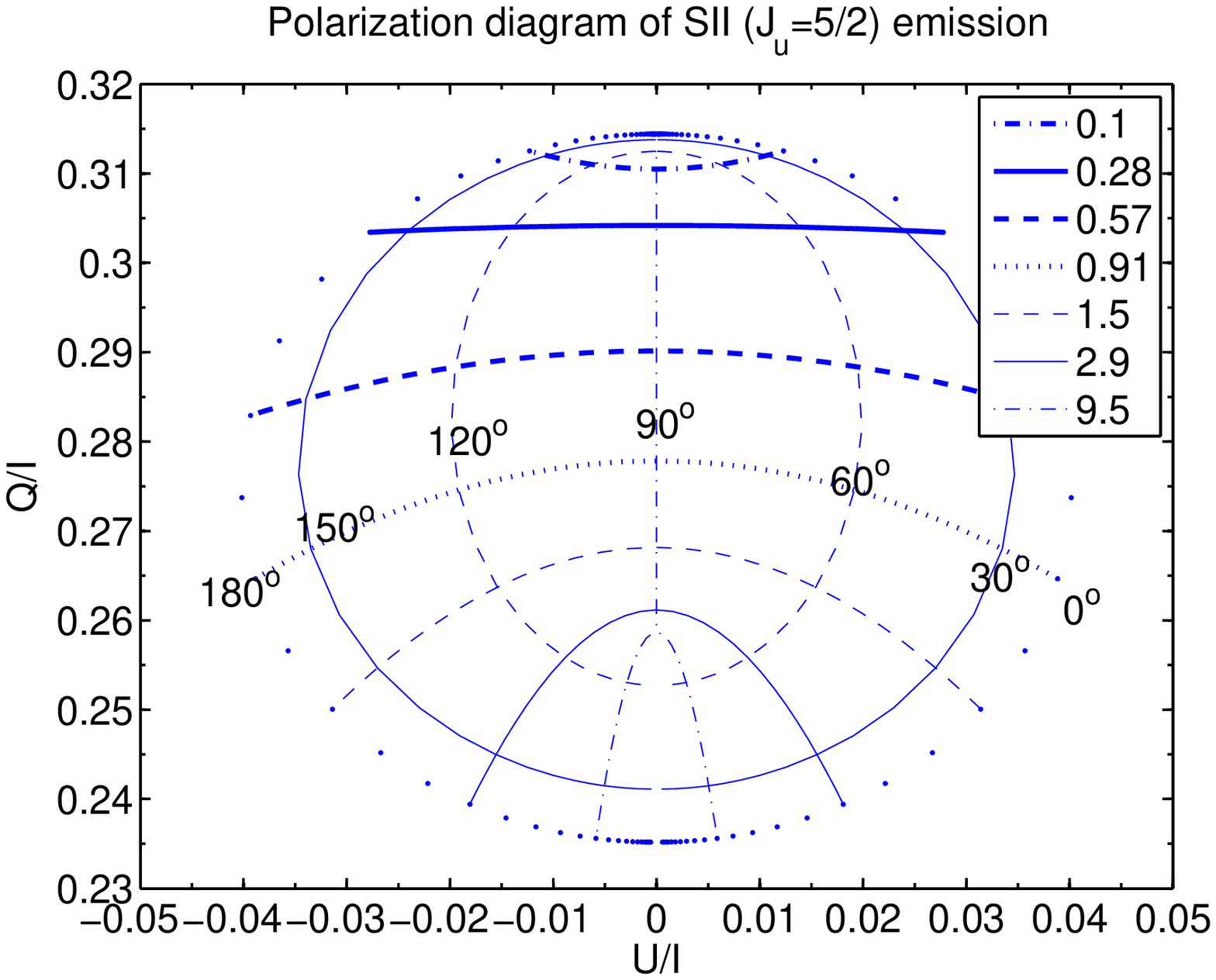}  
\caption{The polarization diagram of SII emission lines at the lower Hanle regime. Same as in Fig.\ref{density},\ref{absorb}, $\theta_0=90^\circ,\, \phi_B=90^\circ$. }
\label{lowemission}
\end{figure}

The upper level density matrix is modulated by the magnetic field through excitation from the ground state (see Fig.\ref{density}). The analytical expressions of the density matrix of the upper level are extremely lengthy and we do not present them here. They can be attained by inserting (Eqs.\ref{maindensity}-\ref{rho00ground}) into Eq.(\ref{uplevel}). Then using Eq.(\ref{emissivity}), one can get the polarization of emissions, which also varies with the magnetic field. Fig.\ref{lowemission} panels shows the corresponding Hanle diagram for the $90^\circ$ scattering. Compared to Hanle regime, the polarizations in ground Hanle regime is larger. This is expected as the magnetic field in the ground Hanle regime is much weaker than that in Hanle regime (see Fig.\ref{regimes}). Therefore, the depolarization caused by magnetic precession is smaller.  

A quantitative comparison can be made with earlier studies, e.g., Landolfi \& Landi Degl¡¯Innocenti (1986). In this paper, there is a polarization diagram for an idealized two-level atom ($J_l=3/2,J_u=5/2$). The atomic species S II we consider has a ground level $J_l=3/2$, but with three upper levels $J_u=1/2,3/2,5/2$. To benchmark our results, we also made a test calculation with only one upper level $J_u=5/2$, and reproduced the result\footnote{Apart from a minus sign of the Stokes parameters Q,U, which is due to different choice of reference.} in the aforementioned paper.

\section{Hanle regime studies: Circumstellar scattering}

The geometry of the system is given in Fig.\ref{diskcart}. We consider two cases. In the first case, a planar equatorial disk is expanding at a constant speed v (see Ignace, Cassinelli, \& Nordsieck 1999). For a point in the disk (r,$\varphi$), the projected speed along the line of sight would be 
\be
v_z=v\sin i \cos\varphi
\label{vz}
\ee
The iso-velocity zone is thus a pair of radial ``spokes" at $\pm \varphi$ since the velocity v and the inclination i are constant. 

In the second case, we consider a Keplerian disk. At a point (r,$\varphi$), the line of sight speed is
\be
v_z=v(R)\sqrt{\frac{R}{r}}\sin i \sin\varphi
\label{vz}
\ee

Because of the differential rotation, the same polarized intensity along the same spoke will be distribution at different frequency (or l.o.s. velocity). Oscillation thus appears in the observed signals (see Fig.\ref{poloiqu},\ref{toroiqu}).   

We discuss two cases of magnetic configurations below: poloidal dipole field and toroidal field. In both cases, the radiation is perpendicular to the magnetic field, i.e., $\theta_r=90^\circ$. Combining Eqs.(\ref{emissivity},\ref{updensity}), we obtain the Stokes parameters for the emission from upper level $J_u=3/2$: 
\bea
I&=&\frac{\lambda^2}{4\pi}AI_*n\varrho^0_0\Psi(\nu-\nu_0) 18.751\Gamma
   ^2+1.1933\sin^2\theta\sin2 (\phi-\phi_r) \Gamma
   +0.5967\sin^2\theta\cos2(\phi-\phi_r)-\cos ^2\theta\left(1.8597\Gamma ^2+0.465\right)+4.6882\nonumber\\
Q&=&\frac{\lambda^2}{4\pi}AI_*n\varrho^0_0\Psi(\nu-\nu_0)\left\{\cos\theta \sin2 \gamma \left[2.3866\Gamma  \cos2( \phi_r-\phi)+1.1933\sin2(\phi_r-\phi)\right]+\cos2 \gamma
  \sin^2\theta(1.8597 \Gamma ^2+0.465)\right.\nonumber\\
&+&\left.\cos 2\theta\cos2 \gamma\left[0.5967\Gamma \sin2(\phi_r-\phi)-0.2983\cos2(\phi-\phi_r)\right]\right\},\nonumber\\
U&=&\frac{\lambda^2}{4\pi}AI_*n\varrho^0_0\Psi(\nu-\nu_0) \left\{1.1933\cos2 \gamma \cos\theta
   \left[\sin2(\phi-\phi_r)-2\Gamma  \cos2 (\phi-\phi_r)\right]+0.2983\cos2\theta \sin2 \gamma \left[\cos2 (\phi-\phi_r)+2\Gamma  \sin2(\phi-\phi_r)\right]\right\}.
\eea


where $\varrho^0_0=\rho^0_0/(127.65 \Gamma^2+31.91)$. The intensity of the incident radiation is $I_*=I_s(r/R_s)^2$, where $I_s, R_s$ are the surface intensity and the radius of the central source.

\subsection{Poloidal field}
\begin{figure}
\includegraphics[%
  width=0.4\textwidth,
  height=0.3\textheight]{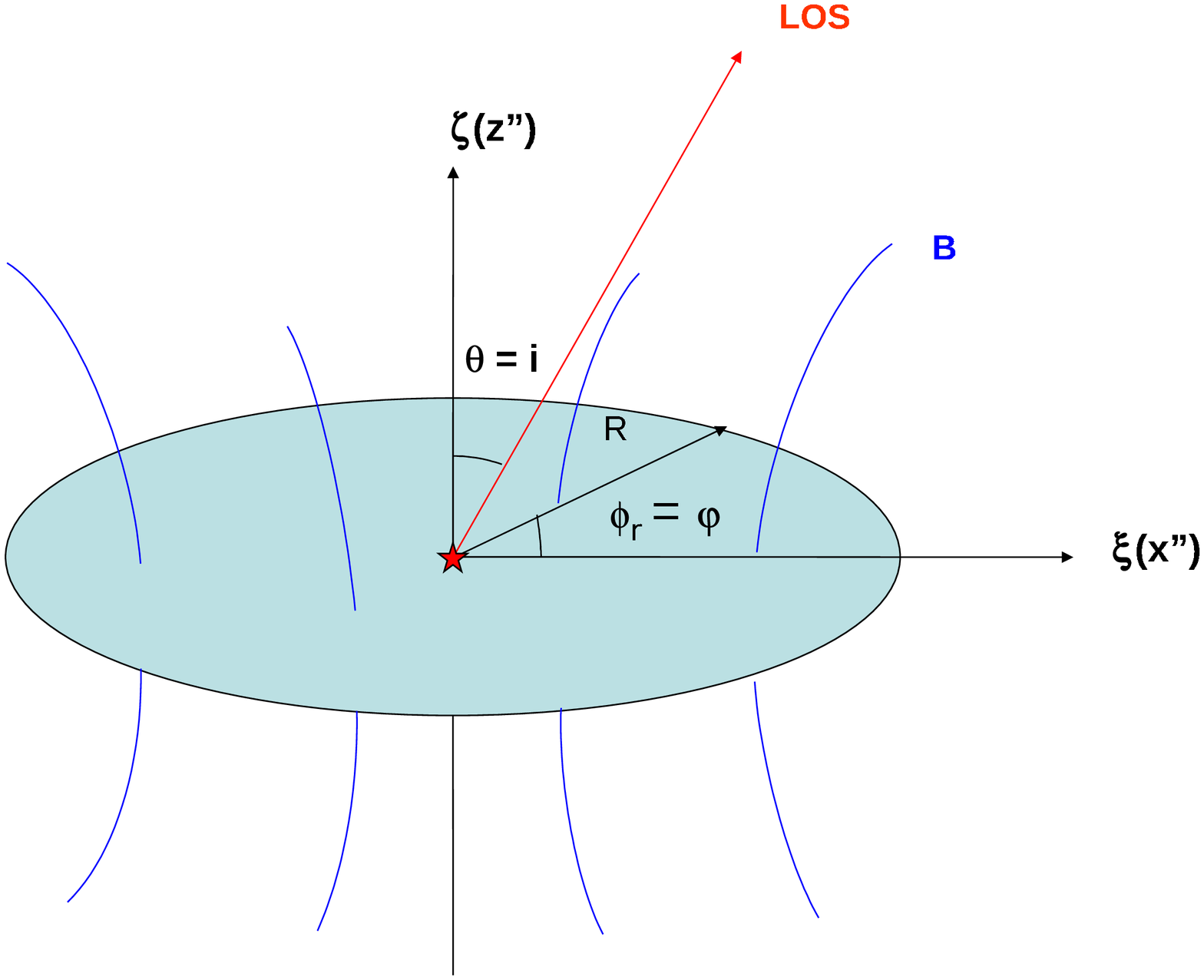}
\includegraphics[%
  width=0.4\textwidth,
  height=0.3\textheight]{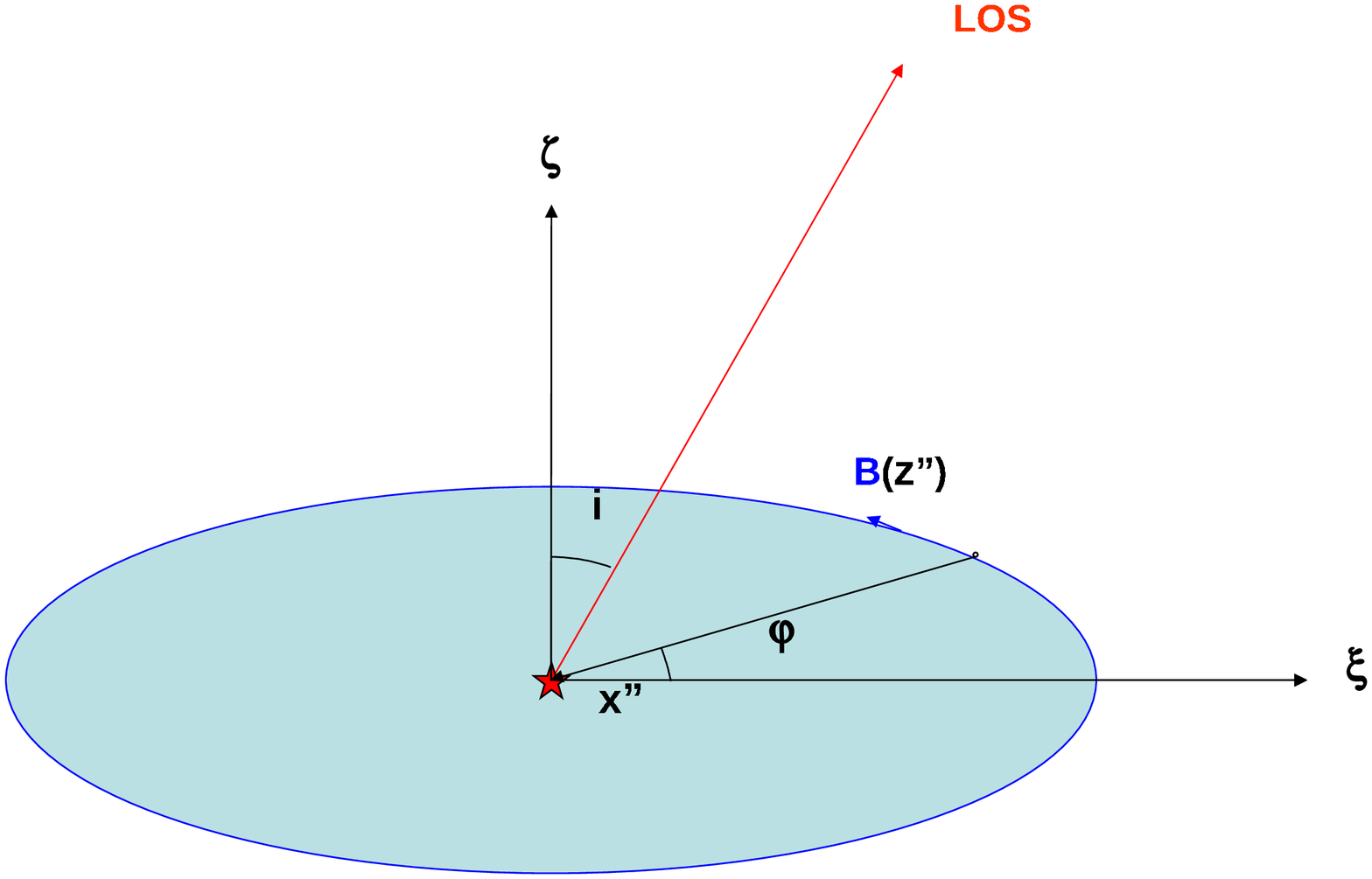}
\caption{Schematic view of a disk with poloidal and toroidal magnetic field. {\em left}: in the case of poloidal field, the field lines are normal to the equatorial plane; {\em right} in the case of toroidal field, the field is tangential. In both cases, the radiation from the central source is normal to the field lines.}
\label{diskcart}
\end{figure}
\begin{figure}  
\includegraphics[%
  width=0.24\textwidth,
  height=0.28\textheight]{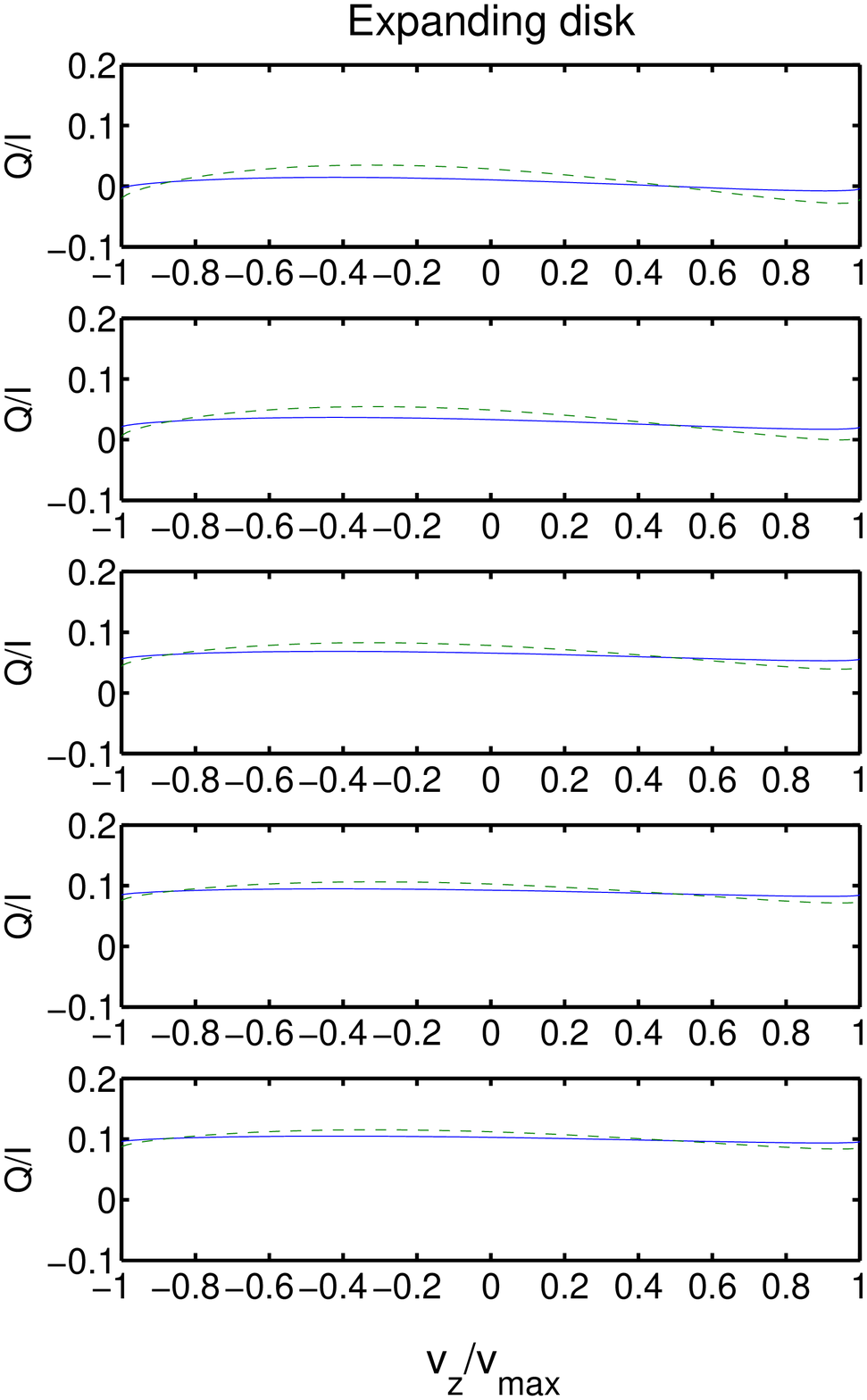}
\includegraphics[%
  width=0.24\textwidth,
  height=0.28\textheight]{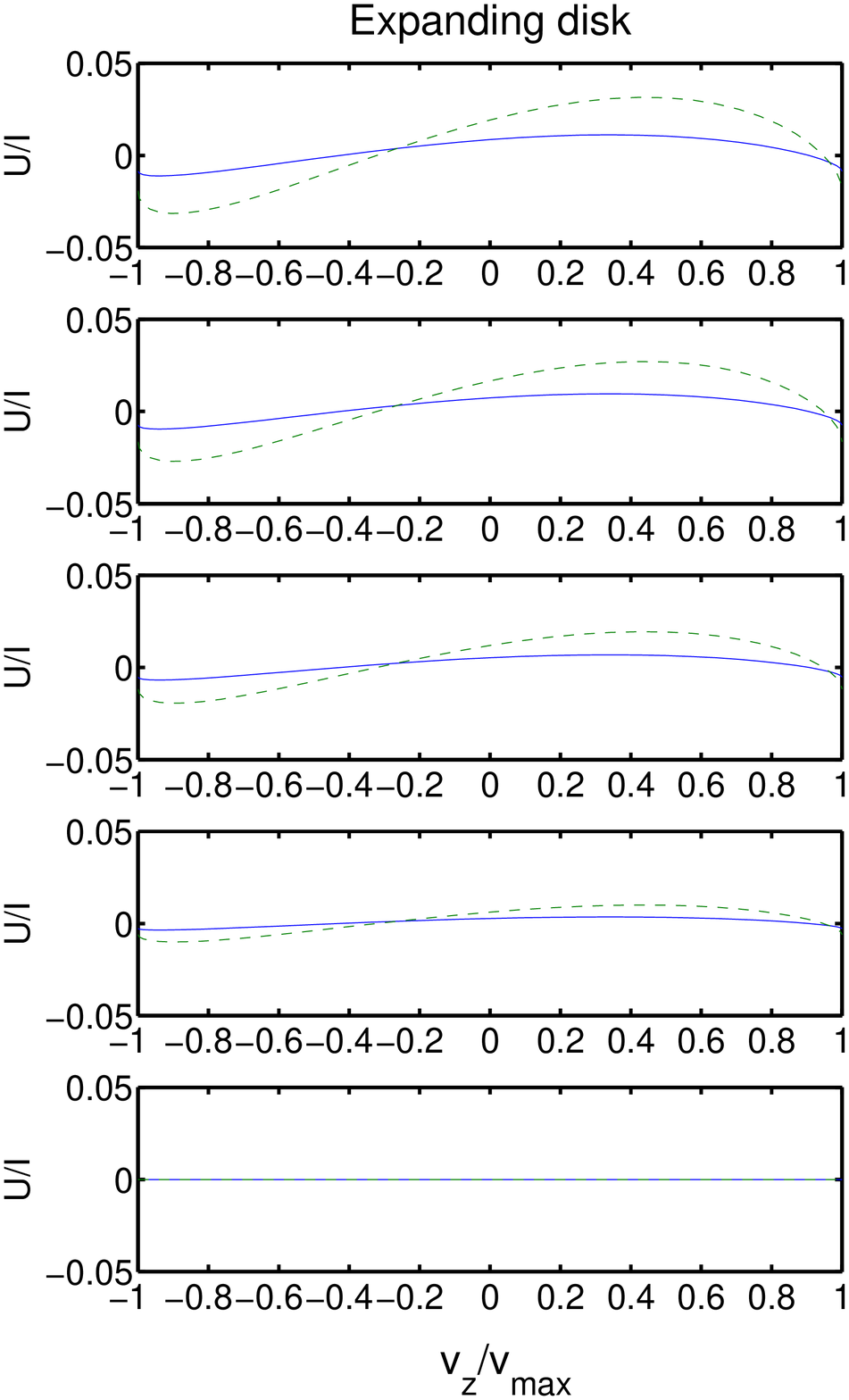}
\includegraphics[%
  width=0.24\textwidth,
  height=0.28\textheight]{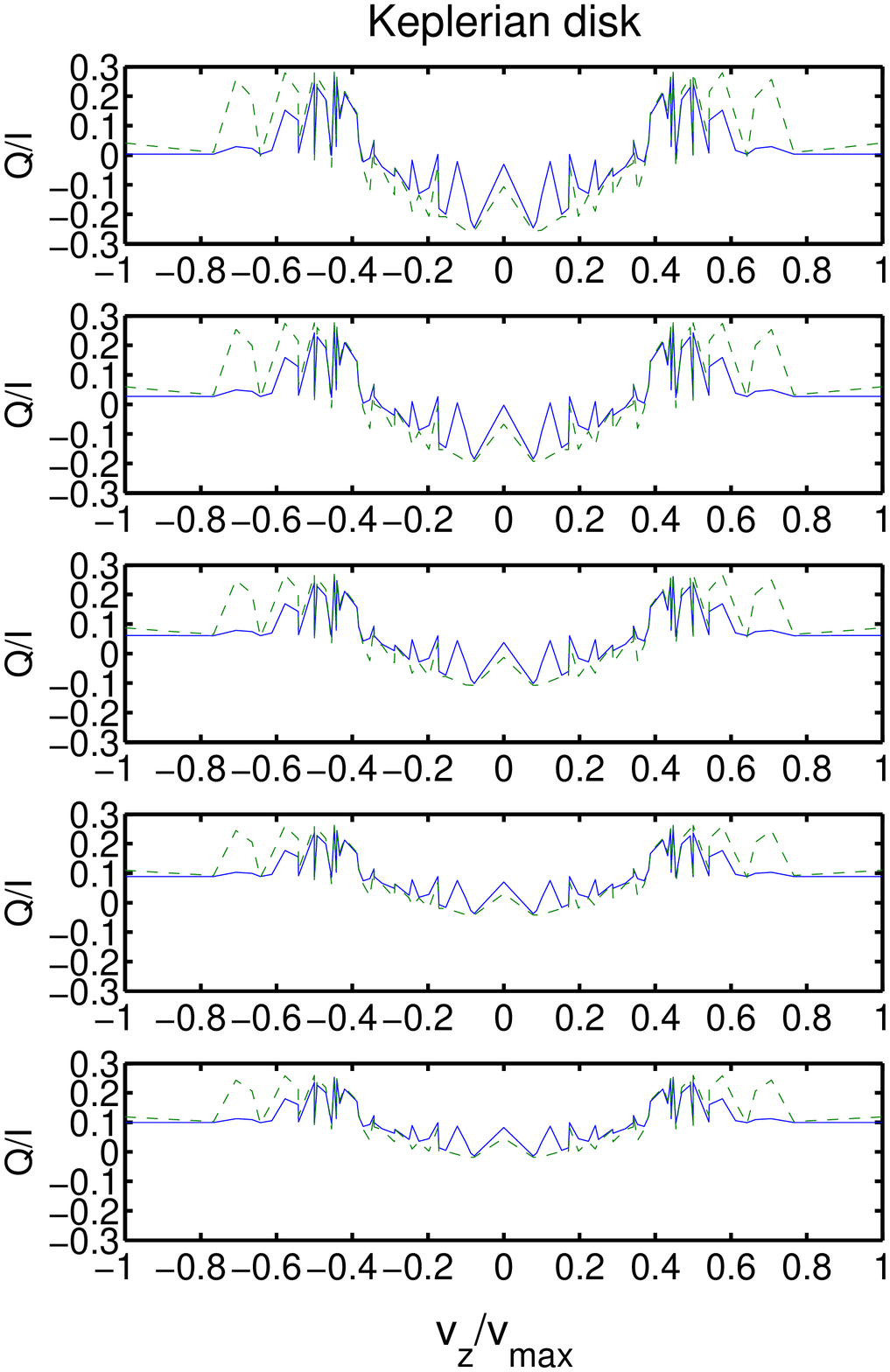}
\includegraphics[%
  width=0.24\textwidth,
  height=0.28\textheight]{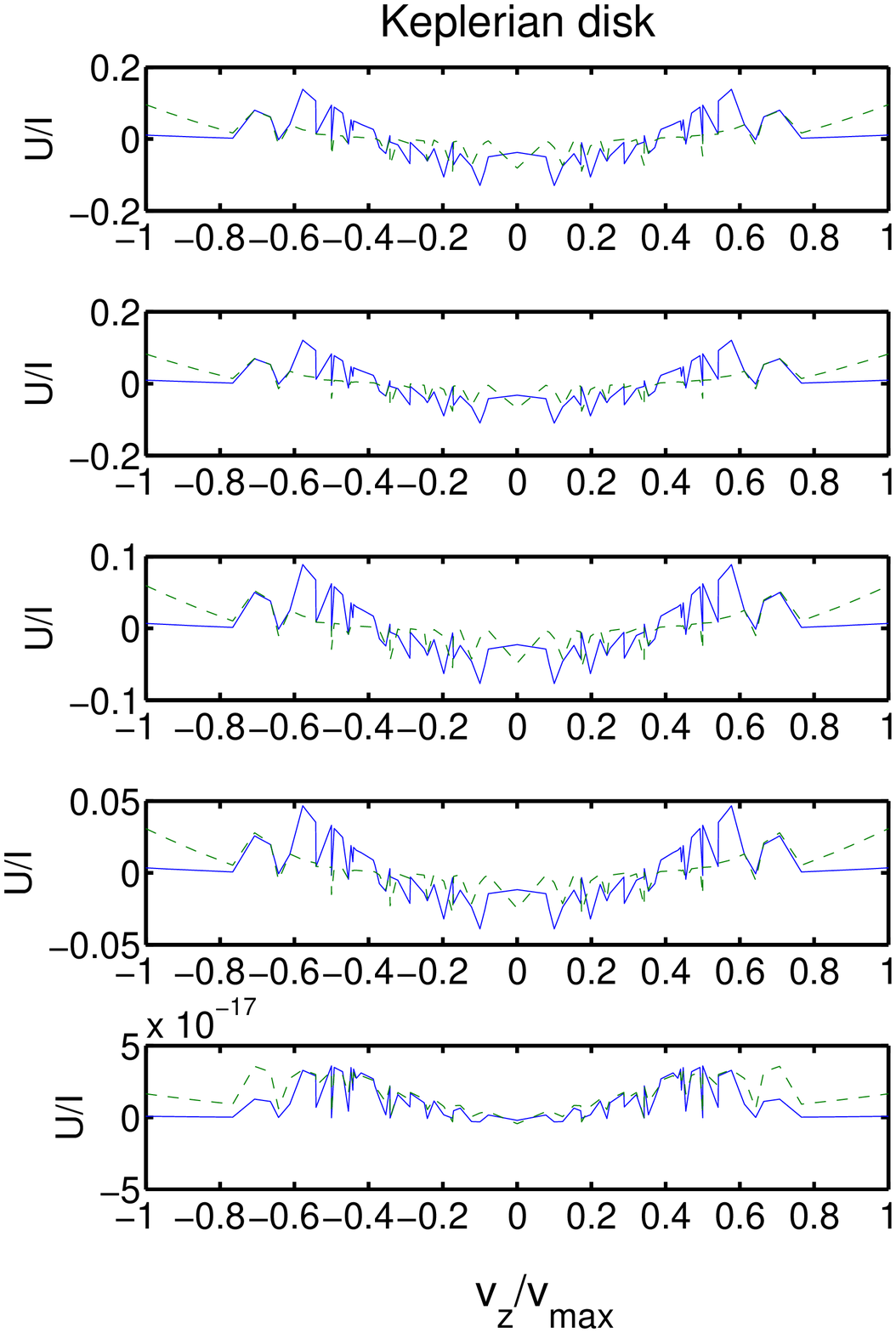}
\caption{Linear polarizations vs. line frequency shift for disk with a dipole field, where $v_{max}=v(R)\sin i$. The {\em left} two panels show the results for an expanding disk; the {\em right} two panels are for the Keplerian disk. The dashed line represents the result for $\Gamma'(R)=0.01$, the solid line is for $\Gamma'(R)=0.1$. The inclinations are $i=10^o, 30^o, 50^o, 70^o, 90^\circ$ in the panels from top to bottom. }
\label{poloiqu}
\end{figure}

First, let us consider the case of a dipole field. In the midplane, the field strength is

\be
B=B_z=\frac{p_m}{r^3}=B(R)\left(\frac{R}{r}\right)^3, 
\ee
where $p_m$ is the magnetic dipole moment. The polarization at each $v_z$(or $\phi$) is the summation of the contributions at different distances (r) along the same ``spoke".  The magnetic field is uniformly directed along the symmetry axis of the plane (see Fig.\ref{diskcart}{\em left}), thus $\theta_r=90^\circ$, $\phi_r=\varphi$. The angle between the line of sight and the magnetic field is equal to the viewing inclination $\theta=i$, and we define $\xi-\zeta$ plane to be parallel to the line of sight and thus the azimuthal angle of l.o.s. is $\phi=0$. We choose the reference plane to be parallel to the $\xi-\zeta$ plane, i.e., $\gamma=0$ (see Fig.\ref{radiageometry}{\em right}). We consider five different viewing inclination angles, $i=10^o, 30^o, 50^o, 70^o, 90^\circ$ and two different magnitudes of magnetic field, $\Gamma'(R)=2\pi g_u \nu_L(R)/A=0.01,0.1$. The results are given in Fig.\ref{poloiqu}.

\subsection{Toroidal field}

\begin{figure} 
\includegraphics[%
  width=0.24\textwidth,
  height=0.28\textheight]{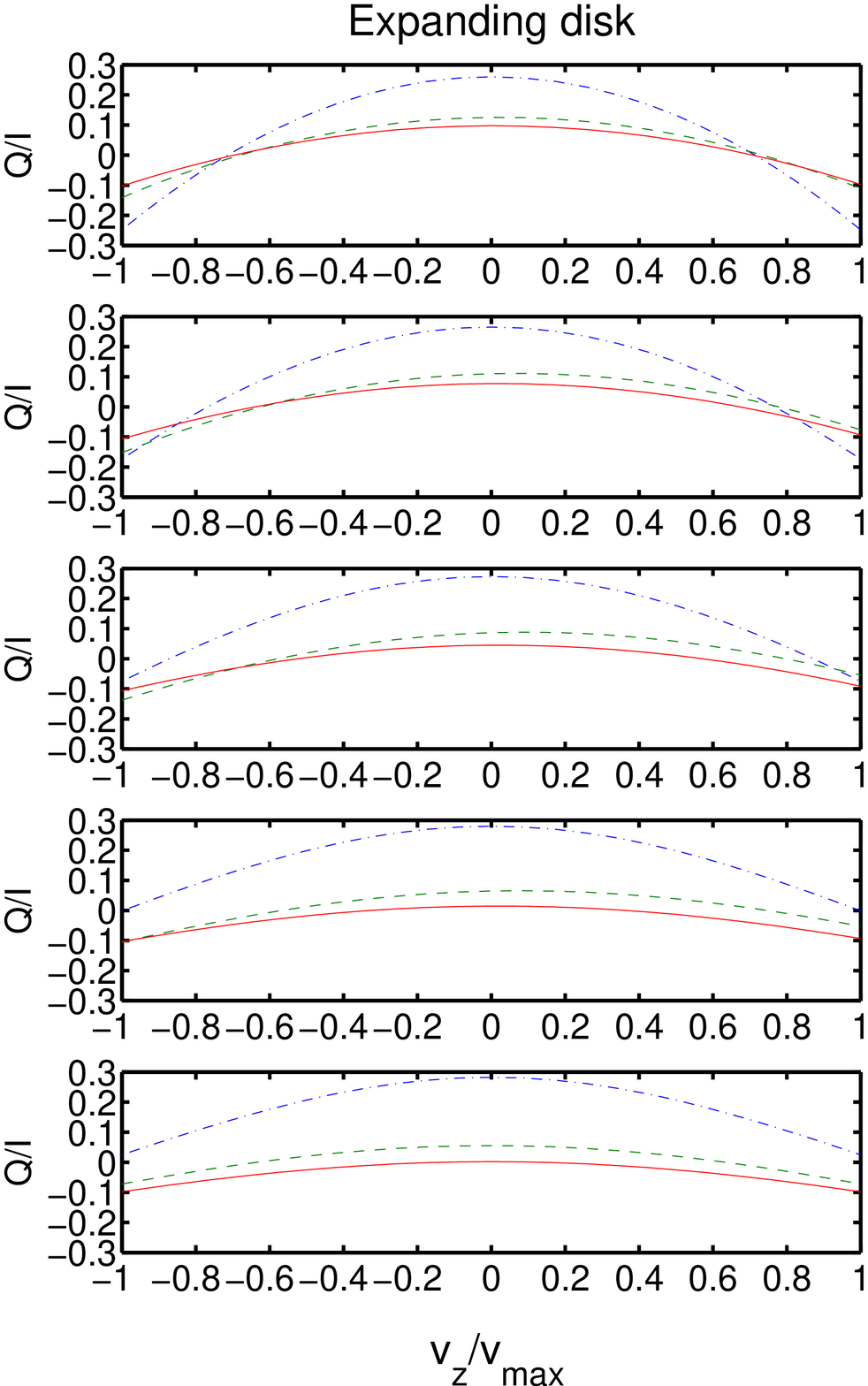}
\includegraphics[%
  width=0.24\textwidth,
  height=0.28\textheight]{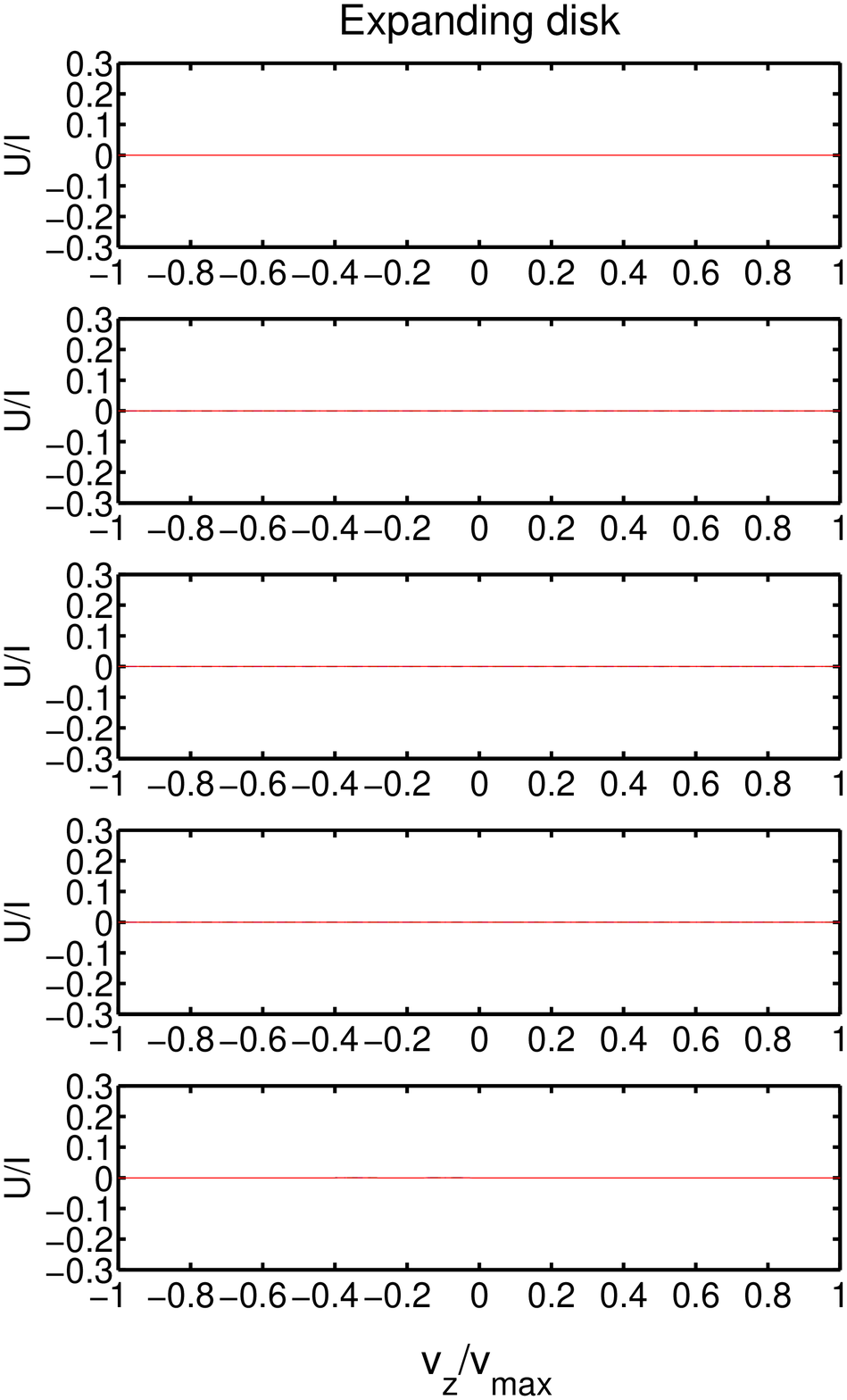}
\includegraphics[%
  width=0.24\textwidth,
  height=0.28\textheight]{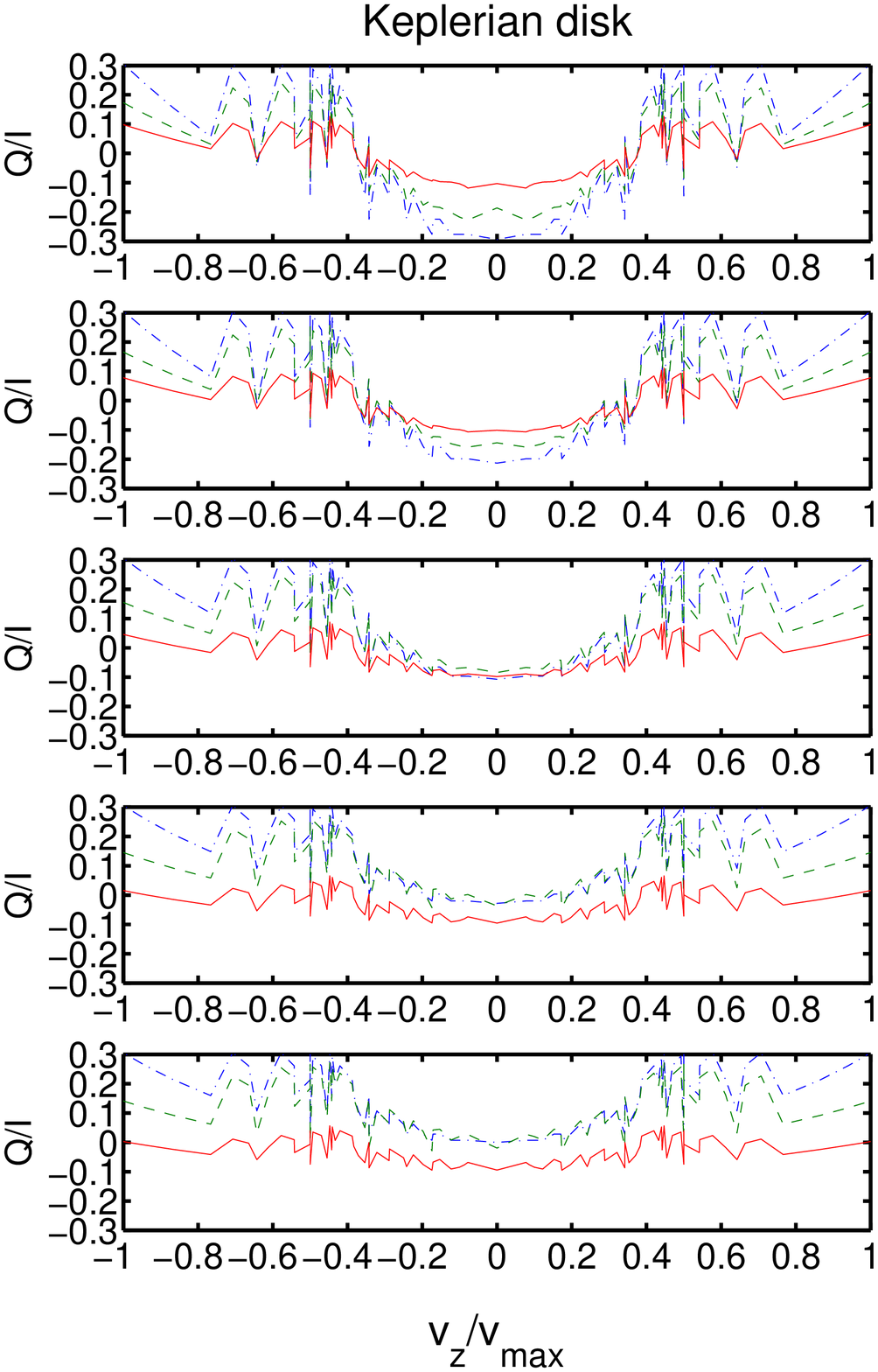}
\includegraphics[%
  width=0.24\textwidth,
  height=0.28\textheight]{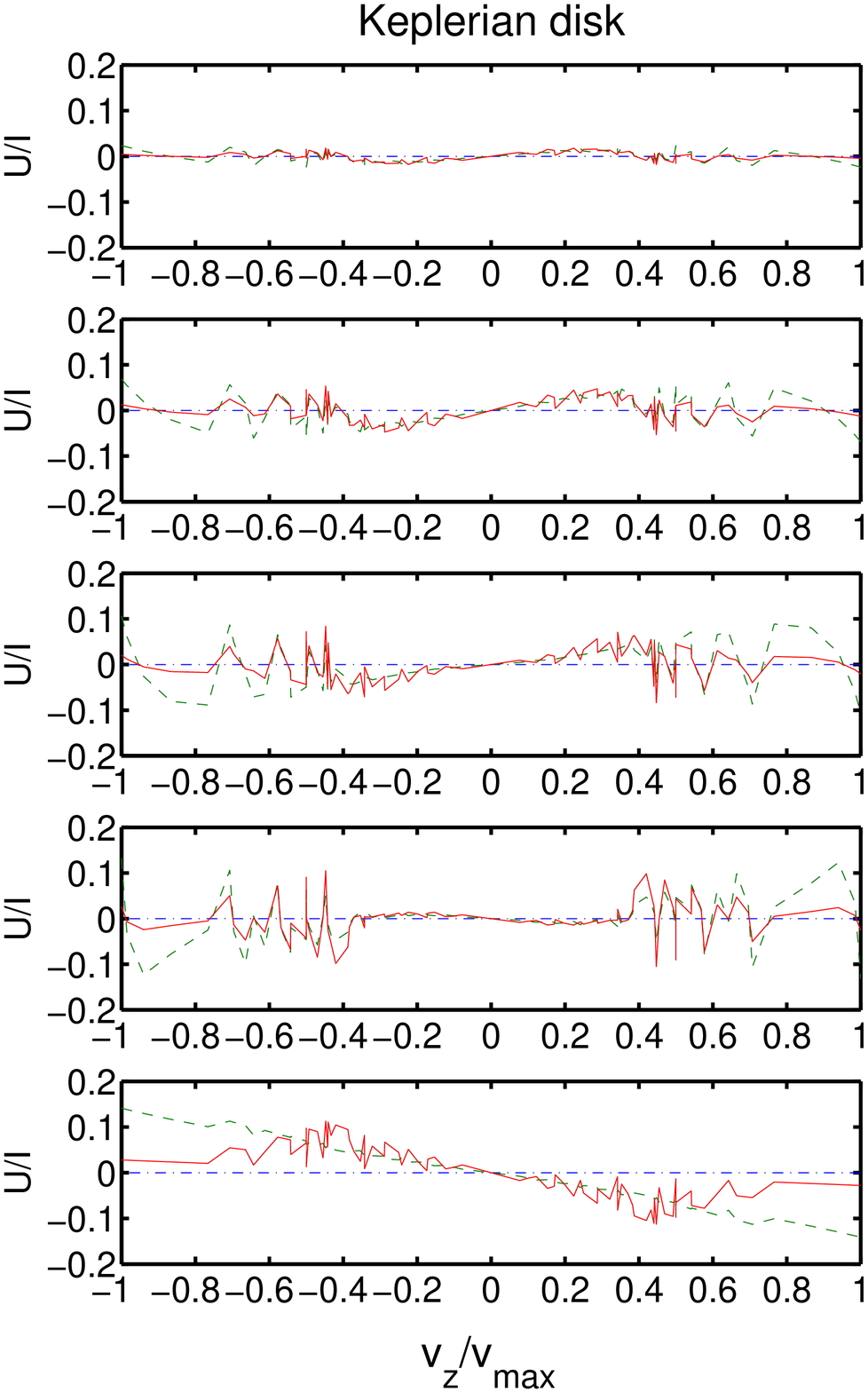}
\caption{Linear polarizations vs. line of sight speed for disk with a toroidal field. The {\em left} two panels show the results for an expanding disk; the {\em right} two panels are for the Keplerian disk. The dashed line represents the result for $\Gamma'(R)=0.1$, the solid line is for $\Gamma'(R)=1$. The inclinations are $i=10^o, 30^o, 50^o, 70^o, 90^\circ$ in the panels from top to bottom.}
\label{toroiqu}
\end{figure}

In the case of a toroidal field $B_\phi=B(R)\times R/r$, the direction of the field depends on the azimuthal angle $\varphi$. The theoretical frame defined by the magnetic field (see Fig.\ref{diskcart}{\em right}) varies from point to point along the ``ring" (see Fig.\ref{diskcart}{\em right}). Consider a point ($r, \varphi$) on the ``ring", B field defines the $z"$ axis and we choose the radial direction to be the $x"$ axis. The radiation is then seen coming from $\theta_r=90^\circ, \phi_r=0^o$. The line of sight seen in the $x"y"z"$ frame is dependant on both the azimuthal angle $\varphi$ and the inclination angle $i$. The detailed derivation of the geometric relation is given in App. A. We only provide the results here: $\theta=\cos^{-1}(-\sin i\sin\varphi), \phi=\cos^{-1}(-\sin i\cos\varphi/\sin\theta)$. Since the magnetic fields are oriented in different directions,  we adopt the plane defined by the symmetry axis and line of sight as the reference plane instead. The angle $\gamma$ (see Fig.\ref{radiageometry}{\em right}) between the two planes, equal to the angle between the magnetic field and the symmetry axis in the plane of sky. It can be proved (see App. A) that for $i\neq 90^\circ$,
\be
\gamma=\cos^{-1}\left(\frac{\cos i\sin\varphi}{\sqrt{\cos^2i\sin^2\varphi+\cos^2\varphi}}\right)\left\{\begin{array}{ll}1, &   0^o\leq\varphi\leq 90^\circ,\\ 
-1, & 90^\circ<\varphi \leq 180^o \end{array}\right.
\ee 
if $i=90^\circ$, $\gamma=0^o$. The results for the five cases with the different inclination angles and two different field strengths $\Gamma'(R)=0.1,1$ are given in Fig.\ref{toroiqu}. Compared to Ignace, Cassinelli \& Nordsieck (1999), the major difference is the symmetry about the line center ($v_z=0$) shift. Because of the inclination, we do not expect symmetry  (see Eq.\ref{vz}).

\section{Effect of atomic alignment on radio and far-infrared magnetic dipole lines}

\subsection{Formalism for magnetic dipole lines}

The alignment on the ground state affects not only the optical (or UV) transitions to the excited state, but also the magnetic dipole transitions within the ground state. We briefly outlined this effect in YLb using HI 21cm and NV 70.7mm lines. We provide here  a more detailed discussion and its implications on the measurements . 21cm line is an important diagnostics, especially for the epoch of reionization. Indeed Ly$\alpha$ pumping of HI has been extensively discussed in literatures and the resulting 21cm line is a powerful tool for studying cosmological earlier epoch. Similar to HI, other species that has a structure within the ground state is also influenced by the optical pumping\footnote{To clarify, we do not distinguish between pumping by optical lines or UV lines, and name them universally optical pumping.} through the Wouthuysen-Field effect (Field 1958; or Furlanetto et al. 2006 for a review).  Recently, the oxygen pumping has been proposed as a probe for the intergalactic metals at the epoch of reionization (Hern\'andez-Monteagudo et al. 2007a,b). 

However, in all these studies, the pumping light is assumed to be isotropic. This is problematic , particularly for the metal lines whose optical depth is small. During the early epoch of reionization, for instance, the ionization sources are localized, which introduces substantial anisotropy. The atomic alignment introduced by the anisotropy of the radiation field can play an important role in many circumstances. We shall show here this oversimplified approach can lead to a substantial error to the predictions.

In some sense, this study is similar the case of weak pumping regime discussed in YLa. But we take into account in addition the absorption and stimulated emission within the ground state owing to the presence of the CMB. The evolution equations of the occupation on the ground state can be obtained correspondingly by adding the absorptions and stimulated emissions to the Eqs.(27,28) in YLa:
\bea
\dot{\rho^k_0}(J^0_l)&=& \sum_{J_l}p_k(J_l,J^0_l)[J_l]\left[A_m+B^s_mI_m\right]\rho^k_0(J_l) +\sum_{J_u}p_k(J_u,J^0_l)[J_u]A\rho^k_0(J_u) \nonumber\\
&-&\sum_{J_l}B_mI_m\rho^{k}_{0}(J^0_l)-\sum_{J_u,K,k'}(\delta_{kk'}B_{lu}\bar{J}^0_0+s_{kk'}(J_u,J^0_l,0,0)B_{lu}\bar{J}^2_0 )\rho^{k'}_{0}(J^0_l).
\label{wpevolutiong}
\eea
For those lower levels other than the ground level, 
\bea
\dot{\rho^k_0}(J_l)&=&\sum_{J'_l}p_k(J'_l,J_l)[J'_l](A_m|_{E'>E}+B^s_m|_{E'>E}I_m+B_m|_{E'<E}I_m)\rho^k_0(J'_l)\nonumber\\
&+&\sum_{J_u}p_k(J_u,J_l)[J_u]A\rho^k_0(J_u)-\sum_{J'_l}\left(A_m|_{E>E'}+B^s_m|_{E>E'}I_m+B_m|_{E<E'}I_m\right)\rho^k_0(J_l).
\eea

In the above equations, $A_m, B_m, B^s_m$ represent respectively the spontaneous emission, absorption and stimulated emission coefficients. $I_m$ is the corresponding radiative intensity. For intergalactic medium, the microwave intensity $I_m$ is usually given by CMB, which by its nature is isotropic\footnote{Therefore it doesn't have the dipole component unlike the optical pumping light.}. If the splitting of the ground state is owing to fine structure, the equivalent temperature $T_*$ of the energy separation of the first metastable level ($J^1_l$) from the ground level ($J^0_l$) is usually much larger than the CMB temperature so that a two-level model involving only the ground level ($J^0_l$) and the first metastable level ($J^1_l$) would be adequate. We also neglected the optical pumping from the level ($J^1_l$), which is much smaller than the pumping from level ($J^0_l$). Combined with the evolution equation of the upper level ($J_u$, eq.\ref{evolution}), we then obtain

\bea
&&\sum_{J_u,k'}BI_\nu\left\{[J_u,J^0_l]p_k(J_u,J^0_l)\frac{A(J^0_l)}{\sum_{J''_l}A''}\left[\delta_{kk'}p_{k}(J_u,J^0_l)+r_{kk'}(J_u,J^0_l,0,0)\frac{\bar{J}^2_0}{\bar{J}^0_0} \right]\right.\nonumber\\
&-&\left.\left[\delta_{kk'}+s_{kk'}(J_u,J^0_l,0,0)\frac{\bar{J}^2_0}{\bar{J}^0_0} \right]\right\}\rho^{k'}_{0}(J^0_l)\nonumber\\
&+&p_k(J^0_l,J_l)[J_l]\left(A_m+B^s_mI_m\right)\rho^{k}_{0}(J_l)-B_mI_m\rho^{k}_{0}(J^0_l) =0,
\label{cmbJl0}
\eea
\bea
&&\left[A_m+B^s_mI_m\right]\rho^k_0(J_l)=p_k(J^0_l, J_l)[J^0_l]\rho^k_0(J^0_l)\nonumber\\
&+&BI_\nu\sum_{J_u,k'}\frac{A(J_l)}{\sum_{J''_l}A''}p_k(J_u,J_l)[J_u,J^0_l]\left[\delta_{kk'}p_{k'}(J_u,J^0_l)+r_{kk'}(J_u,J^0_l,0,0)\frac{\bar{J}^2_0}{\bar{J}^0_0} \right]\rho^{k'}_{0}(J^0_l),
\label{cmbJl}
\eea
where $I_m=B_{\nu(J_l\rightarrow J^0_l)}(T)$ is the blackbody radiative intensity at temperature T. For simplicity, we dropped the index `1' for the first metastable level and refer to it as $J_l$. For many lines $T_*\gg T$, so the excitation and stimulated emission for the magnetic dipole transition within the ground state is much slower than the corresponding emission rate. Define $\beta=BI_\nu/B_mI_m$. When $\beta\gg 1$, the absorption and stimulated emission can be neglected (the terms containing the factor $I_m$) and the above equations reduce to the expressions we obtained in YLa for the weak pumping case. If $\beta$ gets to the order of unity, however, all the magnetic dipole transitions within the ground state should be accounted and the above equations should be applied. The occupation on the ground state is in general determined by the two factors $\beta$ and $\bar{J}^2_0(\theta_r)/\bar{J}^0_0$ (Eq.\ref{irredradia}). If there is no magnetic field, we can choose the quantization axis along the direction of the UV radiation so that $\theta_r=0$ (see Fig.\ref{regimes}{\em right}). In this case the factor $\bar{J}^2_0/\bar{J}^0_0=W_a/(\sqrt{2}W)$ is solely determined by the anisotropy.  We obtain from Eqs.(\ref{cmbJl0},\ref{cmbJl})
\bea
\frac{\rho^0_0(J_l)}{\rho^0_0(J^0_l)}&=&\frac{\sqrt{[J_l^0]}B_mI_m}{\sqrt{[J_l]}(A_m+B^s_mI_m)  }\left\{\beta\left[1-\frac{A(J^0_l)}{\sum_{J''_l}A''}\right] +1\right\},\nonumber\\
\rho^2_0(J_l)&\simeq& \frac{\beta B_mI_m}{A_m+B^s_mI_m}\sum_{J_u}\frac{A(J_l)}{\sum_{J''_l}A''}p_2(J_u,J_l)[J_u,J^0_l]r_{20}(J_u,J^0_l,0,0)\frac{\bar{J}^2_0}{\bar{J}^0_0} \rho^{0}_{0}(J^0_l),\nonumber\\
\rho^2_0(J^0_l)&\simeq& \sum_{J_u}\beta\left[[J_u,J^0_l]\frac{A(J^0_l)}{\sum_{J''_l}A''}r_{20}(J_u,J^0_l,0,0)\frac{\bar{J}^2_0}{\bar{J}^0_0}\left(p_2(J_u,J^0_l)+[J_l]p_2(J^0_l,J_l)p_2(J_u,J_l)\right)\right.\nonumber\\
&-&\left.s_{20}(J_u,J^0_l,0,0)\frac{\bar{J}^2_0}{\bar{J}^0_0} \right]\rho^{0}_{0}(J^0_l)
\label{genrsolution}
\eea
In many cases, it is useful to quantify level population by the spin temperature 
\be
\frac{n(J_l)}{n(J^0_l)}=\frac{[J_l]}{[J_l^0]}\exp\left(-\frac{T_*}{Ts}\right),
\ee
where $n(J)=N\sqrt{[J]}\rho^0_0(J)$. Combined with Eq.(\ref{genrsolution}), the above equation yields
\be
\exp\left(-\frac{T_*}{Ts}\right)= \frac{[J^0_l]B_mI_m}{ [J_l](A_m+B^s_mI_m) }\left\{\beta \left[1-\frac{A(J^0_l)}{\sum_{J''_l}A''}\right] +1\right\}
\label{spintemprt}
\ee

\subsection{Anisotropic Pumping of [O I] 63.2$\micron$}
Here we use O I as an example to demonstrate the effect of atomic alignment on the radio diagnostics based on magnetic dipole transitions within the ground state. A comparison with the recent work on Oxygen pumping in the high-redshift intergalactic medium will be provided. For Oxygen I, their fine structure is illustrated in Fig.\ref{S2OI}. In table~\ref{OIparm}, we list the corresponding coefficients $p_k,~r_{kk'},~ s_{kk'}$ (Eq.\ref{pk}-\ref{skk}).

\begin{figure}
\includegraphics[width=0.33\textwidth,
  height=0.25\textheight]{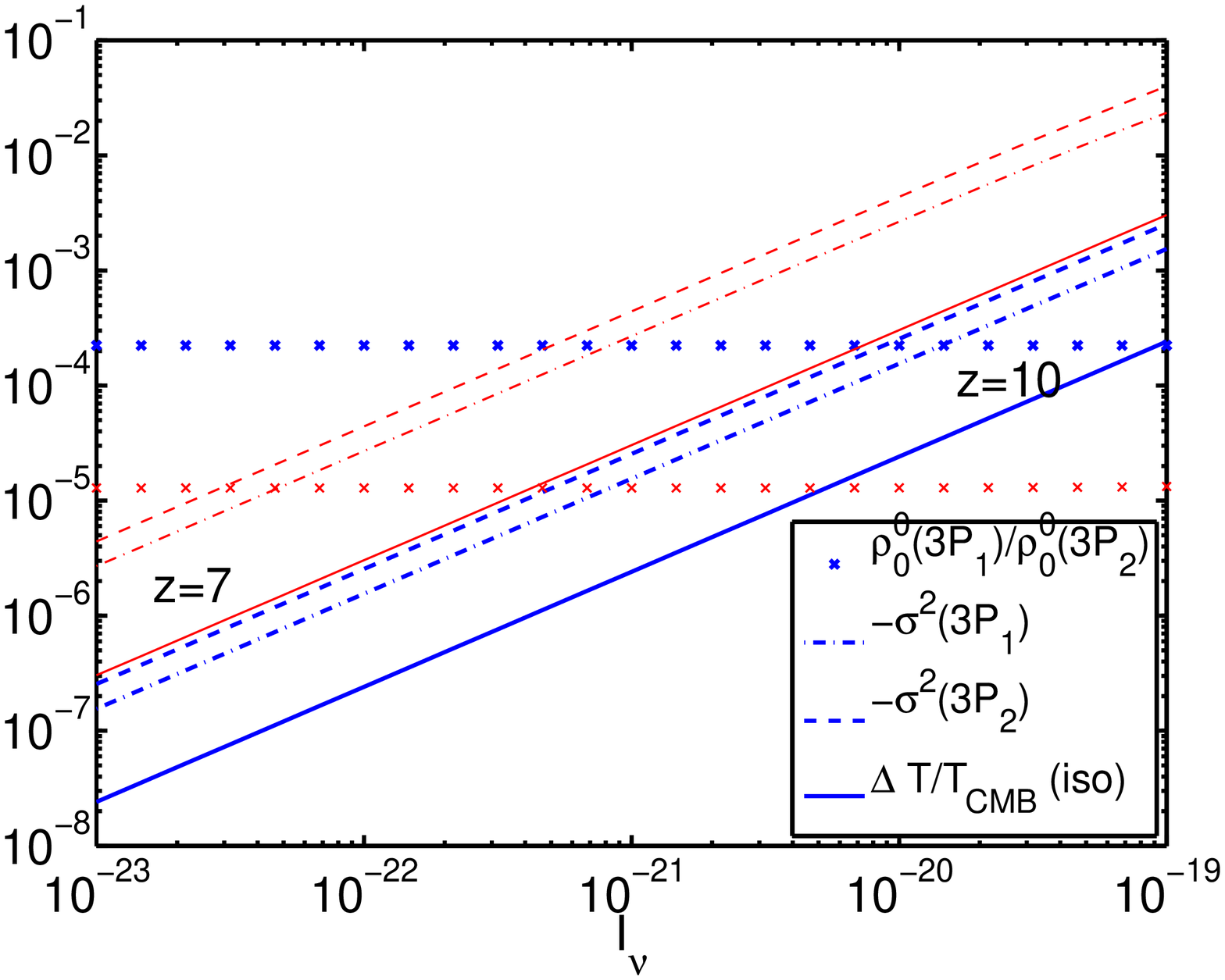}\includegraphics[width=0.33\textwidth,
  height=0.25\textheight]{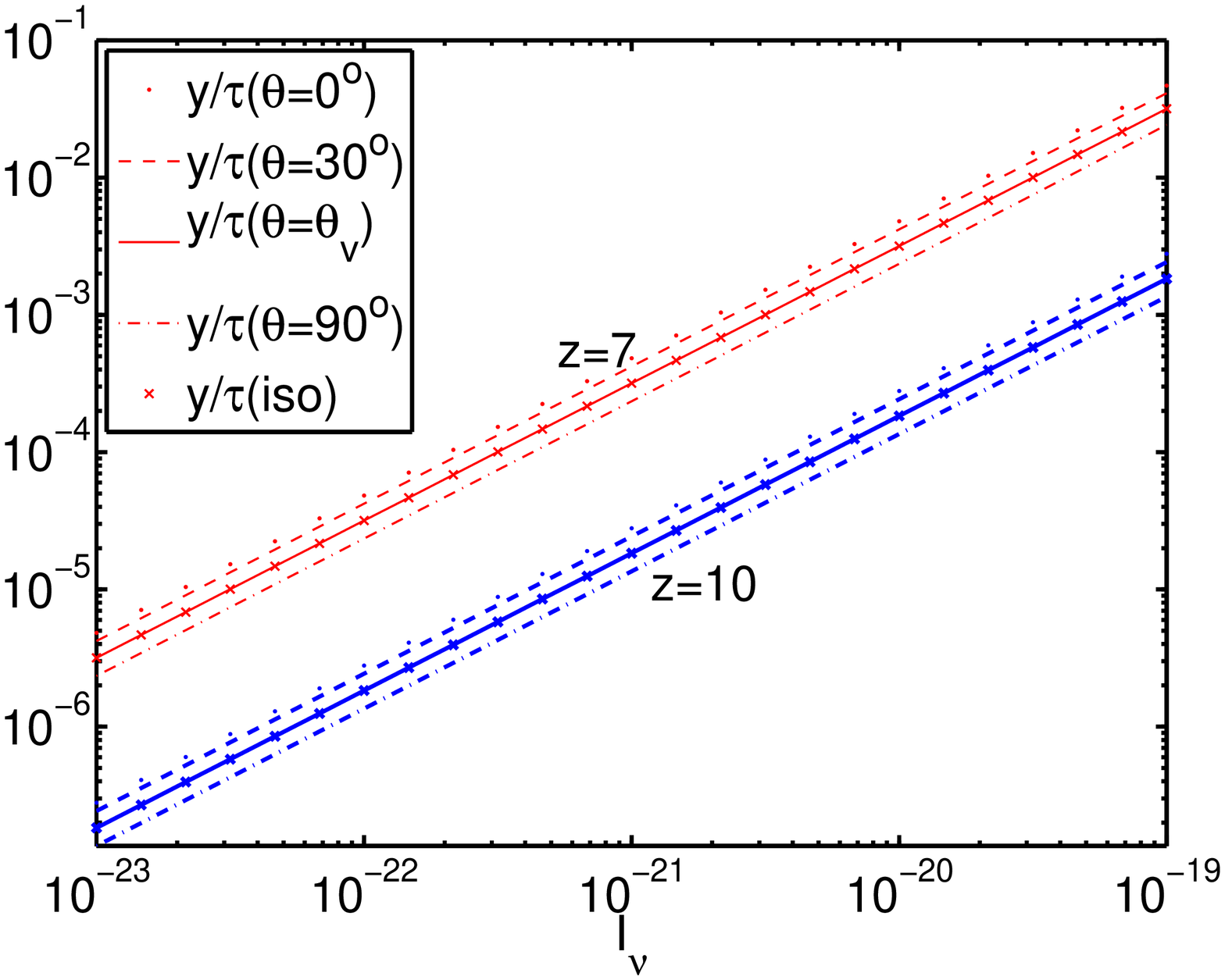}\includegraphics[width=0.33\textwidth,
  height=0.25\textheight]{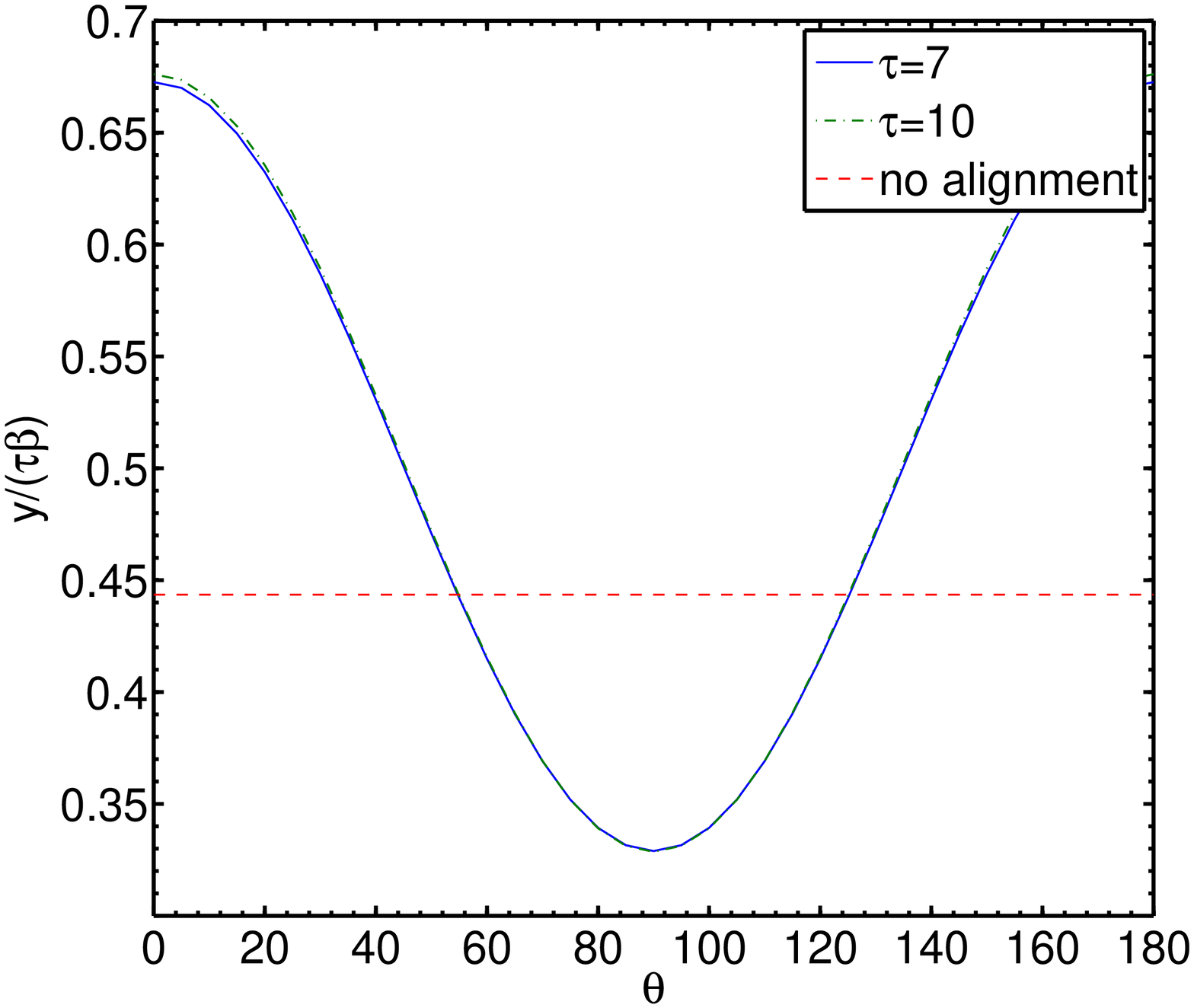}
\caption{{\em Left}: the density matrices of the first metastable level $3P_1$ and the ground level $3P_2$ of O I; {\em middle}: the distortion of CMB due to the anisotropic pumping of O I at redshift z=7, 10. While the temperature distortion (Eq.\ref{deltaT}) almost agrees with the result for isotropic pumping apart from a constant factor, the distortion in the intensity (Eq.\ref{deltaI}) is influenced by the anisotropy of the radiation and the atomic alignment. Only when the angle between the line of sight and the pumping radiation approaches to the Van Vleck angle $\theta_V=54.7^o, 180^o-54.7^o$, the result agrees with the isotropic pumping; {\em right}: the distortion of CMB $y/\tau$ is proportional to $\beta$. The ratio $y/(\tau\beta)$ thus only varies with the angle between line of sight and the pumping radiation due to atomic alignment. The alignment can cause a variation of $y/\tau$ up to a factor of 2. The case without alignment accounted is also plotted for comparison (dashed line). }
\label{cmb}
\end{figure}

Given these parameters, the linear Eqs.(\ref{cmbJl0},\ref{cmbJl}) can be easily solved to get the density matrices for the two sublevels $J_l=1,2$. As shown in Fig.\ref{cmb}{\em left}, the total population among the sublevels is marginally affected by the UV pumping. The spin temperature we obtain, however, differs from the result of Hern\'andez-Monteagudo et al. (2007) by a numerical factor 4/3. This difference arises from the existence of the second metastable level $J_l=0$. In Hern\'andez-Monteagudo et al. (2007), this level is completely neglected. In reality, the atoms pumped from the ground level $J^0_l=2$ to the excited state $^3S_1$ has a finite probability $1/9$ of jumping to the level $J_l=0$ as well as $1/3$ probability of going to level $J_l=1$ . So the total probability of atoms leaving the level $J^0_l=2$ through optical pumping atoms would be $1/9+1/3=4/9$ rather than $1/3$ in the two level model adopted in Hern\'andez-Monteagudo et al. (2007). The deviation of the spin temperature from CMB $\Delta T/T_{cmb}$ is proportional to the probability of the optical pumping rate from the ground level $J^0_l=2$ to other sublevels. Thereby, our result for $\Delta T/T_{cmb}$ is 4/3 larger. From Eq.(\ref{spintemprt}), we get
\be
\frac{T_s-T_{cmb}}{T_{cmb}}=\frac{T_{cmb}}{T_*}\exp\left(\frac{T_*}{T_{cmb}}\right)\left[1-\frac{A(J^0_l)}{\sum_{J_l} A(J_l)}\right]\frac{[J^0_l]\beta B_mI_m}{[J_l](A_m+B^s_mI_m)},
\label{deltaT}
\ee
where $I_m=B_{63.2\micron}(T_{CMB}(z))$ is given by the CMB.

The distortion in the CMB radiation, is, however, influenced by the anisotropy of the radiation and the atomic alignment. This is because both emissivity and absorption coefficient depend also on the dipole component of the density matrix (see Eq.\ref{Mueller0},\ref{emissivity}). The ratio of the optical depth accounting for alignment to that without alignment is,

\bea
\tilde\tau&=&\frac{\tau}{\tau_0}=\frac{\tilde\eta-\tilde\eta_{s,i}\exp(-T_*/T_s)}{1-\exp(-T_*/T_s)}
\label{deltatau}
\eea
According to Eq.(\ref{Mueller0},\ref{emissivity})
\be
\tilde\eta_i=\eta_1/\eta_1^0=1+w_{0l}\sigma^2_0(J^0_l){\cal J}^2_0(i,\Omega),
\label{tdeta}
\ee 
\be
\tilde\eta_{s,i}=\tilde\epsilon_i =\epsilon_i/\epsilon^0_i=1+w_{l0}\sigma^2_0(J_l){\cal J}^2_0(i,\Omega) 
\label{tdetas}
\ee
are the ratios of absorption and stimulated emission coefficients with and without alignment, where ${\cal J}^2_0(i,\Omega)$ is given by Eq.(\ref{incidentrad}).
As we see, unless the angle between the line of sight and the pumping radiation is equal to the 
Van Vleck angle $\theta_V=54.7^o, 180^o-54.7^o$, a finite correction due to atomic alignment $\sigma^2_0$ would
 occur for the optical depth. Eq.(\ref{genrsolution}) indicate that
 $\sigma^2_0(J_l),\sigma^2_0(J^0_l)$ are proportional to $\beta$ and 
$\rho^0_0(J_l)/\rho^0_0(J^0_l)$ is only sensitive to $T_{cmb}$ in the case of $\beta\ll 1$, (see Fig.\ref{cmb}).

The distortion from the CMB can be obtained using the radiative transfer equation $\frac{dI}{ds}=h\nu\Psi(\nu)/4\pi[n_l(A_m+B_{m}I_m)-n_0B^s_{m}I_m]$. In the optically thin case $\tau\ll 1$,

\bea
y=\frac{\Delta I_\nu}{B_\nu(T_{CMB})}&=&\tau\left\{\frac{\tilde\epsilon_0[\exp(T_*/T_{CMB})-1]}{\tilde\eta \exp(T_*/T_s)-\tilde\eta_s}-1\right\}\nonumber\\
&=&\tau_0\left[\frac{\tilde\epsilon_0[\exp(T_*/T_{CMB})-1]}{\exp(T_*/T_s)-1}-\tilde\tau
\label{deltaI}\right]\nonumber\\
&\simeq&\tau_0[\tilde\epsilon_0(1+y_{iso}/\tau_0)-\tilde\tau]\nonumber\\
&=&\tilde\epsilon_0 y_{iso}+\tau_0\frac{[w_{l0}\sigma^2_0(J_l)-w_{0l}\sigma^2_0(J^0_l)](1-1.5\sin^2\theta)/\sqrt{2}}{1-\exp(-T_*/T_s)}
\label{yovertau}
\eea
where $y_{iso}$ is the distortion neglecting the anisotropy of the radiation field and atomic alignment (see Hern\'andez-Monteagudo et al. 2007). Indeed if alignment is not accounted, then 
\be
y=y_{iso}=\tau_0\frac{T_*}{T_s}\frac{T_s-T_{cmb}}{T_{cmb}}=\tau_0\frac{T_{cmb}}{T_s}\exp\left(\frac{T_*}{T_{cmb}}\right)\left[1-\frac{A(J^0_l)}{\sum_{J_l} A(J_l)}\right]\frac{[J^0_l]\beta B_mI_m}{[J_l](A_m+B^s_mI_m)}
\ee
Both $\tilde\eta_s$ and $\tilde\tau$ depend on the line of sight and the atomic alignment (Eqs.\ref{deltatau}, \ref{tdetas}), the resulting distortion in radiation is thus determined by the angle $\theta$ as well as the UV intensity of the OI line $I_\nu$ (or $\beta$, see Eqs.\ref{cmbJl0},\ref{cmbJl}), which determines the atomic density and alignment (see Fig.\ref{cmb}). According to Eq.(\ref{genrsolution}), $\sigma^2_0(J^0_l),\,\sigma^2_0(J_l)\propto \beta$. Since both of the two terms on the right hand side of Eq.\ref{yovertau} are proportional to $\beta$, the resulting distortion $y$ is also proportional to $\beta$. Divide $y\tau$ by $\beta$, we show the dependence on $\theta$ in Fig.\ref{cmb}{\em right}.

\subsection{Magnetic field in the epoch of reionization?}

The issue of magnetic field at the epoch of reionization is a subject of controversies. The fact that the levels of O I ground state can be aligned through anisotropic pumping suggest us a possibility of using atomic alignment to diagnose whether magnetic field exists at that early epoch. 

The degree of polarization in the optically thin case can be obtained in a similar way as above by replacing $\tilde\eta_0, \tilde\epsilon_0$ by $\tilde\eta_1, \tilde\epsilon_1$. In the alignment regime, the Stokes parameters, U=0, and therefore,
\bea
P=\frac{Q_\nu}{B_\nu(T_{CMB})}&=&\tau\left\{\frac{\tilde\epsilon_1[\exp(T_*/T_{CMB})-1]}{\tilde\eta_1 \exp(T_*/T_s)-\tilde\epsilon_1}-1\right\}\nonumber\\
&\simeq&\tau_0[\tilde\epsilon_1(1+y_{iso}/\tau_0)-\frac{\tilde\eta_1-\tilde\epsilon_1}{1-\exp(-T_*/T_s)}]\nonumber\\
&=&\tilde\epsilon_1 y_{iso}-\tau_0\frac{1.5\sin^2\theta [w_{l0}\sigma^2_0(J_l)-w_{0l}\sigma^2_0(J^0_l)]/\sqrt{2} }{1-\exp(-T_*/T_s)}
\label{Povertau}
\eea

\begin{figure} 
\plottwo{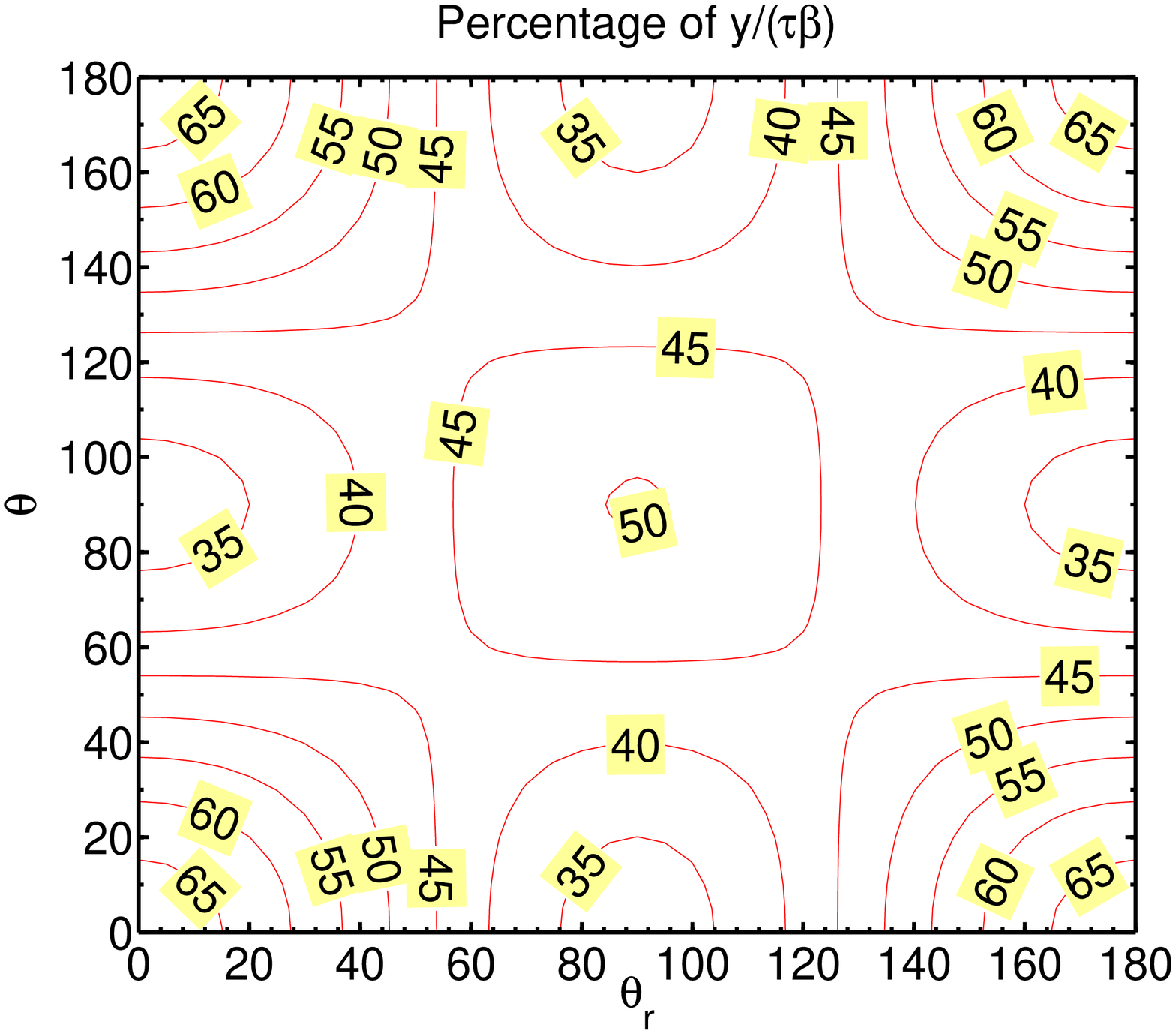}{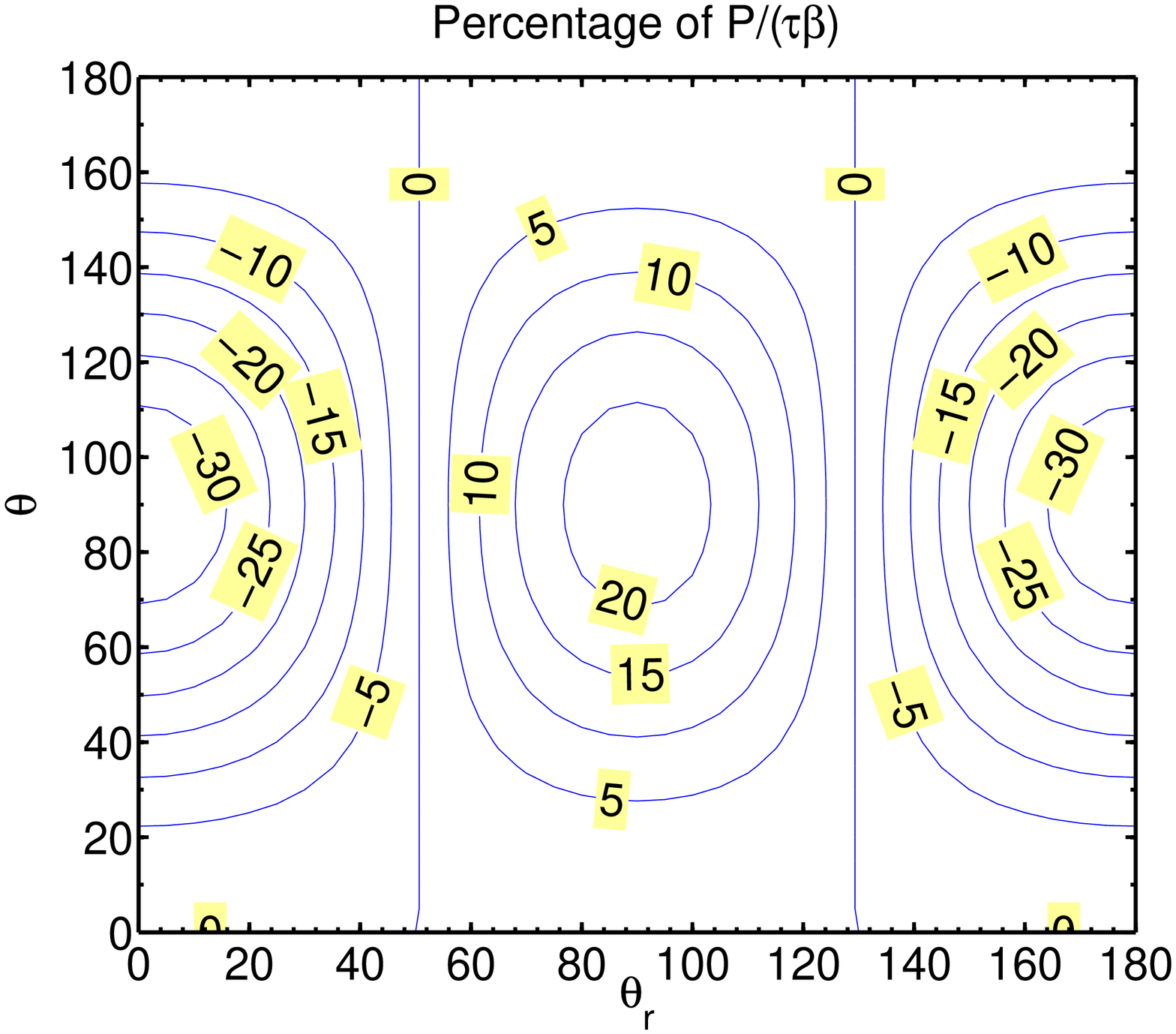}
\caption{The contour plots of equal percentages of the ratios $y/(\tau\beta)$, $P/(\tau\beta)$. $\theta_r,\, \theta$ are respectively the angles of the incident radiation and l.o.s. from the magnetic field.}
\label{OImicroforB}
\end{figure}

In the case of nonzero magnetic field, the density matrices are determined by $\theta_r$, the angle between magnetic field as well as the parameter $\beta$. Similar to $y$, the degree of polarization is also proportional to $\beta$. In Fig.\ref{OImicroforB}, we show the dependence of the ratios $y/(\tau\beta)$, $P/(\tau\beta)$ on $\theta_r$ and $\theta$. Since U=0, the line is polarized either parallel ($P>0$) or perpendicular ($P<0$) to the magnetic field. The switch between the two cases happen at $\theta_r=\theta_V=54.7^o, 180-54.7^o$, which is a common feature of polarization from aligned level (see YLa,b for detailed discussions). 

We discussed pumping of hyperfine lines [H I] 21 cm and [N V] 70.7 mm in YLb and fine line [O I] $63.2\micron$ here. Certainly this effect widely exists in all atoms with some structure on ground state, e.g., Na I, K I, fine structure lines, [C I], [C II], [Si II], [N II], [N III], [O II], [O III],[S II], [S III], [S IV], [Fe II], etc (see Table~4.1 in Lequeux 2005). Many atomic radio lines are affected in the same way and they can be utilized to study the physical conditions, especially in the early universe: abundances, the extent of reionization through the anisotropy (or localization) of the optical pumping sources, and {\em magnetic fields}, etc.

 \section{Discussion}

In this paper we extended our studies of atomic alignment in diffuse medium. We provided more calculations of polarization
of absorption lines arising from aligned species. We filled the gap of our earlier studies by considering polarization of emission lines arising from aligned species with fine structure. In addition, we considered atomic alignment in the presence of 
Hanle effect. By studying both ground and upper state Hanle effect,
 we attempted to give a more comprehensive coverage of the diagnostic of atomic resonant line on weak magnetic field in diffuse circumstellar, interstellar and intergalactic medium. 
 
\subsection{Previous work on atomic alignment}
Let us briefly state some milestones in the studies of atomic alignment. A more detailed discussion is provided in YLa and YLb.
First of all, atomic alignment was studied in laboratory in relation with early 
day maser 
research (see Hawkins 1955).

In astrophysical context atomic alignment was proposed by Varshalovich (1968). The atomic alignment for a idealized atom with two levels and magnetic field directed along the line of sight was considered by Landolfi \& Landi Degl'Innocenti (1986). We {\em benchmarked} our results by their calculations in the restricted geometry of observation. Polarization of emission lines arising from atomic pumping with mixing by magnetic field in the direction perpendicular to the symmetry axis of pumping radiation was considered in Lee, Blandford \& Western (1994). There radiation field, nevertheless, is treated classically and we suspect that 
this cannot provide correct results. Originally, for 
the sake of simplicity, in astro-ph/0412174, we attempted a similar approach motivated by the calculations in Hawkins (1955). However, our further studies proved that the results obtained without full QED treatment are erroneous. As we pointed out in YLa,b, because conservation and transfer of angular momentum is the essence of the problem, it is necessary to quantize the radiation field as the atomic states are quantized.
Our results for the polarization of the Na emission in YLb differ from the corresponding results in Hawkins (1955).  

Most of the research on the atomic alignment is done in Solar physics domain\footnote{Interestingly enough, our study shows that
atomic alignment effects are more important for the diffuse
medium, e.g. around luminous stars, where collisional depopulation is
reduced compared to the Sun.}. This research, however,
dealt with the atomic alignment in a regime, different from that important for diffuse medium.
 The differences include the relative role of collisions, magnetic fields and radiation.  In particular,  Sun  has radiation transfer
effects that complicate the interpretation of the effects arising from
alignment, but are not so important for many circumstellar and
interstellar species. For this reason we developed atomic alignment
theory applicable for the diffuse astrophysical medium. Some of
our results, e.g. polarization of for absorption lines do not have an analogs in Solar
research. Because the alignable species and the regimes of alignment
are different, we performed an extended program of calculation of
alignment for a number of astrophysically important species. However, in the present paper
we also benchmarked our codes by reproducing some results
that were obtained by the Solar community.

An extensive studies of polarization of absorption lines arising from atoms with fine structure submerged in weak magnetic
field of an arbitrary direction are provided in YLa.  The first studies of astrophysically important cases of atomic alignment for atoms with hyperfine structure in weak magnetic field of arbitrary direction are provided in YLb. This paper continues and
extend our studies above. Our approach can be applied to other problems, e.g. the ones discussed in Lee et al. (1994).

We stress the importance of taking into account magnetic fields while calculating the polarization. For instance, polarization of absorption arising from the atomic alignment is suggested as the mechanism of the observed polarization of H $\alpha$ lines (Kuhn et al. 2007). However, the effect of magnetic field is neglected there, while our study present and earlier studies in YLab show that magnetic fields should be accounted for to get correct polarization measures.

\subsection{Studies of stronger magnetic fields}

In our previous papers, we conducted an extensive study of atomic alignment and 
demonstrated their diagnostic power for weak magnetic field in diffuse medium (e.g. in ISM). When field gets stronger to 1G or so, the magnetic mixing becomes comparable to the line-width, this is the domain of Hanle regime. Originally defined as modification of the polarization of resonant scattering, Hanle effect has been known for decades. However, until recently (see Landi Degl'Innocenti 1999, Trujillo Bueno 1999), the alignment of the ground level and the effect of magnetic field on the ground level was not considered. Nevertheless, magnetic field is treated as a destructive factor to the ground level alignment. Magnetic field, in fact, realign the atoms towards (either $\|$ or $\bot$, see YLa) to the magnetic field in addition to reducing the degree of alignment on the ground state. The possibility of using the effect to diagnose the direction of magnetic field as well as the magnetic field strength is not well explored.

In this paper, using SII as an example, we provided a comparative study of Hanle effect with and without atomic alignment on the ground state. A notable effect of atomic alignment on the Hanle effect has been identified. The atomic alignment carries on additional information of the direction of magnetic field, which would be veiled otherwise. And quantitative results are provided for a generic geometry, which enables a possibility of a precise study of both magnetic field strength and its 3D direction. Indeed there are four parameters even for a known radiation source, $\theta_0$, $\theta_B,\,\phi_B$ and B field strength. 

For the study of magnetic field in circumstellar medium, the geometry of the system is clearer than in ISM, namely, the direction of radiation is well defined. Three lines (they can be from the same species) are thus adequate to get the whole information of magnetic field.

We also consider lower level Hanle regime, which occurs when the inverse of the lifetime on the ground level-the photon pumping rate ($9.8\times 10^4 (R_*/r)^2$,\, for pumping of S II $J_l\rightarrow J_u=5/2$ by an o star), becomes comparable to the magnetic precession rate $\nu_L\simeq 88(B/G)$. This can happen for a strong pumping source, e.g., quasars and supernovae, or a relatively weak magnetic field. In this situation, ground state is not aligned in respect to magnetic field. Neither the magnetic field nor the radiation field is a good quantization axis. As the result, coherence appears and the ground state density now resembles the upper state in the Hanle regime. {\em Absorption} from the ground state is modulated the same way as the emission in the Hanle regime. Both the degree and the direction of the polarization varies with the magnetic field. Unlike the solar case, {\em absorption} lines are readily available for circumstellar and interstellar medium. Furthermore, the upper level also gets the imprint of the magnetic field through excitation from the ground state. And therefore, the polarization of {\em emission} also carries the information of the magnetic field. Unfortunately, the polarization of emission in the lower level Hanle regime is not well separated with that in the Hanle regime in the polarization diagram. This entails a complexity for interpreting the observations. However, if we combine emission and absorptions, or different species, we should be able to disentangle the two regimes and obtain the information of the magnetic field and the environment.

{\em Absorption}  from atoms in the ground Hanle regime are, nevertheless, polarized in a {\em unique} way which can distinguish itself from the other two regimes. In the magnetic realignment regime or the Hanle regime, polarization of absorption can only be parallel or perpendicular to ${\bf B}$ in the plane of sky. In the ground Hanle regime, however, the polarization can be in any direction as the case for emission. Thus if the polarizations of multiplets are neither parallel nor perpendicular to each other, the region in study must be in the ground Hanle regime.  The polarization of {\em absorption} lines is thus more informative.

Note that in this paper we consider the region where radiation source is far enough to be treated as a point source. There is thus a substantial dilution of the radiation field so that the dilution factor $W\ll 1$. Thus $B{\bar J}^0_0/A=[J_u]/[J_l]W/(e^{h\nu/k_BT}-1)\ll 1$, meaning that the radiative pumping rate is thus much smaller  than the spontaneous emission rate. The two regimes, namely, Hanle regime ($\Gamma'\simeq 1$) and ground Hanle regime ($\Gamma'\simeq 1$) are well separated physically. For the same reason, we don't include forward scattering when treating the absorptions since the scattering intensity $I_{em}\propto \rho_u\propto {\bar J}^0_0$ is much smaller.

\subsection{Radio Emission Lines Influenced by Optical Pumping}

A new development of the theory here, which is considered neither in YLab, nor in Solar research, is the calculation of the magnetic
dipole transitions between the sublevels of the ground state, provided that the atoms are subject to {\it anisotropic} optical pumping. We provided calculations for OI transitions, but our approach is applicable to various atoms with hyperfine or fine splitting of levels. We showed that the emission arising from such atoms is polarized, which provides a new way of studying magnetic fields. For instance,
in YLb we studied the variations of 21-cm transmission arising from the alignment of HI atoms. Our present paper predicts that the
transmission within fine splitting of levels will be polarized. Apart from apparent galactic and extragalactic applications, this may be an interesting process to
study magnetic fields at the epoch of reionization, which hopefully will be available with the instruments are currently under construction. 

As for our study of OI pumping, we obtained, first of all, a factor of 4/3 correction for the $\Delta T/T_{cmb}$ and $\Delta I_\nu/B_\nu(T_{cmb})$ in the case of isotopic pumping of OI 63.2$\mu{\rm m}$ compared to early study by Hern\'andez-Monteagudo et al. (2007). We also showed in the realistic pumping by OI Balmer $\alpha$ lines, the anisotropy can introduce a factor of 2 variations in the resulting $\Delta T/T_{cmb}$ and $\Delta I_\nu/B_\nu(T_{cmb})$. Only when the angle between the line of sight and the incident radiation is $54.7^\circ$ (so-called Van Vleck angle), the results are coincident with those from isotropic pumping. These are important changes to be taken in modeling of the corresponding processes, as OI is being used to study processes in the early Universe. More importantly, we predicted
that the arising emission is polarized with high degree of polarization (up to 30 percent). This provides a unique opportunity to trace 
magnetic fields in the early Universe and also gauge the degree of reionization that arises from anisotropic optical pumping.

\subsection{Wide extend of alignment effects}

In our studies we were focused on new ways to study magnetic field that atomic alignment of the atoms/ions with fine and hyperfine structure provides. Being alignable in their ground state, these species can be realigned in weak magnetic fields, which is extremely good news for the studies of weak magnetic fields in astrophysical diffuse gas. The latter studies are currently very limited, with polarimetry based on grain
alignment  being the most widely used technique (see Whittet 2005 and ref. therein). However, in spite of the progress of grain alignment
theory (see Lazarian 2007 for a review), the quantitative studies of magnetic fields with aligned atoms are not always possible. Atoms, unlike
dust grains, have much better defined properties, which allows, for instance, tomography of magnetic fields by using different species\footnote{As we discussed in YLa long-lived alignable metastable states that are present for
some atomic species between upper and ground states may act as
proxies of ground states. The life time of the metastable level
may determine the distance from the source over which the atoms
are aligned being on metastable level. 
Absorption from such metastable levels
 can be used as diagnostics for magnetic field in the star vicinity.}, which would be differentially aligned at different distances from the source. In addition, ions can trace magnetic fields in the environments in which 
grains cannot survive. In fact, rather than compare advantages and disadvantages of different ways of studying magnetic fields, we would  stress the complementary nature of different ways of magnetic field studies. The subject is starved for both data and new approaches to getting the data. Note, that atomic realignment happens on the Larmor precession time, which potentially allows to
study the dynamics of fast variations of magnetic fields, e.g., related to MHD turbulence.

An incomplete list of objects where effects of alignment should be accounted for arises from our studies, which include this paper, as well as YLab. These include diffuse medium in the early Universe, quasars, AGNs, reflection nebulae, high and low density ISM, circumstellar regions, accretion disks and comets. One can easily add more astrophysical objects to this list. For instance, Io sodium tail can be studied the same way as sodium tail of comets. In general, in all environment when optical pumping is fast compared with the
collisional processes we expect to see effects of atomic alignment and magnetic realignment of atoms. The wide variety of atoms with fine and hyperfine structure of levels ensures multiple ways that the information can be obtained. Comparing information obtained through different species one can get deep insights into the physics of different astrophysical objects. If the implications of atomic alignment influenced the understanding of particular features of the Solar spectrum, then the studies of atomic alignment in a diffuse
astrophysical media can provide much deeper and yet unforeseen changes in our understanding of a wide variety of physical processes.

We note, that it is also important that some atoms do not demonstrate the effects of alignment.
 A comparison of the properties demonstrated by these atoms with those that 
demonstrate the alignment allows to gauge the importance of the effects of alignment.
In addition, we would like to stress that the effects of atomic alignment are not limited by tracing all important
astrophysical magnetic fields. As we discussed in YLab and this paper, 
the variations of the optical/UV absorptions and emissions induced by atomic alignment may be really appreciable and 
may substantially influence the interpretation of the absorption lines in terms of element abundances. Our present study
of effects of alignment for the radio lines further extends the range of the effects for which a key role can be 
played by atomic alignment. This proves that further study of the effects of atomic alignment are necessary if we
want to correctly interpret numerous pieces of observational data.



\section{Summary}

Our results can be briefly summarized as follows:

$\bullet$ Our calculations of more absorption lines arising from species aligned by weak magnetic field exhibit substantial degrees of polarization. Some of
the atoms exhibit polarization that can be measured by ground-based observations.

$\bullet$ Our calculations of the polarization of the emission lines arising from species aligned by weak magnetic field
show that such studies are promising for magnetic field studies in diffuse medium.

$\bullet$ We discuss the situation when both magnitude and the direction of weak magnetic fields in the diffuse gas can be studied.
We show that this happens when photon pumping rate becomes comparable to the magnetic precession rate, the coherence occurs in the ground level (ground state Hanle effect).

$\bullet$ Dealing with the stronger magnetic field regime, i.e. when upper level Hanle effect is important, we show that the polarizations of emission lines in the Hanle regime are affected by the ground state alignment and must be taken into account.  At the same time, we show that the polarizations of absorption lines remain the same in the Hanle regime as in the atomic alignment regime.

$\bullet$ Our calculations of the intensity ratio of scattered lines or absorption lines also varies in both upper level Hanle and lower level Hanle regime and therefore also carry the information about the direction of magnetic field.

$\bullet$ We show that combining Hanle effect, atomic alignment, and lower level Hanle effect allows one tomography of magnetic field for a variety of environments.

$\bullet$ We demonstrate that atomic alignment can influence the emission of radio lines arising from optical pumping. Therefore the anisotropic nature of the pumping radiation should be taken into account. 

$\bullet$ The polarization of the radio lines can be used to study magnetic fields in the situations when other technique fail, e.g. in 
the Universe at the epoch of reionization.

$\bullet$ With many species affected by atomic alignment and with a wide variety of effects that atomic alignment (and realignment) 
involve, we conclude that both the proper accounting for these effects is essential in modeling, and that studies of the consequences of the
alignment will open new ways to study physical processes in many astrophysical objects.

\begin{acknowledgments}
We are grateful to Ken Nordsieck for valuable comments and suggestions. HY is supported by CITA and the National Science and Engineering Research Council of Canada. AL is supported by the NSF Center for Magnetic Self-Organization
in Astrophysical and Laboratory Plasmas as well as the NSF grant AST 0507164.
\end{acknowledgments}

\appendix

\section{geometric relations in toroidal disk}

In a toroidal disk, the magnetic field is changing direction from place to place. A transformation of the angles from place to place is thus necessary. Basically, three angles are needed, two angles ($\theta,\phi$) (see Fig.\ref{radiageometry}{\em right}) specifying the direction of line of sight in the magnetic frame and the angle $\gamma$ defining the reference plane for the polarization.

\begin{figure}
\plottwo{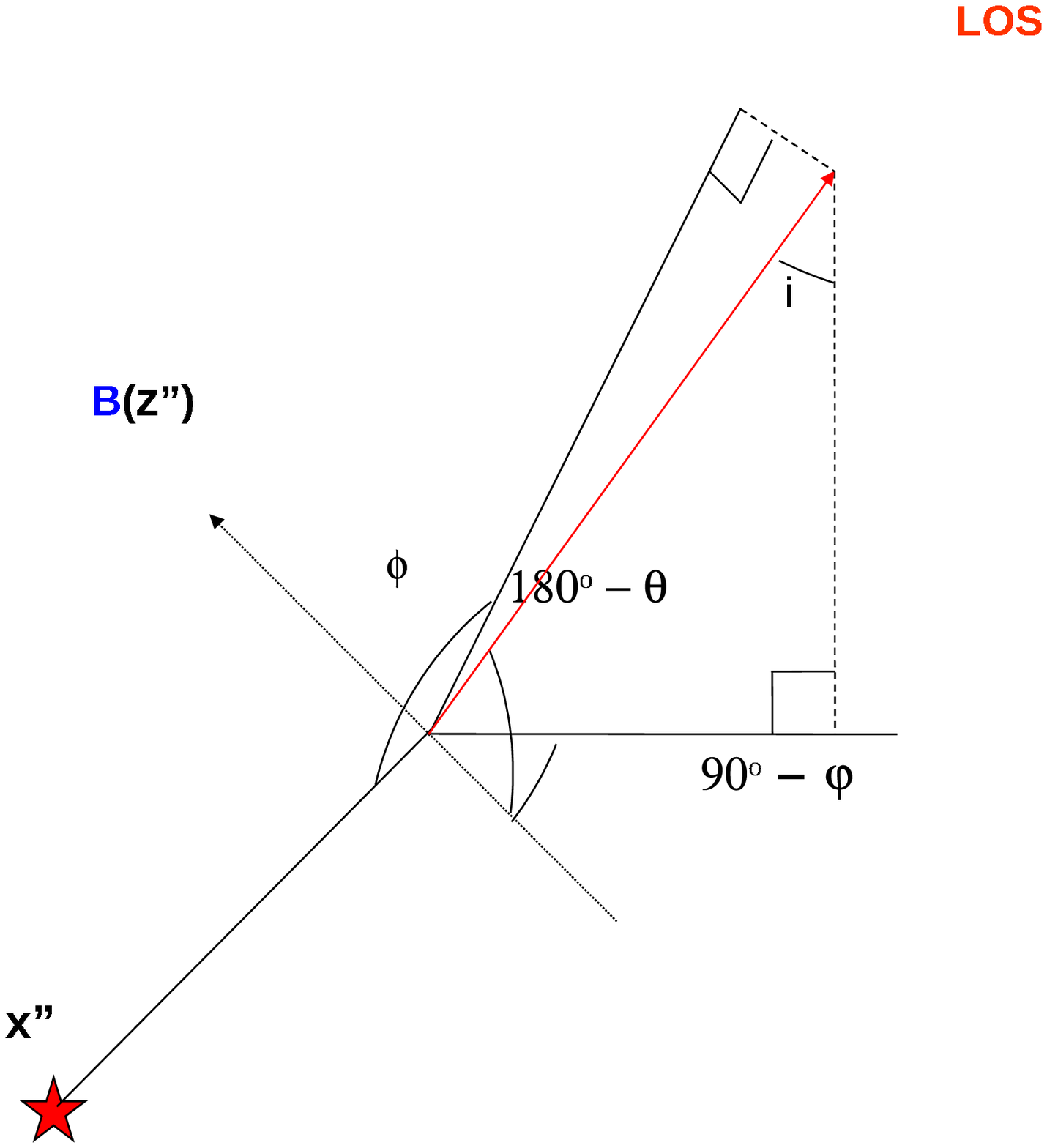}{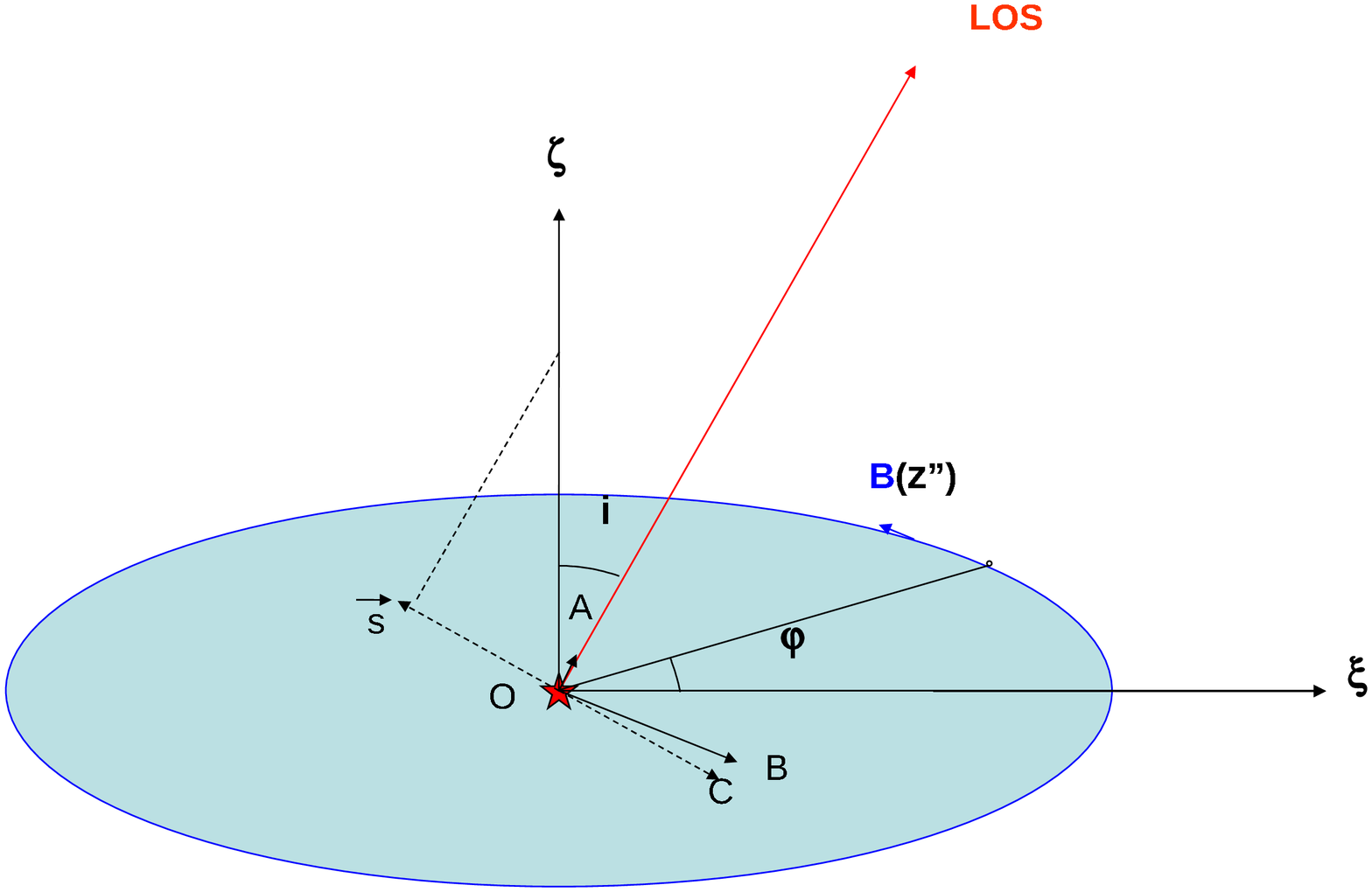}
\caption{{\em Left}: the direction of line of sight seen in the magnetic frame ($\theta,\phi$); {\em right}: $\vec{s}$ is the projection of the $\zeta$ axis in the plane of sky and $-\vec{OC}$ is the projection of magnetic field in the plane of sky. The angle $\gamma$ between the projections of the magnetic field and the disk symmetry axis $\zeta$ in the plane of sky is thus the angle between $\vec s$ and $-\vec{OC}$. }
\label{geometry}
\end{figure}

Since the line of sight is in the $\xi$-$\zeta$ plane (see Fig\ref{diskcart}{\em right}), which is perpendicular to the disk, the angle between the l.o.s. and magnetic field $\theta$ is related to the inclination angle $i$ and the angle $\varphi$ by this relation (see Fig.\ref{geometry}{\em left}) 

\be
\cos\theta=-\cos(90^\circ-i)\cos(90^\circ-\varphi).
\ee
Similarly it can be verified that the cos of the angle between the l.o.s. and $x¡±$ axis is
\be
\sin\theta\cos\phi=-\sin i\cos\varphi.
\ee
thus $\theta=\cos^{-1}(-\sin i\sin\varphi)$, and $\phi=\cos^{-1}(-\sin i\cos\varphi/\sin\theta)$. 

We choose the $\xi$-$\zeta$ plane defined by the l.o.s. and the symmetry axis of the disk as the reference plane for measuring the positional angle of the line polarization, the angle $\gamma$ between this plane and the magnetic field in the plane of sky (see Fig.\ref{radiageometry}{\em right}) is then equal to the angle between the projections of the symmetry axis and magnetic field in the plane of sky. As shown in Fig.\ref{geometry}{\em right}, the unit vector ${\vec s}=(-\cos i,0,\sin i )$ is the projection of the $\zeta$ axis in the plane of sky, $-\vec{OC}$ is the projected magnetic vector in the plane of sky. The angle $\gamma$ between the two vectors is 
\be
\cos^{-1}\left(-\frac{{\vec s}\cdot \vec{OC}}{|OC|}\right).
\label{scalar}
\ee
It can be seen that 
\be
\vec{OC}=\vec{OB}-\vec{OA},
\label{oc}
\ee
where $\vec{OB}$ is the unit vector anti-parallel to the magnetic field $\vec{OB}=(\sin\varphi,-\cos\varphi,0)$, and $\vec{OA}$ is its projection $\vec{OA}=\sin i\sin\varphi(\sin i,0,\cos i)$. Insert them and the expression of ${\vec s}$ into Eqs.(\ref{scalar},\ref{oc}), we obtain
\be
\gamma=\cos^{-1}\left(-\frac{\sin\varphi\cos i}{\sqrt{\sin^2\varphi\cos^2 i+\cos^2\varphi}}\right).  
\ee

\end{document}